\documentclass[pra,twocolumn,aps,showpacs,preprintnumbers,amsmath,amssymb,superscriptaddress]{revtex4-1}

\usepackage{xcolor}
\usepackage{graphicx}% Include figure files
\usepackage{dcolumn}% Align table columns on decimal point
\usepackage{bm}% bold math
\usepackage{amsmath}
\usepackage{cancel}
\usepackage{epstopdf}
\usepackage{tikz}
\usepackage{pgfplots}
\usepackage{makecell} 
\usepackage{booktabs}
\usepgfplotslibrary{fillbetween}
\usepackage{hyperref}
\usepackage{lineno}
\usepackage{dsfont}
\usepackage{upgreek}

\newcommand{\diff}{\mathrm{d}}
\newcommand{\colorDisable}[1]{}

\let\LN@align\align
\let\LN@endalign\endalign
\renewcommand{\align}{\linenomath\LN@align}
\renewcommand{\endalign}{\LN@endalign\endlinenomath}

\let\LN@gather\gather
\let\LN@endgather\endgather
\renewcommand{\gather}{\linenomath\LN@gather}
\renewcommand{\endgather}{\LN@endgather\endlinenomath}
\makeatother

\begin{document}

\title{Quantum-enhanced Doppler lidar}

\author{Maximilian Reichert}
\email[Corresponding author: ]{\qquad maximilian.reichert@ehu.eus}
\affiliation{Department of Physical Chemistry, University of the Basque Country UPV/EHU, Apartado 644, 48080 Bilbao, Spain}
\affiliation{EHU Quantum Center, University of the Basque Country UPV/EHU, Bilbao, Spain}

\author{Roberto Di Candia}
\affiliation{Department of Information and Communications Engineering, Aalto University, Espoo, 02150 Finland}
\affiliation{Department of Microtechnology and Nanoscience (MC2), Chalmers University of Technology, SE-41296 G\"oteborg, Sweden}

\author{Moe Z. Win}
\affiliation{Laboratory for Information and Decision Systems, Massachusetts Institute of Technology, Cambridge, MA 02139, USA}

\author{Mikel Sanz}
\email[Corresponding author: ]{\qquad mikel.sanz@ehu.eus}
\affiliation{Department of Physical Chemistry, University of the Basque Country UPV/EHU, Apartado 644, 48080 Bilbao, Spain}
\affiliation{EHU Quantum Center, University of the Basque Country UPV/EHU, Bilbao, Spain}
\affiliation{Basque Center for Applied Mathematics (BCAM), Alameda de Mazarredo 14, 48009 Bilbao, Spain}
\affiliation{IKERBASQUE, Basque Foundation for Science, Plaza Euskadi 5, 48009, Bilbao, Spain}

\begin{abstract}
We propose a quantum-enhanced lidar system to estimate a target's radial velocity which employs squeezed and frequency entangled signal and idler beams. We compare its performance against a classical protocol using a coherent state with the same pulse duration and energy, showing that quantum resources provide a precision enhancement in the estimation of the velocity of the object. We identify three distinct parameter regimes characterized by the amount of squeezing and frequency entanglement. In two of them, a quantum advantage exceeding the standard quantum limit is achieved assuming no photon losses. Additionally, we show that an optimal measurement to attain these results in the lossless case is frequency-resolved photon counting. Finally, we consider the effect of photon losses for the high-squeezing regime, which leads to a constant factor quantum advantage higher than $3$ dB in the variance of the estimator, given a roundtrip lidar-to-target-to-lidar transmissivity larger than $50\%$. 
\end{abstract}

\maketitle

\section{Introduction}
Quantum metrology exploits quantum mechanical resources, such as entanglement and squeezing, to measure a physical parameter with higher resolution than any strategy with classical resources. Many quantum metrology protocols in the photonic regime~\cite{pirandola2018} have been proposed such as quantum illumination (QI)~\cite{lloyd2008,Tan2008, barzanjeh2015,casariego2020,zhuangfading}, quantum enhanced position and velocity estimation~\cite{giovannetti2001, liu2019enhancing, maccone2020,zhuang2021ranging, zhuang2022ultimate,zhuang2017,huang2021}, quantum phase estimation~\cite{huang2016,lasheras2017},  transmission parameter estimation~\cite{woodworth2020transmission,woodworth2022transmission, spedalieri2018,spedalieri2020,shi2020}, noise estimation~\cite{pirandola2017ultimate}, and estimation of separation between objects~\cite{lupo2016ultimate,kose2021quantum}, among others. In these protocols, information about an object is retrieved by interrogating it with a signal beam. In the most general strategy, this signal is correlated or entangled with an idler beam, which is retained in the lab to perform a joint measurement at the end of the protocol. Indeed, the scheme can be seen as an interferometer setup, in which a channel depending on the parameter of interest is only applied to the signal mode.

Of particular interest for remote sensing applications is the QI protocol, where the aim is to detect the presence of a weakly reflecting target with an error probability smaller than using the best classical strategy. Here, a quantum advantage in the error probability exponent can be achieved by using a global measurement (up to $6$~dB)~\cite{Zhuang20171, Nair2020, DiCandia2021,pirandola2019fundamental,zhuang2020ultimate}, or by using  local measurements (up to $3$~dB)~\cite{Guha2009, Sanz2017, Jonsson2022}. This advantage is only achieved in a very noisy environment, such as the case of room-temperature microwave band, by a large bandwidth two-mode squeezed-vacuum state~\cite{Tan2008}. This requires a signal with a very low photon number per mode, which in the microwave regime is challenging to transmit open-air. Since amplifying the signal has been shown to break the quantum advantage~\cite{Shapiro2020,Jonsson2020}, QI as originally thought remains an elusive achievement so far,  even though recent progress has been made on relaxing the requirements for quantum advantage~\cite{zhuang2022}.

Once the presence of a target is established, properties like its location and velocity are also of interest. These can be estimated via signal arrival time and frequency measurement making use of the Doppler effect. Giovannetti, Lloyd and Maccone showed in~\cite{giovannetti2001}, that the GLM states, named after them, defined in the frequency domain, can attain the Heisenberg limit (HL), which is a $1/N$ scaling of the estimation error of the arrival time, where $N$ is the total number of photons. Equivalently, GLM states defined in the time domain reach the HL for the estimation error of frequency. This constitutes a quadratic improvement compared with the standard quantum limit (SQL)  achieved by the classical protocol.  In~\cite{zhuang2017}, the simultaneous estimation of location and radial velocity was considered using two GLM states in the frequency and time domain, respectively, that are transformed into two entangled signal and idler beams via a beam splitter. It was shown that the velocity and the location can simultaneously be estimated achieving the Heisenberg limit. This proves that frequency entanglement lifts the Arthurs-Kelly relation~\cite{arthurs1965}, which states that the location and velocity of an object can not be estimated with arbitrary precision using unentangled light. The work~\cite{huang2021} further extended this by addressing the simultaneous estimation of relative location and velocity of two targets by means of two-photon entangled states. The main drawbacks of these previous works are the use of two-photon states, which does not allow for a photon-number-dependent analysis, and the use of the multiphoton GLM states, which are non-normalizable and thus not physical. In~\cite{shapiro2007quantum}, a normalized version of the GLM state was introduced for range estimation. Here, it was shown that the Heisenberg scaling persists for the normalized version. However, these GLM-type states are fragile in lossy channels. The loss of a single photon renders the state useless for retrieving information about the parameter. Although the robustness against losses of these GLM-type states may be improved by reducing their entanglement, this comes at the cost of decreasing the enhancement in the scaling of the error estimation. Furthermore, it is challenging to produce GLM states in the laboratory for photon numbers $N>2$~\cite{maccone2020}.

In this article, we propose a protocol for a quantum Doppler lidar, which estimates the radial velocity of a reflecting object using quantum light. As a probe state,  frequency-entangled twin-beams are used. The signal beam is sent against the moving object, which causes a frequency shift due to the Doppler effect. Finally, a measurement of the returned signal and the idler is performed. We propose for the protocol  a  multimode probe state that can be generated by a parametric downconverter. The state is composed of photon pairs that share frequency entanglement. This photon-pair structure is resilient against losses, since the loss of a single photon only effects its partner, but not the other photon pairs. This is a crucial difference with GLM states, where the loss of a single photon means the loss of all the information about the parameter of interest due to the global entanglement. The quantum protocol is benchmarked against a classical protocol shining the object with the same energy and for the same time duration to make the comparison fair. We employ the Quantum Fisher information (QFI) as the figure of merit in the comparison, since it gives the maximal amount of extractable information about the parameter of interest. Calculating the QFI for this multimode state is challenging, but by using properties of Gaussian states and introducing Schmidt modes, which effectively discretizes the frequency-continuous problem, we derive an analytical expression for the QFI. Two quantum resources can be identified in our resource quantum state, namely, squeezing and frequency entanglement. The performance of the quantum protocol is studied as a function of the photon number in three different parameter regimes, called high-frequency entanglement, high-squeezing, and mixed regime. The latter, for which a remarkable Heisenberg scaling can be attained, is called in this manner because neither squeezing nor frequency entanglement are dominant. We propose a measurement setup that attains the QFI, consequently achieving the highest estimation accuracy of the velocity. It is noteworthy that the measurement setup can be performed separately in the signal and the idler, facilitating the experimental requirements.

The paper is structured as follows. In Section~\ref{sec:estimation} and \ref{sec:gaussianstates},  the fundamentals of quantum estimation theory  and Gaussian states are introduced. In Section~\ref{sec:model}, we model the moving target as a perfectly-reflective mirror boosted at a relative constant velocity. Afterwards in Sections~\ref{sec:classical} and~\ref{sec:quantum}, we introduce the probe states employed in both the quantum and classical protocols. As a figure of merit to benchmark their performance, we make use of the QFI. Then, in Section~\ref{sec:advantage}, we discuss the different parameter regimes obtained and study when quantum advantage exists and how it behaves as a function of the signal photon number. In Section \ref{sec:loss},  the protocol is studied in the presence of losses in the signal beam. Finally, in section~\ref{sec:loss} an optimal measurement attaining the ultimate precision set by the quantum Cram\'er-Rao bound is provided.

\section{Results}

\subsection{Model of the moving target}\label{sec:model}
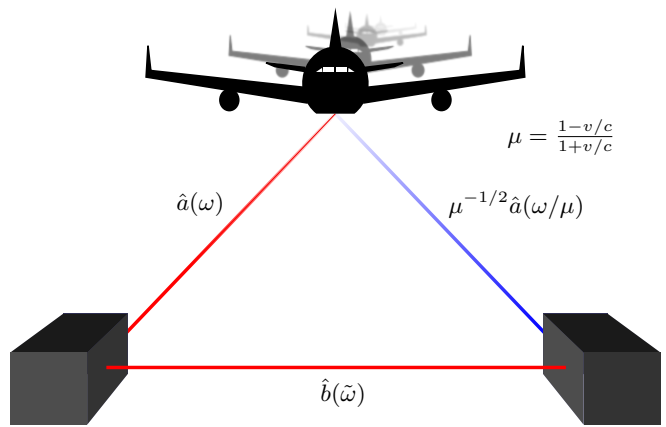
\begin{figure}
\centering
\begin{tikzpicture}
    \node[anchor=south west,inner sep=-1] (image) at (0,0) {\includegraphics[width=\columnwidth]{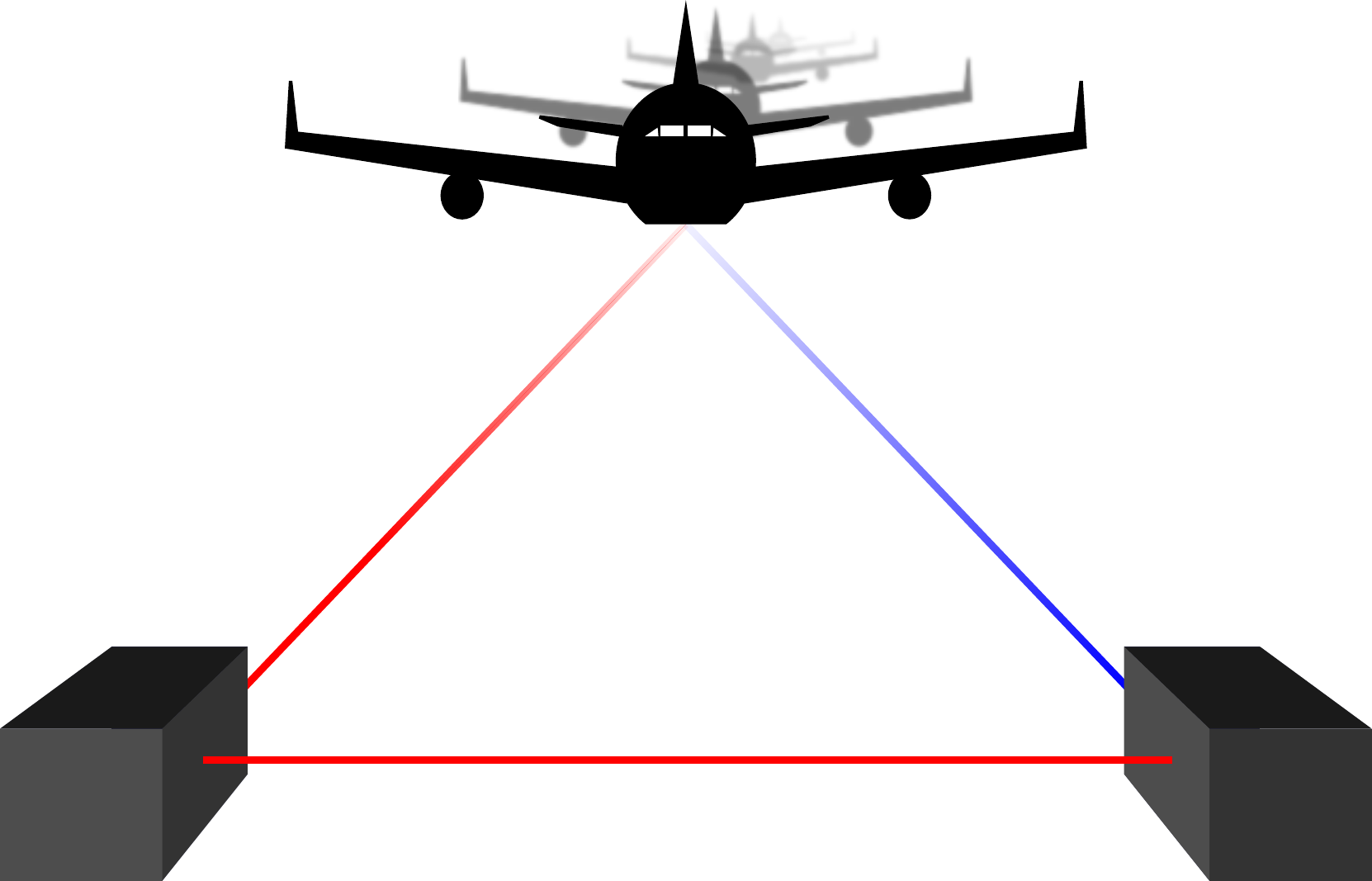}};
        \node at (4.1,4.2) {$v$};
         \node at (2.5,2.9) {$\hat{a}(\omega)$};
             \node at (4.4,0.4) {$\hat{b}(\tilde{\omega})$};
                      \node at (6.7,2.9) {$\mu^{-1/2} \hat{a}(\omega/\mu)$};
                      \node at (7.3,3.8) {$\mu = \frac{1-v/c}{1+v/c}$};
%    \begin{scope}[x={(image.south east)},y={(image.north west)}]
%%        \draw[help lines,xstep=.1,ystep=.1] (0,0) grid (1,1);
%      \node at (1,0.5) {yes};
%    \end{scope}
\end{tikzpicture}
\caption{Scheme of a quantum Doppler lidar. A twin-beam multimode squeezed vacuum state is produced by the transmitter on the bottom left. The signal beam is sent towards the moving target where it is reflected and its frequency Doppler shifted. The idler beam does not interact with the moving target and is retained. Both the reflected signal beam and the idler beam are measured at the receiver on the bottom right.} \label{fig:M1}
\end{figure}
We model the object of which we wish to estimate its constant radial velocity $v$ relative to emitter as a perfect mirror in a $(1+1)$-dimensional spacetime. For now, we assume an absence of noise and loss. As can be seen in Fig.~\ref{fig:M1}, the quantum Doppler lidar emits a signal beam towards the moving object, while also emitting an idler beam which is retained in the laboratory, such that a  measurement can be performed of the returned signal and the idler. The electromagnetic field of the signal beam obeys the wave equation $(\partial_t^2 - c^2\partial_x^2 ) \phi (t,x) = 0$, where $c$ is the speed of light and we only consider one polarisation of the field for the sake of simplicity. The presence of the target which is modelled as a perfect mirror imposes the boundary condition $\phi(t,x_m)=0$, where $x_m = v t$ is the location of the mirror. We assume the emitter to be to the right of the mirror, which corresponds to its spatial coordinate $> x_m$. The general solution of the wave equation satisfying the boundary condition  is given by
\begin{linenomath}
\begin{equation} \label{fieldoperator}
\phi (x,t) = \int_0^\infty \frac{\diff \omega}{\sqrt{4\uppi  \omega}} \left( \text{e}^{-\text{i}\omega(ct+x)}-\text{e}^{-\text{i} \frac{\omega}{\mu} (ct-x)} \right) a (\omega)  +h.c. 
\end{equation} 
\end{linenomath}
where $\mu = (1-v/c)/(1+v/c)$ is the Doppler parameter. We choose to estimate the parameter $\mu$ instead of $v$, as it naturally arises in the Doppler effect. The estimation error of $v$ is related to the one of $\mu$ via the error propagation formula for the QFI
$J(v)=(\partial_v \mu(v))^2 J(\mu(v))$. The Fourier coefficients  $a(\omega)$ and their complex conjugates get promoted by canonical quantization to annihilation  and creation operators,  which we denote by $\hat{a}(\omega)$ and $\hat{a}^\dag (\omega)$. They satisfy the relations $[\hat{a}(\omega), \hat{a}(\tilde{\omega})] = [\hat{a}^\dag(\omega), \hat{a}^\dag(\tilde{\omega})] = 0$ and $[\hat{a}(\omega),\hat{a}^\dagger (\tilde{\omega})]=\delta (\omega-\tilde{\omega})$. The idler frequency mode is referred to as $\hat{b}(\omega)$ and satisfies the same commutation relations. It commutes with the signal mode as both beams are spatially separated. The first term in Eq.~\eqref{fieldoperator} in brackets represents the incoming wave, while the second term is the outgoing wave which is Doppler shifted $\omega \rightarrow \omega/\mu$. 
Now, let us derive the Bogoliubov transformation $\hat{U}_\mu$ which maps the incoming modes $\hat{a}(\omega)$ to the Doppler reflected outgoing modes, denoted as $\hat{a}(- \omega)$. For this, a change of integration variables is performed in the second term in Eq.~\eqref{fieldoperator}, leading to
\begin{linenomath}
\begin{multline*}
\hat{\phi} (x,t) = \\ \int_0^\infty \frac{\diff \omega}{\sqrt{4\uppi \omega}} \left (\text{e}^{-\text{i}\omega(ct+x)} \hat{a}(\omega) + \text{e}^{-\text{i}\omega (ct-x)}  \hat{a}(-\omega)  +h.c. \right ),
\end{multline*}
\end{linenomath}
with the  operator $ \hat{a}(-\omega) \equiv -\mu^{1/2} \hat{a}(\mu \omega)$. Thus, the process of reflection is described by the unitary transformation $\hat{U}_\mu \hat{a}(\omega) \hat{U}_\mu^\dagger =- \mu^{-1/2} \hat{a}(\omega/\mu)$. The prefactor $\mu^{-1/2}$ ensures a proper normalization and the change of sign is the $\pi$ phase shift that radiation experiences when reflected. The vacuum state $|0\rangle$, which satisfies $\hat{a}(\omega)|0\rangle = \hat{b}(\omega)|0\rangle =0$, remains unchanged after Doppler reflection, that is $\hat{U}_\mu |0\rangle = |0\rangle$. In the most general framework, the outgoing mode also picks up a phase factor  $\exp(\text{i} 2\omega x_m / (c-v))$ depending on the velocity and location $x_m$ of the object. Therefore, this phase  could in principle also be used to estimate the velocity, but generally at the cost of an additional knowledge about the location.  Furthermore, in real world applications the phase often is randomized due to surface properties of the object and information about $v$ is lost. Hence, as a first step, we will neglect the information from the phase and we will only consider the information about the velocity that is encoded in the frequency spectrum of the light beams. The QFI $J_q$ derived here is a lower bound of the QFI in which phases are also taken into account. 

\subsection{Classical protocol}\label{sec:classical}
In the classical protocol we take a coherent signal as the probe state. For a continuum of frequency modes, a coherent state is defined as $|\psi\rangle = \exp[\alpha \int \diff \omega f(\omega) (\hat{a}(\omega)-\hat{a}^\dagger (\omega))]|0\rangle$, where  we take the displacement constant  $\alpha$ to be a real number for the sake of simplicity. The spectral amplitude $f(\omega)$ shall be an arbitrary differentiable and normalized function, i.e. $\int \diff \omega \vert f(\omega)\vert^2 = 1$. We assume that the carrier frequency $\omega_c = \int \diff \omega \vert f(\omega)\vert ^2 \omega$ is much larger than the bandwidth $\Delta \omega$ defined as $\Delta\omega^2 =\int \diff \omega \vert f(\omega)\vert^2 (\omega- \omega_c)^2$, the so-called narrow-bandwidth approximation. This allows us to change the limits of integration to $(-\infty,\infty)$. The reflected state is given by $\hat{U}_\mu |\psi \rangle = |\psi_\mu \rangle =  \exp[\alpha \int \diff \omega \mu^{1/2}f(\mu\omega) (\hat{a}^\dag(\omega)-\hat{a} (\omega))]|0\rangle$, where we have used $\hat{U}_\mu \text{e}^{\hat{A}}\hat{U}_\mu^\dag = \text{e}^{\hat{U}_\mu \hat{A} \hat{U}_\mu^\dag} = \text{e}^{\hat{A}_\mu}$, and $\hat{A}$ is the exponent of the coherent state. Thus, the state is still a coherent state after the reflection but with an amplitude $f(\omega) \rightarrow -\mu^{1/2} f(\mu\omega)$. The mean frequency is shifted to $\omega_c/\mu$ and the spectral bandwidth is stretched or compressed by a factor of $1/\mu$. Therefore, estimating the frequency and the variance provides information about the parameter $\mu$.  The calculation of the QFI is straightforward, we need to compute $|\partial_\mu \psi_\mu\rangle$. As $[\partial_\mu \hat{A}_\mu, \hat{A}_\mu] = 0$, we can write $| \partial_\mu \psi_\mu\rangle = \text{e}^{\hat{A}_\mu} \partial_\mu \hat{A}_\mu |0\rangle$. As a consequence, it follows that $\langle \psi_\mu | \partial_\mu \psi_\mu \rangle = 0$ and $\langle \partial_\mu \psi_\mu |\partial_\mu \psi_\mu \rangle = \langle 0| \partial_\mu \hat{A}_\mu^\dag \partial_\mu \hat{A}_\mu |0 \rangle$. This leads to the expression for the QFI 
\begin{align}
J_c (\mu) &= \frac{4\alpha^2}{\mu^2} \int \diff \omega \left( \frac{1}{2} f (\omega)+ \omega \partial_\omega f(\omega) \right)^2.  \label{QFIc}
\end{align}
This is the general expression for the QFI. Let us now make some approximations to gain physical insights. First, we consider small velocities compared to the speed of light $v/c \ll 1$, for which the frequency Doppler shift is approximately $2\omega_c v/c$.  
Let us now introduce the spectral amplitudes' Fourier transform $g(t) = \int \diff \omega f(\omega) \text{e}^{i \omega t}$. The time duration $\Delta T$ of the pulse is given by $\Delta T^2 =  \int \diff t \, t^2 \vert g(t)\vert^2 - \left( \int \diff t \, t \vert g(t)\vert^2 \right)^2 $. Using the further approximation $\Delta T \Delta \omega v/c \ll 1$, which is standard in the classical literature~\cite{trees}, we obtain
\begin{align}
J_c (\mu) \approx \frac{4}{\mu^2} \omega_c^2 N_c \Delta T^2 .
\end{align}
We see that the classical protocol follows the SQL scaling expected for a classical strategy. Furthermore, we note that three parameters completely define the optimal performance of a classical lidar, the photon number $\alpha^2 =N_c$, the carrier frequency $\omega_c$ and the time duration $\Delta T$ of the pulse.

\subsection{Quantum protocol} \label{sec:quantum}
For the quantum protocol,  we use a  twin-beam multimode squeezed vacuum state. This state  can be produced in the laboratory by non-linear optical processes, such as spontaneous parametric down-conversion (SPDC).  In this process of SPDC, a  pump beam, which is considered to be classical, interacts with a $\chi^{(2)}$ non-linear optical medium. Photons of the pump field decay into  signal and idler photon pairs. The use of a waveguide for SPDC allows for reducing the number of spatial modes to one for each beam~\cite{mosley2009,christ2009, christ2011,horn2012monolithic,francesconi2020} given by $\hat{a}(\omega)$ (signal) and $\hat{b}(\tilde{\omega})$ (idler). The effective Hamiltonian describing the process is given by~\cite{eckstein2011} 
\begin{equation}
\hat{H}_I = i \hbar \xi \int \diff \omega \int \diff \tilde{\omega} f(\omega,\tilde{\omega}) \hat{a}(\omega) \hat{b}(\tilde{\omega}) + h.c., \label{Hamilton}
\end{equation}
where the coupling constant $\xi$, referred to as the squeezing parameter, is chosen to be real for simplicity, and proportional to the intensity of the classical pump beam and the strength of the interaction. The normalized joint spectral amplitude $f(\omega,\tilde{\omega})$ depends on the specifics of the non-linear process and on the pump beam.  In the case of SPDC, the joint spectral amplitude can be in many cases approximated as a double Gaussian~\cite{merkouche2022} which also simplifies analytic calculations
\begin{multline}
f(\omega,\tilde{\omega}) = \sqrt{\frac{2}{\uppi \sigma\epsilon}} \exp \left(- \frac{(\omega+\tilde{\omega}-\omega_0)^2}{2\sigma^2} \right) \\
\times\exp \left( - \frac{(\omega-\tilde{\omega})^2}{2\epsilon^2} \right) . \label{doubleGaussian}
\end{multline}
The first exponential function in Eq.~\eqref{doubleGaussian} with argument $\omega+\tilde{\omega}$ comprises energy conservation of the photon decay process and it is inherited by the frequency mode spectrum of the pump beam, which is assumed to be Gaussian with mean frequency $\omega_0$ and variance $\sigma^2/2$. The second exponential function with argument $\omega-\tilde{\omega}$ corresponds to the phase matching condition, i.e. momentum conservation of the photon decay process, and depends on the spatial properties of the pump beam and the non-linear medium. Thus, by modifying the pump beam, both functions composing $f(\omega,\tilde{\omega})$ can independently be tailored~\cite{francesconi2020}. We again assume the narrow-bandwidth approximation $\omega_0 \gg \sigma$ and $\omega_0 \gg \epsilon$. The double Gaussian can be decomposed into its Schmidt modes~\cite{fedorov2009} as $f(\omega,\tilde{\omega}) = \sum_{n=0}^{\infty} r_n \psi_n (\omega-\omega_0/2) \psi_n(\tilde{\omega}-\omega_0/2)$, where $\lbrace \psi_n (\omega) \rbrace$ is an orthonormal set closely related to the Hermite functions (further details in Supplementary Material~\ref{App:doubleGaussian}). The relative weight $r_n^2$ of each individual mode is given by $r_n=\frac{2\sqrt{\sigma\epsilon}}{\sigma+\epsilon}(\frac{\sigma-\epsilon}{\sigma+\epsilon})^{n}$ with $\sum r_n^2 =1$. The number of active modes is given by the Schmidt number $K= (\sum_n r_n^4)^{-1} = \frac{\sigma^2+\epsilon^2}{2\sigma\epsilon}$, which we interpret as a measure of frequency entanglement within the signal and idler photon pair. For $K=1$, only one pair of modes is necessary to describe the state and the double Gaussian factorizes, that is $f(\omega,\tilde{\omega}) = \psi_0 (\omega-\omega_0/2) \psi_0(\tilde{\omega}-\omega_0/2)$, which implies no frequency entanglement. For $K>1$, the state is frequency entangled and the degree of entanglement grows monotonically with $K$. In Refs.~\cite{francesconi2020, xie2015},  techniques were proposed to generate Schmidt numbers  in the range of $K\sim 400-5000$, which corresponds to an extremely high-frequency entanglement of the photon pair. The Schmidt modes capture the spectral structure of $f(\omega,\tilde{\omega})$ in a discrete manner, and thus it is natural to introduce  discrete annihilation and creation operators $\hat{a}_n = \int \diff \omega \psi_n (\omega-\omega_0/2) \hat{a}(\omega)$ and $\hat{b}_n = \int \diff \omega \psi_n (\omega-\omega_0/2) \hat{b}(\omega)$ which are smeared out versions of $\hat{a}(\omega)$ and $\hat{b}(\tilde{\omega})$~\cite{blow1990}. The modes satisfy the commutation relations $[\hat{a}_n,\hat{a}_m]= [\hat{b}_n,\hat{b}_m] = [\hat{a}_n,\hat{b}^\dag_m] = 0$ and $[\hat{a}_n,\hat{a}_m^\dagger]=[\hat{b}_n,\hat{b}_m^\dagger]=\delta_{nm}$ due to the orthonormality of  $\lbrace\psi_n (\omega) \rbrace$. The discrete description of the problem substantially facilitates the calculation of the QFI. The Hamiltonian in Eq.~\eqref{Hamilton} is given in the discrete description by
\begin{linenomath}
\begin{equation}
\hat{H}_I = i \hbar \xi \sum_{n=0}^{\infty} r_n \left( \hat{a}_n \hat{b}_n - \hat{a}_n^\dagger \hat{b}_n^\dagger \right) \equiv i \hbar \xi \sum_{n=0}^{\infty} \hat{H}_n .
\end{equation}
\end{linenomath}
As the Hamiltonians for the individual modes commute $[\hat{H}_n,\hat{H}_m]=0$, the total squeezing operator $\hat{S}= e^{-i \hat{H}_I/\hbar}$ of the SPDC process can be written as a tensor product of squeezing operators for each individual mode $\hat{S} = \bigotimes_{n=0}^{\infty} \hat{S}_n$ with $\hat{S}_n = e^{\xi \hat{H}_n}$. The squeezing parameter of the squeezer corresponding to the mode $n$ is given by $\xi r_n$. Finally, we are able to express the probe state of the quantum protocol using discrete creation operators. Using the normal ordered representation of squeezing operators~\cite{barnett2002}, we find (see Supplementary Material~\ref{App:twinbeam} for details)
\begin{linenomath}
\begin{multline}
\hat{S}|0\rangle = \bigotimes_{n=0}^{\infty} \frac{1}{\cosh (\xi r_n)} \exp \left(- \tanh (\xi r_n) \hat{a}_n^\dagger \hat{b}_n^\dagger \right)|0\rangle . \label{probestate}
\end{multline}
\end{linenomath}
Thus, the twin-beam multimode squeezed vacuum state is just the product state of independent two-mode squeezed vacuum states. Now, the reflected state $|\psi_\mu\rangle = \hat{U}_\mu \hat{S} |0\rangle$ is 
\begin{equation} \label{refelectedQstate}
|\psi_\mu\rangle = \mathcal{N} \exp \left(- \sum_{n=0}^\infty \tanh (\xi r_n) \hat{a}_{n\mu}^\dagger \hat{b}_n^\dagger \right)|0\rangle ,
\end{equation}
where we have transformed the product in Eq.~\eqref{probestate} into a sum in the exponent and we have introduced the normalization constant $ \mathcal{N} = \prod_n 1/\cosh (\xi r_n)$, which is independent of $\mu$. The operator $\hat{a}_n^\dag$ transforms into $\hat{U}_\mu \hat{a}_n \hat{U}_\mu^\dag =\hat{a}_{n\mu}^\dagger  =- \int \diff \omega \mu^{1/2} \psi_n (\mu \omega-\omega_0/2) \hat{a}^\dagger(\omega)$, picking up a phase shift and a $\mu$-dependence, whereas $\xi$, $r_n$, and the idler modes $\hat{b}_n$ remain $\mu$-independent. The mean frequency of the transformed mode is given by $\omega_0/2\mu = \overline{\omega}$, as one would expect from the Doppler effect. The bandwidth of each mode is proportional to $ \sqrt{\sigma \epsilon/2 }$, and it transforms into $\sqrt{\sigma \epsilon/ 2\mu }  \equiv \overline{\sigma}$ after the reflection. In the continuous formalism, the joint spectral amplitude converts into $f(\omega,\tilde{\omega}) \rightarrow - \mu^{1/2} f(\mu\omega,\tilde{\omega})$.  Now, in order to calculate the QFI, we need to first evaluate the derivative $|\partial_\mu \psi_\mu\rangle$. The only component of the state that depends on $\mu$ is $\hat{a}_{n \mu}^\dagger$. The derivative can be calculated using the properties of the Hermite functions and we find that $\partial_{\mu} \hat{a}_{n \mu}^\dagger$ is a linear combination of creation operators $\hat{a}_{n\mu}^\dagger$ ranging from modes $n-2$ to $n+2$. As the derivative of the exponent in Eq.~\eqref{refelectedQstate} commutes with the exponent itself, we find $
|\partial_\mu \psi_\mu\rangle = - \sum_n \tanh(\xi r_n) (\partial_\mu \hat{a}_n^\dagger) \hat{b}_n^\dagger S |0\rangle$, see Supplementary Material~\ref{App:QFIquantum}. By using the transformation rule $\hat{S}^\dag \hat{a}_{n\mu} \hat{S} = \hat{a}_{n\mu} \cosh (\xi r_n) - \hat{b}_{n}^\dag \sinh (\xi r_n)$ and the analogous rule for the idler mode, whose derivation is discussed in Supplementary Material~\ref{App:transformationrules}, we finally find the analytic expression for the QFI (see Supplementary Material~\ref{App:QFIquantum} for the full derivation). This splits up into frequency and mode-bandwidth contributions as 
$J_q(\mu) = (\partial_\mu \overline{\omega})^2 J_q (\overline{\omega}) +(\partial_\mu \overline{\sigma})^2 J_q (\overline{\sigma})$ with  
\begin{equation}
J_q (\mu) = \frac{1}{\mu^2} \frac{\omega_0^2}{\sigma\epsilon} \left(Z_{\overline{\omega}} + \frac{\sigma\epsilon}{\omega_0^2}Z_{\overline{\sigma}} \right), 
\label{Jq}
\end{equation}
with the frequency term defined as
\begin{multline}
Z_{\overline{\omega}}=\sum_{n=0}^{\infty} \sinh^2 (\xi r_n) \Big( n \cosh^2 (\xi r_{n-1})  \\+ (n+1)\cosh^2 (\xi r_{n+1}) \Big) 
\end{multline}
and the mode-bandwidth term as
\begin{multline}
Z_{\overline{\sigma}} =\sum_{n=0}^{\infty} \sinh^2 (\xi r_n) \Big( n(n-1) \cosh^2 (\xi r_{n-2}) \\ + (n+1)(n+2) \cosh^2 (\xi r_{n+2}) \Big) .
\end{multline}
The bandwidth contribution is suppressed by the factor $\sigma \epsilon /\omega_0^2$ as can be seen in Eq.~\eqref{Jq}, which is small due to the narrow-bandwidth approximation. For a typical SPDC process in potassium dihydrogen phosphate crystal pumped by a frequency doubled titanium-sapphire laser, this factor is approximately $\sqrt{\sigma \epsilon / \omega_0^2} \sim 0.01$~\cite{davis2017}.
\subsection{A fair comparison}\label{sec:advantage}
Let us now compare the performance of the quantum and the classical protocols and find out under which conditions quantum advantage is achieved.  For that, we examine the quantum advantage ratio $ J_q / J_c $, where we have omitted the dependence on $\mu$ for the sake of readability. In the case $J_q / J_c  > 1$, the quantum strategy outperforms the classical one assuming that an optimal measurement is performed and the Cram\'er-Rao bound is attained, which is usually the case in the absence of thermal noise photons. We already pointed out in Sec. \ref{sec:classical}. that the classical lidar is solely characterized by the three parameters photon number, carrier frequency and time duration. So to fairly compare both protocols, we set these three parameters equal for both signal beams. For photon number and mean frequency, this corresponds to $\omega_c = \omega_0/2$ and $\alpha^2 = \sum_n \sinh^2 (\xi r_n)$. Now, let us calculate the time duration of the quantum signal beam. For that we introduce the time-domain version of the creation and annihilation operators via $\hat{E}^\dag(t) = \int \diff \omega\,  \text{e}^{\text{i}\omega t} \hat{a}^\dag (\omega)$, which is the operator creating a photon at time $t$ at the transmitter. The normalized power of the signal beam is defined as $\vert s(t) \vert^2 = \langle \psi\vert \hat{E}^\dag(t) \hat{E}(t) \vert \psi\rangle / N_S$. The time duration can then be calculated and we find
\begin{align}
\Delta T^2 &= \int \diff t \, t^2 \vert s(t)\vert^2 - \left( \int \diff t \, t \vert s(t)\vert^2 \right)^2 \\
 &= \frac{2}{\sigma\epsilon} \left( \frac{\sum_{n=0}^\infty \sinh^2 (\xi r_n) n}{\sum_{m=0}^\infty \sinh^2 (\xi r_m)} + \frac{1}{2} \right) .
\end{align}
The detailed calculations can be found in Supplementary Material~\ref{supp:time}. As both $J_q$ and $\Delta T$ are given by infinite series containing hyperbolic trigonometric functions, we will in the following study parameter regimes in which simple analytic expression for the respective quantities can be obtained, which helps to interpret the results.

\subsection{No frequency entanglement} 
Let us first study the case in which no frequency entanglement is present between signal and idler beams. In this case,  we have $K=1$, i.e. $\sigma = \epsilon$. The state reduces to the well-known two-mode  squeezed vacuum state $|\psi_\mu\rangle = \exp (\xi (\hat{a}_{0\mu}\hat{b}_0-\hat{a}_{0\mu}^\dag \hat{b}_0^\dag))|0\rangle$ with signal photon number $N_S= \sinh^2 (\xi)$. We find that
\begin{equation}
\frac{J_q}{J_c} = 1
\end{equation}
for all values of the squeezing parameter $\xi$ (Supplementary Material~\ref{App:QFInoEntanglement}). Thus, no quantum advantage is achieved with a two-mode squeezed vacuum state. Both protocols obey the SQL $J_q, J_c \sim N_S$.  In similar interferometric phase estimation protocols, Heisenberg scaling is achieved with the two-mode squeezed state. But due to our ignorance of the target's position $x_m$, the information about the velocity contained in the phase $\exp(\text{i} 2\omega x_m /(c-v))$ cannot be accessed and thus Heisenberg scaling is not achievable in our case. Thus, frequency entanglement $K>1$ is necessary for quantum advantage in our protocol with pure probe states given in Eq.~\eqref{probestate}. 
\subsection{High frequency-entanglement regime}
Let us now consider the case in which the frequency entanglement is the dominant quantum resource. We specify this regime by the condition  $\xi \ll  K^{1/2}$, which allows us to approximate the hyperbolic functions as $\sinh^2(\xi r_n) \approx \xi^2 r_n^2$ and $\cosh^2(\xi r_n) \approx 1$. The number of photons  in mode $n$ is given by $N_{Sn} = \sinh^2 (\xi r_n) \ll 1$ and the total photon number can be approximated as $N_S \approx \xi^2$, where we have only taken the first term of the approximation into account (Supplementary Material~\ref{App:photonnumber}). With this, the ratio of QFIs is
\begin{equation}
\frac{J_q}{J_c} \approx 1 + \frac{\sigma^2 +\epsilon^2}{2\omega_0^2} \left( 1+ \frac{1}{K^2} + \frac{1}{K(K^2+K)} \right). 
\end{equation}
The first term is the frequency contribution and is equal to $1$. The remaining terms correspond to the bandwidth contribution
which is small due to the narrow bandwidth approximation. Thus, in the high-frequency entanglement regime no  quantum advantage can be obtained, the classical and quantum protocol perform equally well. With the further constraint $\xi \ll 1$, the state becomes a superposition of the vacuum and a two-photon state, the same state used in Ref.~\cite{zhuang2017,huang2021}. Even though these states yield no quantum advantage in estimating the velocity alone, they yield  advantage in jointly estimating the position and velocity of a target.

\subsection{High-squeezing regime}\label{sec:high-squeezing}
\begin{figure}
\begin{center}
\begin{tikzpicture}
\pgfplotsset{%
    width=0.9\linewidth,
    height=0.9\linewidth}
  \begin{axis} [ylabel near ticks,ytick={0.5,1},legend pos=north east,domain=0:200,ymin=0, ymax= 1.3, xlabel = $\xi$, ylabel = $8 Z_{\overline{\omega},\overline{\sigma}}/(4 N_S)^{1+\sqrt{\frac{K-1}{K+1}}} $, xmax=200, xmin=0,legend style={empty legend} ]
  \addlegendentry{\tikz\draw[blue,fill=blue] (0,0) circle (.5ex); $K=10 $}
  \addlegendentry{\tikz\draw[red,fill=red] (0,0) circle (.5ex); $K=20$}
%K=10
  \addplot [smooth,mark=none,thick, smooth, blue] coordinates{(1., 5.7286) (1.5, 2.77006) (2., 1.66891) (2.5, 1.14172) (3., 0.85265) (3.5, 0.681354) (4., 0.575556) (4.5, 0.509309) (5., 0.468378) (5.5, 0.44432) (6., 0.431804) (6.5, 0.427306) (7., 0.428409) (7.5, 0.43342) (8., 0.441135) (8.5, 0.450696) (9., 0.461489) (9.5, 0.473075) (10., 0.485145) (10.5, 0.497483) (11., 0.509938) (11.5, 0.522404) (12., 0.534811) (12.5, 0.547111) (13., 0.55927) (13.5, 0.571267) (14., 0.583086) (14.5, 0.594718) (15., 0.606156) (15.5, 0.617395) (16., 0.62843) (16.5, 0.63926) (17., 0.649882) (17.5, 0.660295) (18., 0.670496) (18.5, 0.680486) (19., 0.690264) (19.5, 0.699828) (20., 0.709179) (20.5, 0.718316) (21., 0.727241) (21.5, 0.735953) (22., 0.744452) (22.5, 0.752742) (23., 0.760821) (23.5, 0.768693) (24., 0.776358) (24.5, 0.783819) (25., 0.791078) (25.5, 0.798137) (26., 0.804999) (26.5, 0.811667) (27., 0.818143) (27.5, 0.824431) (28., 0.830533) (28.5, 0.836454) (29., 0.842195) (29.5, 0.847761) (30., 0.853156) (30.5, 0.858382) (31., 0.863444) (31.5, 0.868345) (32., 0.873089) (32.5, 0.877679) (33., 0.88212) (33.5, 0.886414) (34., 0.890567) (34.5, 0.89458) (35., 0.898459) (35.5, 0.902206) (36., 0.905826) (36.5, 0.909322) (37., 0.912697) (37.5, 0.915955) (38., 0.9191) (38.5, 0.922134) (39., 0.925061) (39.5, 0.927885) (40., 0.930608) (40.5, 0.933234) (41., 0.935765) (41.5, 0.938205) (42., 0.940557) (42.5, 0.942824) (43., 0.945008) (43.5, 0.947112) (44., 0.949139) (44.5, 0.951091) (45., 0.952971) (45.5, 0.954781) (46., 0.956524) (46.5, 0.958202) (47., 0.959817) (47.5, 0.961372) (48., 0.962868) (48.5, 0.964309) (49., 0.965694) (49.5, 0.967028) (50., 0.96831) (50.5, 0.969544) (51., 0.970732) (51.5, 0.971873) (52., 0.972972) (52.5, 0.974028) (53., 0.975044) (53.5, 0.976021) (54., 0.97696) (54.5, 0.977863) (55., 0.978731) (55.5, 0.979566) (56., 0.980369) (56.5, 0.98114) (57., 0.981882) (57.5, 0.982595) (58., 0.98328) (58.5, 0.983938) (59., 0.984571) (59.5, 0.985179) (60., 0.985764) (60.5, 0.986326) (61., 0.986866) (61.5, 0.987384) (62., 0.987883) (62.5, 0.988362) (63., 0.988822) (63.5, 0.989264) (64., 0.989689) (64.5, 0.990097) (65., 0.990489) (65.5, 0.990865) (66., 0.991227) (66.5, 0.991575) (67., 0.991909) (67.5, 0.99223) (68., 0.992538) (68.5, 0.992834) (69., 0.993118) (69.5, 0.993391) (70., 0.993653) (70.5, 0.993905) (71., 0.994147) (71.5, 0.99438) (72., 0.994603) (72.5, 0.994817) (73., 0.995023) (73.5, 0.995221) (74., 0.995411) (74.5, 0.995593) (75., 0.995769) (75.5, 0.995937) (76., 0.996098) (76.5, 0.996254) (77., 0.996403) (77.5, 0.996546) (78., 0.996683) (78.5, 0.996815) (79., 0.996942) (79.5, 0.997063) (80., 0.99718) (80.5, 0.997293) (81., 0.9974) (81.5, 0.997504) (82., 0.997603) (82.5, 0.997699) (83., 0.99779) (83.5, 0.997878) (84., 0.997963) (84.5, 0.998044) (85., 0.998122) (85.5, 0.998197) (86., 0.998268) (86.5, 0.998337) (87., 0.998404) (87.5, 0.998467) (88., 0.998528) (88.5, 0.998587) (89., 0.998643) (89.5, 0.998697) (90., 0.998749) (90.5, 0.998799) (91., 0.998847) (91.5, 0.998893) (92., 0.998937) (92.5, 0.998979) (93., 0.99902) (93.5, 0.999059) (94., 0.999097) (94.5, 0.999133) (95., 0.999167) (95.5, 0.9992) (96., 0.999232) (96.5, 0.999263) (97., 0.999292) (97.5, 0.99932) (98., 0.999348) (98.5, 0.999374) (99., 0.999399) (99.5, 0.999423) (100., 0.999446) (100.5, 0.999468) ((101., 0.999489) (101.5, 0.999509) (102., 0.999529) (102.5, 0.999548) (103., 0.999566) (103.5, 0.999583) (104., 0.9996) (104.5, 0.999616) (105., 0.999631) (105.5, 0.999646) (106., 0.99966) (106.5, 0.999673) (107., 0.999686) (107.5, 0.999699) (108., 0.999711) (108.5, 0.999722) (109., 0.999733) (109.5, 0.999744) (110., 0.999754) (110.5, 0.999764) (111., 0.999774) (111.5, 0.999783) (112., 0.999791) (112.5, 0.9998) (113., 0.999808) (113.5, 0.999815) (114., 0.999823) (114.5, 0.99983) (115., 0.999836) (115.5, 0.999843) (116., 0.999849) (116.5, 0.999855) (117., 0.999861) (117.5, 0.999867) (118., 0.999872) (118.5, 0.999877) (119., 0.999882) (119.5, 0.999887) (120., 0.999891) (120.5, 0.999895) (121., 0.9999) (121.5, 0.999904) (122., 0.999908) (122.5, 0.999911) (123., 0.999915) (123.5, 0.999918) (124., 0.999921) (124.5, 0.999925) (125., 0.999928) (125.5, 0.99993) (126., 0.999933) (126.5, 0.999936) (127., 0.999938) (127.5, 0.999941) (128., 0.999943) (128.5, 0.999946) (129., 0.999948) (129.5, 0.99995) (130., 0.999952) (130.5, 0.999954) (131., 0.999956) (131.5, 0.999957) (132., 0.999959) (132.5, 0.999961) (133., 0.999962) (133.5, 0.999964) (134., 0.999965) (134.5, 0.999967) (135., 0.999968) (135.5, 0.999969) (136., 0.99997) (136.5, 0.999972) (137., 0.999973) (137.5, 0.999974) (138., 0.999975) (138.5, 0.999976) (139., 0.999977) (139.5, 0.999978) (140., 0.999979) (140.5, 0.999979) (141., 0.99998) (141.5, 0.999981) (142., 0.999982) (142.5, 0.999983) (143., 0.999983) (143.5, 0.999984) (144., 0.999985) (144.5, 0.999985) (145., 0.999986) (145.5, 0.999986) (146., 0.999987) (146.5, 0.999987) (147., 0.999988) (147.5, 0.999988) (148., 0.999989) (148.5, 0.999989) (149., 0.99999) (149.5, 0.99999) (150., 0.999991) (150.5, 0.999991) (151., 0.999991) (151.5, 0.999992) (152., 0.999992) (152.5, 0.999992) (153., 0.999993) (153.5, 0.999993) (154., 0.999993) (154.5, 0.999993) (155., 0.999994) (155.5, 0.999994) (156., 0.999994) (156.5, 0.999994) (157., 0.999995) (157.5, 0.999995) (158., 0.999995) (158.5, 0.999995) (159., 0.999995) (159.5, 0.999996) (160., 0.999996) (160.5, 0.999996) (161., 0.999996) (161.5, 0.999996) (162., 0.999996) (162.5, 0.999997) (163., 0.999997) (163.5, 0.999997) (164., 0.999997) (164.5, 0.999997) (165., 0.999997) (165.5, 0.999997) (166., 0.999997) (166.5, 0.999998) (167., 0.999998) (167.5, 0.999998) (168., 0.999998) (168.5, 0.999998) (169., 0.999998) (169.5, 0.999998) (170., 0.999998) (170.5, 0.999998) (171., 0.999998) (171.5, 0.999998) (172., 0.999998) (172.5, 0.999998) (173., 0.999999) (173.5, 0.999999) (174., 0.999999) (174.5, 0.999999) (175., 0.999999) (175.5, 0.999999) (176., 0.999999) (176.5, 0.999999) (177., 0.999999) (177.5, 0.999999) (178., 0.999999) (178.5, 0.999999) (179., 0.999999) (179.5, 0.999999) (180., 0.999999) (180.5, 0.999999) (181., 0.999999) (181.5, 0.999999) (182., 0.999999) (182.5, 0.999999) (183., 0.999999) (183.5, 0.999999) (184., 0.999999) (184.5, 0.999999) (185., 0.999999) (185.5, 0.999999) (186., 0.999999) (186.5, 1.) (187., 1.) (187.5, 1.) (188., 1.) (188.5, 1.) (189., 1.) (189.5, 1.) (190., 1.) (190.5, 1.) (191., 1.) (191.5, 1.) (192., 1.) (192.5, 1.) (193., 1.) (193.5, 1.) (194., 1.) (194.5, 1.) (195., 1.) (195.5, 1.) (196., 1.) (196.5, 1.) (197., 1.) (197.5, 1.) (198., 1.) (198.5, 1.) (199., 1.) (199.5, 1.) (200., 1.)
  };    
  %K=20
    \addplot [smooth,mark=none, thick, smooth, red] coordinates{(1., 10.7106) (1.5, 4.96154) (2., 2.88084) (2.5, 1.8966) (3., 1.35495) (3.5, 1.02678) (4., 0.814641) (4.5, 0.671211) (5., 0.571192) (5.5, 0.500003) (6., 0.448727) (6.5, 0.411637) (7., 0.384895) (7.5, 0.365843) (8., 0.35258) (8.5, 0.343719) (9., 0.338224) (9.5, 0.335309) (10., 0.334376) (10.5, 0.334959) (11., 0.336701) (11.5, 0.339322) (12., 0.342606) (12.5, 0.346385) (13., 0.35053) (13.5, 0.354941) (14., 0.359542) (14.5, 0.364276) (15., 0.369099) (15.5, 0.373978) (16., 0.378891) (16.5, 0.38382) (17., 0.388752) (17.5, 0.393678) (18., 0.398593) (18.5, 0.403493) (19., 0.408375) (19.5, 0.413239) (20., 0.418082) (20.5, 0.422906) (21., 0.427711) (21.5, 0.432496) (22., 0.437263) (22.5, 0.442012) (23., 0.446743) (23.5, 0.451457) (24., 0.456155) (24.5, 0.460836) (25., 0.465502) (25.5, 0.470152) (26., 0.474787) (26.5, 0.479407) (27., 0.484011) (27.5, 0.488601) (28., 0.493175) (28.5, 0.497735) (29., 0.502279) (29.5, 0.506808) (30., 0.511321) (30.5, 0.515818) (31., 0.5203) (31.5, 0.524765) (32., 0.529214) (32.5, 0.533646) (33., 0.53806) (33.5, 0.542458) (34., 0.546838) (34.5, 0.5512) (35., 0.555543) (35.5, 0.559868) (36., 0.564174) (36.5, 0.56846) (37., 0.572727) (37.5, 0.576974) (38., 0.581201) (38.5, 0.585407) (39., 0.589592) (39.5, 0.593755) (40., 0.597898) (40.5, 0.602018) (41., 0.606116) (41.5, 0.610192) (42., 0.614245) (42.5, 0.618275) (43., 0.622281) (43.5, 0.626264) (44., 0.630224) (44.5, 0.634159) (45., 0.63807) (45.5, 0.641956) (46., 0.645817) (46.5, 0.649654) (47., 0.653465) (47.5, 0.657251) (48., 0.661011) (48.5, 0.664745) (49., 0.668454) (49.5, 0.672136) (50., 0.675792) (50.5, 0.679421) (51., 0.683024) (51.5, 0.686601) (52., 0.69015) (52.5, 0.693672) (53., 0.697168) (53.5, 0.700636) (54., 0.704077) (54.5, 0.707491) (55., 0.710877) (55.5, 0.714236) (56., 0.717567) (56.5, 0.72087) (57., 0.724146) (57.5, 0.727394) (58., 0.730615) (58.5, 0.733808) (59., 0.736973) (59.5, 0.74011) (60., 0.74322) (60.5, 0.746302) (61., 0.749356) (61.5, 0.752382) (62., 0.755381) (62.5, 0.758352) (63., 0.761296) (63.5, 0.764212) (64., 0.7671) (64.5, 0.769961) (65., 0.772794) (65.5, 0.775601) (66., 0.77838) (66.5, 0.781131) (67., 0.783856) (67.5, 0.786554) (68., 0.789224) (68.5, 0.791868) (69., 0.794485) (69.5, 0.797076) (70., 0.79964) (70.5, 0.802177) (71., 0.804688) (71.5, 0.807174) (72., 0.809633) (72.5, 0.812066) (73., 0.814473) (73.5, 0.816855) (74., 0.819211) (74.5, 0.821542) (75., 0.823847) (75.5, 0.826128) (76., 0.828383) (76.5, 0.830614) (77., 0.83282) (77.5, 0.835002) (78., 0.837159) (78.5, 0.839293) (79., 0.841402) (79.5, 0.843487) (80., 0.845549) (80.5, 0.847587) (81., 0.849602) (81.5, 0.851594) (82., 0.853563) (82.5, 0.855509) (83., 0.857432) (83.5, 0.859333) (84., 0.861212) (84.5, 0.863068) (85., 0.864903) (85.5, 0.866716) (86., 0.868507) (86.5, 0.870277) (87., 0.872026) (87.5, 0.873753) (88., 0.87546) (88.5, 0.877146) (89., 0.878812) (89.5, 0.880458) (90., 0.882083) (90.5, 0.883689) (91., 0.885274) (91.5, 0.88684) (92., 0.888387) (92.5, 0.889915) (93., 0.891424) (93.5, 0.892914) (94., 0.894385) (94.5, 0.895838) (95., 0.897272) (95.5, 0.898689) (96., 0.900087) (96.5, 0.901468) (97., 0.902831) (97.5, 0.904177) (98., 0.905506) (98.5, 0.906818) (99., 0.908113) (99.5, 0.909391) (100., 0.910653) (100.5, 0.911898) (101., 0.913128) (101.5, 0.914341) (102., 0.915539) (102.5, 0.916721) (103., 0.917888) (103.5, 0.919039) (104., 0.920175) (104.5, 0.921297) (105., 0.922404) (105.5, 0.923496) (106., 0.924573) (106.5, 0.925637) (107., 0.926686) (107.5, 0.927721) (108., 0.928743) (108.5, 0.929751) (109., 0.930745) (109.5, 0.931726) (110., 0.932695) (110.5, 0.93365) (111., 0.934592) (111.5, 0.935521) (112., 0.936438) (112.5, 0.937343) (113., 0.938235) (113.5, 0.939116) (114., 0.939984) (114.5, 0.940841) (115., 0.941686) (115.5, 0.942519) (116., 0.943341) (116.5, 0.944152) (117., 0.944952) (117.5, 0.94574) (118., 0.946518) (118.5, 0.947286) (119., 0.948042) (119.5, 0.948789) (120., 0.949525) (120.5, 0.95025) (121., 0.950966) (121.5, 0.951672) (122., 0.952368) (122.5, 0.953055) (123., 0.953732) (123.5, 0.954399) (124., 0.955058) (124.5, 0.955707) (125., 0.956347) (125.5, 0.956978) (126., 0.9576) (126.5, 0.958214) (127., 0.958819) (127.5, 0.959415) (128., 0.960003) (128.5, 0.960583) (129., 0.961155) (129.5, 0.961719) (130., 0.962274) (130.5, 0.962822) (131., 0.963363) (131.5, 0.963895) (132., 0.96442) (132.5, 0.964938) (133., 0.965448) (133.5, 0.965951) (134., 0.966447) (134.5, 0.966936) (135., 0.967418) (135.5, 0.967893) (136., 0.968361) (136.5, 0.968823) (137., 0.969278) (137.5, 0.969727) (138., 0.970169) (138.5, 0.970605) (139., 0.971035) (139.5, 0.971458) (140., 0.971876) (140.5, 0.972287) (141., 0.972693) (141.5, 0.973093) (142., 0.973487) (142.5, 0.973875) (143., 0.974258) (143.5, 0.974635) (144., 0.975007) (144.5, 0.975374) (145., 0.975735) (145.5, 0.976091) (146., 0.976442) (146.5, 0.976788) (147., 0.977129) (147.5, 0.977465) (148., 0.977796) (148.5, 0.978123) (149., 0.978444) (149.5, 0.978761) (150., 0.979074) (150.5, 0.979382) (151., 0.979685) (151.5, 0.979984) (152., 0.980279) (152.5, 0.980569) (153., 0.980856) (153.5, 0.981138) (154., 0.981416) (154.5, 0.98169) (155., 0.98196) (155.5, 0.982226) (156., 0.982488) (156.5, 0.982746) (157., 0.983001) (157.5, 0.983252) (158., 0.983499) (158.5, 0.983742) (159., 0.983982) (159.5, 0.984219) (160., 0.984452) (160.5, 0.984682) (161., 0.984908) (161.5, 0.985131) (162., 0.985351) (162.5, 0.985568) (163., 0.985781) (163.5, 0.985991) (164., 0.986198) (164.5, 0.986403) (165., 0.986604) (165.5, 0.986802) (166., 0.986997) (166.5, 0.98719) (167., 0.98738) (167.5, 0.987566) (168., 0.987751) (168.5, 0.987932) (169., 0.988111) (169.5, 0.988287) (170., 0.98846) (170.5, 0.988631) (171., 0.9888) (171.5, 0.988966) (172., 0.98913) (172.5, 0.989291) (173., 0.98945) (173.5, 0.989606) (174., 0.98976) (174.5, 0.989912) (175., 0.990062) (175.5, 0.990209) (176., 0.990355) (176.5, 0.990498) (177., 0.990639) (177.5, 0.990778) (178., 0.990915) (178.5, 0.991049) (179., 0.991182) (179.5, 0.991313) (180., 0.991442) (180.5, 0.991569) (181., 0.991695) (181.5, 0.991818) (182., 0.99194) (182.5, 0.992059) (183., 0.992177) (183.5, 0.992294) (184., 0.992408) (184.5, 0.992521) (185., 0.992632) (185.5, 0.992742) (186., 0.99285) (186.5, 0.992956) (187., 0.993061) (187.5, 0.993164) (188., 0.993266) (188.5, 0.993366) (189., 0.993464) (189.5, 0.993562) (190., 0.993657) (190.5, 0.993752) (191., 0.993845) (191.5, 0.993936) (192., 0.994027) (192.5, 0.994116) (193., 0.994203) (193.5, 0.994289) (194., 0.994374) (194.5, 0.994458) (195., 0.994541) (195.5, 0.994622) (196., 0.994702) (196.5, 0.994781) (197., 0.994859) (197.5, 0.994935) (198., 0.995011) (198.5, 0.995085) (199., 0.995158) (199.5, 0.99523) (200., 0.995301)
};    
% K=10 bandwidth
    \addplot [smooth,mark=none,thick, smooth, blue,dashed] coordinates{  (1., 55.9392) (1.5, 25.9043) (2., 14.6647) (2.5, 9.24009) (3.,  6.22854) (3.5, 4.40797) (4., 3.24565) (4.5, 2.47619) (5.,  1.95378) (5.5, 1.59223) (6., 1.3378) (6.5, 1.15572) (7.,  1.02298) (7.5, 0.924099) (8., 0.848628) (8.5, 0.789468) (9.,  0.741795) (9.5, 0.702322) (10., 0.6688) (10.5, 0.639682) (11.,  0.613891) (11.5, 0.59067) (12., 0.569479) (12.5, 0.549925) (13.,  0.531716) (13.5, 0.514632) (14., 0.498505) (14.5, 0.483202) (15.,  0.468618) (15.5, 0.454668) (16., 0.441283) (16.5, 0.428405) (17.,  0.415986) (17.5, 0.403988) (18., 0.392376) (18.5, 0.381121) (19.,  0.3702) (19.5, 0.359591) (20., 0.349277) (20.5, 0.339242) (21.,  0.329473) (21.5, 0.319959) (22., 0.310689) (22.5, 0.301654) (23.,  0.292847) (23.5, 0.28426) (24., 0.275888) (24.5, 0.267725) (25.,  0.259765) (25.5, 0.252004) (26., 0.244439) (26.5, 0.237064) (27.,  0.229875) (27.5, 0.22287) (28., 0.216045) (28.5, 0.209397) (29.,  0.202922) (29.5, 0.196617) (30., 0.190479) (30.5, 0.184505) (31.,  0.178692) (31.5, 0.173037) (32., 0.167538) (32.5, 0.16219) (33.,  0.156992) (33.5, 0.151939) (34., 0.14703) (34.5, 0.142261) (35.,  0.13763) (35.5, 0.133133) (36., 0.128768) (36.5, 0.124531) (37.,  0.12042) (37.5, 0.116432) (38., 0.112563) (38.5, 0.108812) (39.,  0.105176) (39.5, 0.101651) (40., 0.0982344) (40.5, 0.0949241)  (41., 0.0917171) (41.5, 0.0886107) (42., 0.0856024) (42.5,  0.0826894) (43., 0.0798693) (43.5, 0.0771394) (44., 0.0744973)  (44.5, 0.0719406) (45., 0.0694668) (45.5, 0.0670736) (46.,  0.0647587) (46.5, 0.0625197) (47., 0.0603545) (47.5, 0.0582609)  (48., 0.0562368) (48.5, 0.05428) (49., 0.0523886) (49.5,  0.0505605) (50., 0.0487938) (50.5, 0.0470867) (51., 0.0454372)  (51.5, 0.0438435) (52., 0.042304) (52.5, 0.0408169) (53.,  0.0393805) (53.5, 0.0379932) (54., 0.0366535) (54.5, 0.0353597)  (55., 0.0341105) (55.5, 0.0329043) (56., 0.0317398) (56.5,  0.0306156) (57., 0.0295303) (57.5, 0.0284827) (58., 0.0274715)  (58.5, 0.0264956) (59., 0.0255537) (59.5, 0.0246446) (60.,  0.0237674) (60.5, 0.0229208) (61., 0.022104) (61.5, 0.0213158)  (62., 0.0205553) (62.5, 0.0198216) (63., 0.0191137) (63.5,  0.0184308) (64., 0.0177719) (64.5, 0.0171363) (65., 0.0165233)  (65.5, 0.0159318) (66., 0.0153614) (66.5, 0.0148111) (67.,  0.0142804) (67.5, 0.0137685) (68., 0.0132748) (68.5, 0.0127987)  (69., 0.0123395) (69.5, 0.0118966) (70., 0.0114695) (70.5,  0.0110576) (71., 0.0106605) (71.5, 0.0102774) (72., 0.00990808)  (72.5, 0.00955193) (73., 0.00920849) (73.5, 0.00887734) (74.,  0.00855803) (74.5, 0.00825014) (75., 0.00795327) (75.5,  0.00766703) (76., 0.00739105) (76.5, 0.00712495) (77.,  0.00686839) (77.5, 0.00662103) (78., 0.00638255) (78.5,  0.00615262) (79., 0.00593094) (79.5, 0.00571722) (80.,  0.00551118) (80.5, 0.00531254) (81., 0.00512103) (81.5,  0.00493641) (82., 0.00475842) (82.5, 0.00458684) (83.,  0.00442142) (83.5, 0.00426196) (84., 0.00410823) (84.5,  0.00396003) (85., 0.00381717) (85.5, 0.00367945) (86.,  0.00354668) (86.5, 0.0034187) (87., 0.00329533) (87.5, 0.0031764)  (88., 0.00306176) (88.5, 0.00295124) (89., 0.00284471) (89.5,  0.00274202) (90., 0.00264303) (90.5, 0.00254761) (91.,  0.00245563) (91.5, 0.00236696) (92., 0.0022815) (92.5,  0.00219911) (93., 0.0021197) (93.5, 0.00204315) (94., 0.00196936)  (94.5, 0.00189823) (95., 0.00182967) (95.5, 0.00176359) (96.,  0.00169988) (96.5, 0.00163848) (97., 0.00157929) (97.5,  0.00152224) (98., 0.00146725) (98.5, 0.00141425) (99.,  0.00136315) (99.5, 0.00131391) (100., 0.00126644) (100.5,  0.00122068) (101., 0.00117658) (101.5, 0.00113406) (102.,  0.00109309) (102.5, 0.00105359) (103., 0.00101552) (103.5,  0.000978827) (104., 0.000943457) (104.5, 0.000909364) (105.,  0.000876503) (105.5, 0.000844829) (106., 0.000814299) (106.5,  0.000784872) (107., 0.000756508) (107.5, 0.000729168) (108.,  0.000702816) (108.5, 0.000677417) (109., 0.000652935) (109.5,  0.000629337) (110., 0.000606592) (110.5, 0.000584669) (111.,  0.000563538) (111.5, 0.00054317) (112., 0.000523538) (112.5,  0.000504616) (113., 0.000486378) (113.5, 0.000468798) (114.,  0.000451854) (114.5, 0.000435522) (115., 0.00041978) (115.5,  0.000404607) (116., 0.000389983) (116.5, 0.000375887) (117.,  0.0003623) (117.5, 0.000349205) (118., 0.000336582) (118.5,  0.000324416) (119., 0.00031269) (119.5, 0.000301387) (120.,  0.000290493) (120.5, 0.000279992) (121., 0.000269872) (121.5,  0.000260116) (122., 0.000250714) (122.5, 0.000241651) (123.,  0.000232916) (123.5, 0.000224497) (124., 0.000216382) (124.5,  0.00020856) (125., 0.000201021) (125.5, 0.000193754) (126.,  0.00018675) (126.5, 0.00018) (127., 0.000173493) (127.5,  0.000167221) (128., 0.000161176) (128.5, 0.00015535) (129.,  0.000149734) (129.5, 0.000144322) (130., 0.000139104) (130.5,  0.000134076) (131., 0.000129229) (131.5, 0.000124558) (132.,  0.000120055) (132.5, 0.000115715) (133., 0.000111532) (133.5,  0.0001075) (134., 0.000103614) (134.5, 0.0000998684) (135.,  0.0000962582) (135.5, 0.0000927785) (136., 0.0000894246) (136.5,  0.0000861919) (137., 0.000083076) (137.5, 0.0000800728) (138.,  0.0000771782) (138.5, 0.0000743882) (139., 0.0000716991) (139.5,  0.0000691071) (140., 0.0000666089) (140.5, 0.000064201) (141.,  0.0000618801) (141.5, 0.0000596431) (142., 0.000057487) (142.5,  0.0000554088) (143., 0.0000534057) (143.5, 0.0000514751) (144.,  0.0000496143) (144.5, 0.0000478207) (145., 0.0000460919) (145.5,  0.0000444257) (146., 0.0000428197) (146.5, 0.0000412717) (147.,  0.0000397797) (147.5, 0.0000383417) (148., 0.0000369556) (148.5,  0.0000356196) (149., 0.0000343319) (149.5, 0.0000330908) (150.,  0.0000318946) (150.5, 0.0000307416) (151., 0.0000296302) (151.5,  0.0000285591) (152., 0.0000275266) (152.5, 0.0000265315) (153.,  0.0000255724) (153.5, 0.0000246479) (154., 0.0000237569) (154.5,  0.0000228981) (155., 0.0000220703) (155.5, 0.0000212724) (156.,  0.0000205034) (156.5, 0.0000197622) (157., 0.0000190478) (157.5,  0.0000183592) (158., 0.0000176955) (158.5, 0.0000170558) (159.,  0.0000164392) (159.5, 0.0000158449) (160., 0.0000152721) (160.5,  0.00001472) (161., 0.0000141878) (161.5, 0.0000136749) (162.,  0.0000131806) (162.5, 0.0000127041) (163., 0.0000122448) (163.5,  0.0000118022) (164., 0.0000113755) (164.5, 0.0000109643) (165.,  0.0000105679) (165.5, 0.0000101859) (166., 9.81763*10^-6) (166.5, 9.46272*10^-6) (167., 9.12063*10^-6) (167.5, 8.79091*10^-6) (168., 8.47311*10^-6) (168.5, 8.1668*10^-6) (169., 7.87156*10^-6) (169.5, 7.587*10^-6) (170., 7.31272*10^-6) (170.5, 7.04836*10^-6) (171., 6.79355*10^-6) (171.5, 6.54796*10^-6) (172., 6.31124*10^-6) (172.5, 6.08308*10^-6) (173., 5.86317*10^-6) (173.5, 5.65121*10^-6) (174., 5.44692*10^-6) (174.5, 5.25001*10^-6) (175., 5.06021*10^-6) (175.5, 4.87728*10^-6) (176., 4.70096*10^-6) (176.5, 4.53102*10^-6) (177., 4.36722*10^-6) (177.5, 4.20934*10^-6) (178., 4.05717*10^-6) (178.5, 3.91049*10^-6) (179., 3.76913*10^-6) (179.5, 3.63287*10^-6) (180., 3.50154*10^-6) (180.5, 3.37495*10^-6) (181., 3.25294*10^-6) (181.5, 3.13535*10^-6) (182., 3.022*10^-6) (182.5, 2.91275*10^-6) (183., 2.80745*10^-6) (183.5, 2.70596*10^-6) (184., 2.60814*10^-6) (184.5, 2.51385*10^-6) (185., 2.42297*10^-6) (185.5, 2.33538*10^-6) (186., 2.25095*10^-6) (186.5, 2.16958*10^-6) (187., 2.09115*10^-6) (187.5, 2.01555*10^-6) (188., 1.94268*10^-6) (188.5, 1.87245*10^-6) (189., 1.80476*10^-6) (189.5, 1.73952*10^-6) (190., 1.67663*10^-6) (190.5, 1.61602*10^-6) (191., 1.5576*10^-6) (191.5, 1.50129*10^-6) (192., 1.44702*10^-6) (192.5, 1.39471*10^-6) (193., 1.34429*10^-6) (193.5, 1.29569*10^-6) (194., 1.24885*10^-6) (194.5, 1.2037*10^-6) (195., 1.16019*10^-6) (195.5, 1.11825*10^-6) (196., 1.07782*10^-6) (196.5, 1.03885*10^-6) (197., 1.0013*10^-6) (197.5, 9.65101*10^-7) (198., 9.30212*10^-7) (198.5, 8.96584*10^-7) (199., 8.64171*10^-7) (199.5, 8.3293*10^-7) (200., 8.02819*10^-7)
    };    
    %K=20 bandwidth
     \addplot [smooth,mark=none, thick, smooth, red,dashed] coordinates{ (8., 2.38491) (8.5, 2.0999) (9.,  1.87602) (9.5, 1.69841) (10., 1.55601) (10.5, 1.44051) (11.,  1.34567) (11.5, 1.26678) (12., 1.20028) (12.5, 1.14349) (13.,  1.09437) (13.5, 1.05136) (14., 1.01329) (14.5, 0.979236) (15.,  0.948501) (15.5, 0.920536) (16., 0.894911) (16.5, 0.871284) (17.,  0.849382) (17.5, 0.828983) (18., 0.809904) (18.5, 0.791996) (19.,  0.775131) (19.5, 0.759204) (20., 0.744122) (20.5, 0.729807) (21.,  0.71619) (21.5, 0.703211) (22., 0.690817) (22.5, 0.678962) (23.,  0.667602) (23.5, 0.656702) (24., 0.646226) (24.5, 0.636144) (25.,  0.626429) (25.5, 0.617056) (26., 0.608001) (26.5, 0.599243) (27.,  0.590763) (27.5, 0.582545) (28., 0.57457) (28.5, 0.566825) (29.,  0.559297) (29.5, 0.551971) (30., 0.544836) (30.5, 0.537883) (31.,  0.5311) (31.5, 0.524478) (32., 0.518008) (32.5, 0.511683) (33.,  0.505495) (33.5, 0.499437) (34., 0.493503) (34.5, 0.487686) (35.,  0.481981) (35.5, 0.476382) (36., 0.470885) (36.5, 0.465485) (37.,  0.460178) (37.5, 0.454959) (38., 0.449825) (38.5, 0.444773) (39.,  0.439799) (39.5, 0.434899) (40., 0.430071) (40.5, 0.425313) (41.,  0.420621) (41.5, 0.415994) (42., 0.411428) (42.5, 0.406923) (43.,  0.402475) (43.5, 0.398083) (44., 0.393745) (44.5, 0.38946) (45.,  0.385225) (45.5, 0.38104) (46., 0.376903) (46.5, 0.372813) (47.,  0.368769) (47.5, 0.364768) (48., 0.360811) (48.5, 0.356896) (49.,  0.353023) (49.5, 0.349189) (50., 0.345395) (50.5, 0.34164) (51.,  0.337922) (51.5, 0.334242) (52., 0.330598) (52.5, 0.326989) (53.,  0.323416) (53.5, 0.319877) (54., 0.316373) (54.5, 0.312901) (55.,  0.309463) (55.5, 0.306057) (56., 0.302683) (56.5, 0.299341) (57.,  0.29603) (57.5, 0.29275) (58., 0.2895) (58.5, 0.286281) (59.,  0.283091) (59.5, 0.27993) (60., 0.276799) (60.5, 0.273697) (61.,  0.270623) (61.5, 0.267577) (62., 0.26456) (62.5, 0.26157) (63.,  0.258608) (63.5, 0.255673) (64., 0.252765) (64.5, 0.249885) (65.,  0.24703) (65.5, 0.244203) (66., 0.241402) (66.5, 0.238626) (67.,  0.235877) (67.5, 0.233154) (68., 0.230456) (68.5, 0.227783) (69.,  0.225136) (69.5, 0.222514) (70., 0.219917) (70.5, 0.217344) (71.,  0.214796) (71.5, 0.212273) (72., 0.209773) (72.5, 0.207298) (73.,  0.204847) (73.5, 0.20242) (74., 0.200017) (74.5, 0.197637) (75.,  0.19528) (75.5, 0.192947) (76., 0.190637) (76.5, 0.18835) (77.,  0.186085) (77.5, 0.183844) (78., 0.181625) (78.5, 0.179428) (79.,  0.177254) (79.5, 0.175101) (80., 0.172971) (80.5, 0.170863) (81.,  0.168776) (81.5, 0.166711) (82., 0.164667) (82.5, 0.162645) (83.,  0.160643) (83.5, 0.158663) (84., 0.156704) (84.5, 0.154765) (85.,  0.152847) (85.5, 0.150949) (86., 0.149072) (86.5, 0.147214) (87.,  0.145377) (87.5, 0.143559) (88., 0.141762) (88.5, 0.139983) (89.,  0.138225) (89.5, 0.136485) (90., 0.134765) (90.5, 0.133063) (91.,  0.131381) (91.5, 0.129717) (92., 0.128071) (92.5, 0.126444) (93.,  0.124836) (93.5, 0.123245) (94., 0.121672) (94.5, 0.120117) (95.,  0.11858) (95.5, 0.117061) (96., 0.115558) (96.5, 0.114073) (97.,  0.112605) (97.5, 0.111154) (98., 0.109719) (98.5, 0.108302) (99.,  0.106901) (99.5, 0.105516) (100., 0.104147) (100.5, 0.102794)  (101., 0.101457) (101.5, 0.100136) (102., 0.098831) (102.5,  0.0975411) (103., 0.0962665) (103.5, 0.095007) (104., 0.0937626)  (104.5, 0.0925331) (105., 0.0913184) (105.5, 0.0901182) (106.,  0.0889326) (106.5, 0.0877612) (107., 0.086604) (107.5, 0.0854609)  (108., 0.0843317) (108.5, 0.0832163) (109., 0.0821145) (109.5,  0.0810262) (110., 0.0799512) (110.5, 0.0788895) (111., 0.0778409)  (111.5, 0.0768052) (112., 0.0757823) (112.5, 0.0747721) (113.,  0.0737745) (113.5, 0.0727893) (114., 0.0718163) (114.5,  0.0708556) (115., 0.0699068) (115.5, 0.06897) (116., 0.0680449)  (116.5, 0.0671315) (117., 0.0662296) (117.5, 0.0653391) (118.,  0.0644598) (118.5, 0.0635917) (119., 0.0627346) (119.5,  0.0618885) (120., 0.0610531) (120.5, 0.0602283) (121., 0.0594141)  (121.5, 0.0586104) (122., 0.0578169) (122.5, 0.0570336) (123.,  0.0562604) (123.5, 0.0554971) (124., 0.0547437) (124.5, 0.054)  (125., 0.053266) (125.5, 0.0525414) (126., 0.0518262) (126.5,  0.0511204) (127., 0.0504237) (127.5, 0.049736) (128., 0.0490574)  (128.5, 0.0483876) (129., 0.0477265) (129.5, 0.0470741) (130.,  0.0464302) (130.5, 0.0457948) (131., 0.0451678) (131.5,  0.0445489) (132., 0.0439383) (132.5, 0.0433356) (133., 0.042741)  (133.5, 0.0421541) (134., 0.0415751) (134.5, 0.0410037) (135.,  0.0404399) (135.5, 0.0398835) (136., 0.0393345) (136.5,  0.0387929) (137., 0.0382584) (137.5, 0.0377311) (138., 0.0372108)  (138.5, 0.0366974) (139., 0.0361909) (139.5, 0.0356912) (140.,  0.0351981) (140.5, 0.0347117) (141., 0.0342317) (141.5,  0.0337582) (142., 0.0332911) (142.5, 0.0328303) (143., 0.0323756)  (143.5, 0.0319271) (144., 0.0314846) (144.5, 0.0310481) (145.,  0.0306174) (145.5, 0.0301926) (146., 0.0297735) (146.5,  0.0293601) (147., 0.0289523) (147.5, 0.02855) (148., 0.0281532)  (148.5, 0.0277617) (149., 0.0273756) (149.5, 0.0269947) (150.,  0.026619) (150.5, 0.0262484) (151., 0.0258828) (151.5, 0.0255222)  (152., 0.0251666) (152.5, 0.0248158) (153., 0.0244697) (153.5,  0.0241284) (154., 0.0237918) (154.5, 0.0234597) (155., 0.0231322)  (155.5, 0.0228092) (156., 0.0224906) (156.5, 0.0221764) (157.,  0.0218665) (157.5, 0.0215608) (158., 0.0212593) (158.5, 0.020962)  (159., 0.0206688) (159.5, 0.0203795) (160., 0.0200943) (160.5,  0.019813) (161., 0.0195355) (161.5, 0.0192619) (162., 0.0189921)  (162.5, 0.018726) (163., 0.0184635) (163.5, 0.0182047) (164.,  0.0179494) (164.5, 0.0176977) (165., 0.0174494) (165.5,  0.0172046) (166., 0.0169631) (166.5, 0.016725) (167., 0.0164902)  (167.5, 0.0162587) (168., 0.0160303) (168.5, 0.0158051) (169.,  0.0155831) (169.5, 0.0153641) (170., 0.0151482) (170.5,  0.0149352) (171., 0.0147252) (171.5, 0.0145182) (172., 0.014314)  (172.5, 0.0141126) (173., 0.0139141) (173.5, 0.0137183) (174.,  0.0135252) (174.5, 0.0133348) (175., 0.0131471) (175.5,  0.0129619) (176., 0.0127794) (176.5, 0.0125994) (177., 0.0124219)  (177.5, 0.0122469) (178., 0.0120743) (178.5, 0.0119042) (179.,  0.0117364) (179.5, 0.0115709) (180., 0.0114078) (180.5,  0.0112469) (181., 0.0110883) (181.5, 0.0109319) (182., 0.0107777)  (182.5, 0.0106257) (183., 0.0104758) (183.5, 0.0103279) (184.,  0.0101822) (184.5, 0.0100385) (185., 0.00989675) (185.5,  0.00975702) (186., 0.00961926) (186.5, 0.00948342) (187.,  0.00934949) (187.5, 0.00921744) (188., 0.00908723) (188.5,  0.00895886) (189., 0.00883228) (189.5, 0.00870748) (190.,  0.00858443) (190.5, 0.00846311) (191., 0.00834349) (191.5,  0.00822555) (192., 0.00810926) (192.5, 0.00799461) (193.,  0.00788157) (193.5, 0.00777012) (194., 0.00766023) (194.5,  0.00755189) (195., 0.00744507) (195.5, 0.00733976) (196.,  0.00723592) (196.5, 0.00713355) (197., 0.00703261) (197.5,  0.0069331) (198., 0.00683498) (198.5, 0.00673825) (199.,  0.00664288) (199.5, 0.00654885) (200., 0.00645615)};    
  \addplot [mark=none , thick] {1};
 \end{axis}
% \draw [red] (current bounding box.south east) rectangle (current bounding box.north west);
\end{tikzpicture}%
\caption{The normalized frequency (solid lines) and bandwidth (dashed lines) contributions $8 Z_{\overline{\omega},\overline{\sigma}}/(4 N_S)^{\sqrt{\frac{K-1}{K+1}}} $ are plotted against the squeezing parameter $\xi$ for  Schmidt numbers $K=10$ and $K=20$. The normalised QFI approaches $1$, which indicates a scaling above the SQL in the limit $K^{3/2} \ll \xi $. The normalized bandwidth contribution is much smaller and goes to $0$ for $K^{3/2} \ll \xi $.  It  is even further suppressed by the factor $\sigma \epsilon /\omega_0^2$.}  \label{highsqueezingfig}
\end{center}
\end{figure}
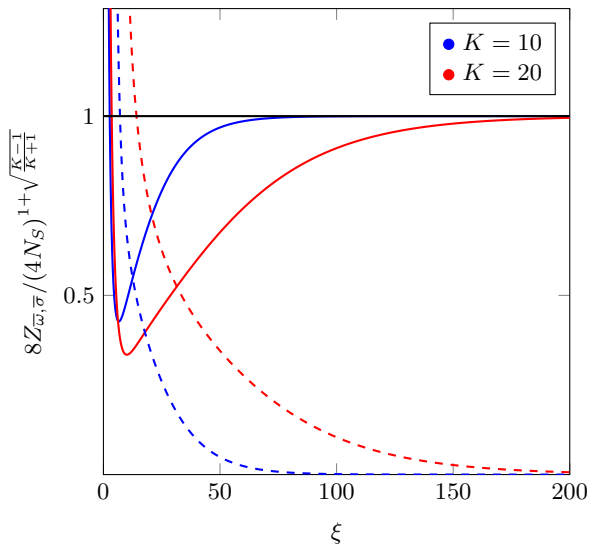

Let us now consider a regime in which squeezing is the dominant quantum resource and the frequency entanglement is relatively weak.  We specify the parameter conditions as $\xi \gg K^{3/2}$ and $K \gtrsim 1.5$.  These conditions helps us to put the QFI into a concise analytic form. Additionally, by requiring  $K \gtrsim 1.5$, the high-squeezing regime is sufficiently distinct from the no-entanglement regime with $K=1$. The details about the calculations performed in this Subsection can be found in Supplementary Material~\ref{App:highsqueezing}.  The fraction of photons in the mode $n$ is given by  $N_{Sn}/N_S$, where $N_{Sn} = \sinh^2 (\xi r_n)$ is the  photon number of the mode $n$ of the signal beam. By increasing $\xi$ for a fixed $K$, the relative contribution of higher modes $n >0$ decreases. In the high-squeezing regime, almost all the photons reside in the $0$ mode, that is $N_{S} \approx N_{S0} \gg N_{S1} \gg 1$, implying a high photon number per mode but a low number of active modes, contrary to the high-frequency entanglement regime. With this, we arrive in the asymptotic limit at the result 
\begin{equation}\label{Eq:QFIhighsqueezing}
\frac{J_q}{J_c} \approx \frac{1}{3} \left(4N_S \right)^{\sqrt{\frac{K-1}{K+1}}} ,
\end{equation} 
where it was used that $N_{Sn} =\sinh^2(\xi r_n) \approx \cosh^2(\xi r_n)$ for $n=0,1$ and only terms of order $N_{S0}N_{S1}$ in Eq.~\eqref{Jq} contribute significantly. This is the reason why the bandwidth terms are negligible. In Fig.~\ref{highsqueezingfig}, both the normalized frequency (solid lines) and bandwidth (dashed lines) terms are plotted against $\xi$ for Schmidt numbers $K=10$ and $K=20$, confirming our analytical results.  Interestingly, the QFI in terms of the mode photon numbers is $J_q \sim N_{S0}N_{S1}$, which indeed indicates a scaling better than the SQL, and nearly reaches the HL for $K$ big enough. As a conclusion, increasing the squeezing for a fixed $K$ also increases the quantum advantage, so squeezing can be seen as a sensitivity-enhancing resource of the protocol.  Up to this point, photon loss and noise has been neglected. In Section \ref{sec:loss}, the impact of loss, but not noise, on the high-squeezing regime will be examined.

\subsection{The mixed regime}

Now, let us study the intermediate parameter regime $K^{1/2} \ll \xi \ll K^{3/2}$.  Under these conditions, multiple modes are active like in the high-frequency entanglement regime and the photon number per mode is high $N_{Sn} \gg 1$ like in the high-squeezing regime, hence the name mixed regime. Using these conditions, we can derive in the asymptotic limit an analytic expression of the QFI 
\begin{equation}
\frac{J_q}{J_c}\approx  \left( \frac{ \xi}{2^{1/2}K^{3/2}} + \frac{\sigma \epsilon}{4\omega_0^2 }\right) N_S,
\end{equation}where the first term is again the frequency contribution and the second term the bandwidth contribution, following both a Heisenberg scaling $J_q \sim N_S^2$.  For details about the calculations in this Subsection, see Supplementary Material~\ref{App:mixed}.  The factor $\xi /2^{1/2}K^{3/2}$ is smaller than $1$, but we still have $(\xi /2^{1/2} K^{3/2}) N_S \gg 1$ thus guaranteeing quantum advantage. 
Since both terms  $\sigma\epsilon/4\omega_0^2$  and  $\xi /2^{1/2} K^{3/2}$ are smaller than $1$, we cannot generally decide which contribution is dominant. For instance, in the experimental setup referred to in Section~\ref{sec:quantum}, we had that $\sqrt{\sigma\epsilon/} \omega_0 \sim 0.01$, the bandwidth can be safely neglected in the mixed regime, at least for values of $\xi$ and $K$ up to $100$ as can be seen in Supplementary Material~\ref{App:mixed}. 
Therefore, we will neglect the bandwidth contribution from this point on. 
\begin{figure}[h]
\includegraphics[width=\linewidth]{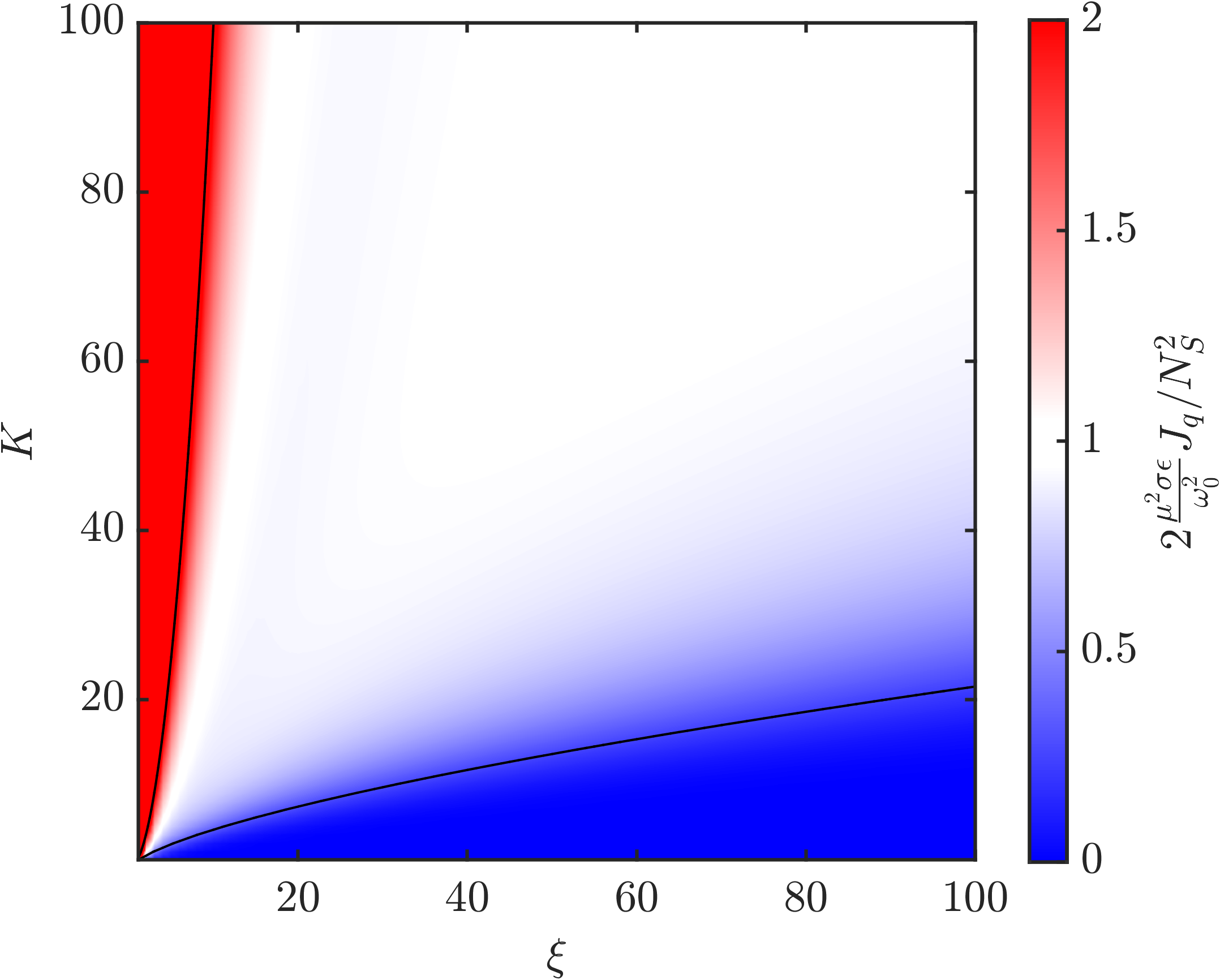}% Here is how to import EPS art
\caption{\label{HeisenbergPlot} We plot the normalized QFI $2 \frac{\mu^2 \sigma\epsilon}{\omega_0^2} J_q / N_S^2$, not to be confused with the quantum advantage ratio. The plot shows the three parameter regimes and their corresponding borders given by the contours $\xi = K^{1/2}$ and $\xi = K^{3/2}$. The mixed regime is characterized by the value of $1$, depicted as white, and thus shows Heisenberg scaling  and validates our analytical expression for the QFI. The high-squeezing regime, in which quantum advantage above the SQL is achieved, is depicted as blue with values below $1$. The red area is the high-entanglement regime, where the values range far above $2$ but were cut off. In this regime, no quantum advantage is achieved.}
\end{figure}
 In  Fig.~\ref{HeisenbergPlot}, the ratio $ \frac{ 2\mu^2 \sigma\epsilon}{\omega_0^2}J_q/N_S^2 \approx 2 Z_{\overline{\omega}}/N_S^2$ is plotted for both $\xi$ and $K$ up to the values of $100$.  Three distinct regions corresponding to the three parameter regimes can be appreciated. In the white area, which corresponds to a value of $1$ for the ratio, we observe a behavior of $J_q \sim N_S^2$ for the QFI and thus Heisenberg scaling. The red area is the high frequency entanglement regime and the blue area is the high-squeezing regime.  Curiously, quantum advantage is achieved  in the two regimes with high photon number per mode, and not in the high-frequency entanglement regime with a low photon number per mode. This in contrast to the quantum illumination protocol, where small photon number per mode is necessary to achieve quantum advantage.
 \begin{table}[h]%The best place to locate the table environment is directly after its first reference in text
\caption{\label{table}%
Listed are the quantum advantages for the different parameter regimes. Regime 1,2 and 3 correspond to the high-frequency entanglement, the high-squeezing and the mixed regime respectively.   Here, the contributions due to the bandwidth shift are neglected.}
\begin{ruledtabular}
\begin{tabular}{lll}
Regime 1& Regime 2 & Regime 3\\
$\xi \ll K^{1/2}$ & $\xi \gg K^{3/2}$ &  $ K^{1/2} \ll \xi \ll K^{3/2}$\\
\colrule
$\frac{J_q}{J_c} \approx 1$ & $\frac{J_q}{J_c} \sim N_S^{\sqrt{\frac{K-1}{K+1}}}$  &  $\frac{J_q}{J_c} \sim \frac{\xi}{K^{3/2}}N_S$ \\
\end{tabular}
\end{ruledtabular}
\end{table}
The parameter conditions of the three regimes and their corresponding quantum advantages are summerized in TABLE~\ref{table}. 

\subsection{A loss analysis for the high-squeezing regime} \label{sec:loss}
So far we have considered the ideal scenario in which no photons are lost and the returned state is pure. In realistic scenarios, the probe state at the receiver will be mixed due to photon loss and thermal noise. In Ref.~\cite{zhuang2022ultimate}, time-of-flight estimation in the microwave regime was studied, in which the thermal photon number per mode is much larger than $1$. A similar state was used, a continuous wave squeezed state, and a  considerable quantum advantage was proved at a certain threshold of the signal-to-noise ratio. This so-called threshold effect arises only in the presence of thermal noise and requires an analysis that goes beyond the calculation of the QFI. In our protocol, we assume operation in the optical regime, in which thermal noise can be neglected and solely relying on the QFI suffices. Photon loss, however, has to be considered to assess if the protocol shows quantum advantage in more realistic scenarios. Because photon loss mixes the state, the calculation of the QFI is significantly more complicated. Thus, we will only study the high-squeezing regime in which the state can be described sufficiently well by only a couple of Schmidt modes and thus allows us to derive analytical expressions. We assume no losses in the idler beam. Photon loss in the signal beam can occur on the way to and from the target and/or during the interaction with the object (which generalizes the protocol to non-perfectly reflecting objects). The probability of losing a signal-photon is assumed to be frequency independent and it is modeled by a beam splitter 
\begin{equation}
\hat{U}_B \hat{a}(\omega) \hat{U}_B^\dag = \sqrt{\eta} \hat{a}(\omega) + \sqrt{1-\eta} \hat{c}(\omega) ,
\end{equation}
where $\hat{c}(\omega)$ is an auxiliary mode which cannot be accessed by the experimenter and will be traced out at the end. In this framework, the beam splitter commutes with the Doppler reflection operation, so only one beam splitter with effective  transmissivity $\eta$ is required  for the lidar-to-target-to-lidar roundtrip
We choose to apply this beam splitter operation after the reflection at the receiver level. The final state is a Gaussian state. Gaussian states are fully described by their first two moments $\textbf{d}$ and $\Upsigma$, the definitions and an introduction to Gaussian states can be found in Section \ref{sec:gaussianstates} and Ref.~\cite{safranek2018}. As we discussed in Section \ref{sec:high-squeezing}, only the first two pairs of modes $\hat{a}_{0\mu}, \hat{a}_{1\mu}$ and $\hat{b}_0, \hat{b}_1$ are populated with a significant amount of photons. This allows us to omit the rest of the modes by tracing them out and thus derive a lower bound for the QFI. Alternatively, we could justify the neglect of the higher modes by tailoring the joint-spectral amplitude $f(\omega,\tilde{\omega})$, such that only the first two modes are active. To use the formula for the QFI of Gaussian states given in~\cite{safranek2018}, we need to change the basis (i.e. the modes $\hat{a}_{0\mu}$ and $\hat{a}_{1\mu}$) to make it parameter independent.  To do so, we assume that a prior estimate $\mu_0$ of the parameter is known and we only want to estimate the small deviation $\delta$ with $\mu = \mu_0 + \delta$, which is standard in most parameter estimation protocols.  We expand the Schmidt modes around $\mu_0$ up to the first order and find
\begin{align}
\hat{a}_{0\mu} & \approx \hat{a}_{0\mu_0} - \delta \frac{\omega_0}{2\mu \sqrt{\sigma\epsilon}} \hat{a}_{1\mu_0} \\
\hat{a}_{1\mu} & \approx \hat{a}_{1\mu_0} + \delta \frac{\omega_0}{2\mu \sqrt{\sigma\epsilon}}  \left( \hat{a}_{0\mu_0} - \sqrt{2} \hat{a}_{2\mu_0} \right) ,
\end{align}
where we have neglected the terms corresponding to the bandwidth contribution, as they are small in this regime,  which we have previously established in Section~\ref{sec:high-squeezing}.
Now, we have the modes $\hat{a}_{0\mu_0},\hat{a}_{1\mu_0},\hat{a}_{2\mu_0},\hat{b}_{0},\hat{b}_{1},\hat{c}_{0\mu_0},\hat{c}_{1\mu_0},\hat{c}_{2\mu_0}$, where $\hat{c}_{n\mu_0} =\int \diff \omega \mu_0^{1/2} \psi_n (\mu_0 \omega-\omega_0/2) \hat{c}(\omega)$ are the auxiliary Schmidt modes. The resulting covariance matrix and QFI are calculated in Supplementary Material~\ref{supp:loss}. We recover the result from Eq.~\eqref{Eq:QFIhighsqueezing} for the lossless case $\eta =1$, which confirms the validity of our approach and our approximations. To make a fair comparison, we also have to consider the classical protocol under the effect of photon loss. The QFI of the classical strategy is simply reduced by the factor $\eta$, that is $J_c \approx \eta \omega_0^2 N_S \Delta T^2/\mu^2$. We arrive at the ratio
\begin{equation}
\frac{J_q}{J_c} \approx  \frac{1}{1-\eta} . \label{eq:QFIloss1}
\end{equation}
where we assumed $N_{S1}(1-\eta) \gg 1$ to obtain a compact result. This assumption causes the divergence in Eq.~\eqref{eq:QFIloss1} because as $\eta\rightarrow 1$, we have $N_{S1} \rightarrow \infty$. Without this assumption, we recover the result of the lossless scenario in the limit of $\eta \rightarrow 1$.
The quantum advantage ratio in Eq.~\eqref{eq:QFIloss1} does not depend on the photon number, and thus, photon loss destroys the near HL scaling and brings it down to the SQL, that is $J_q \sim N_{S}$.  A constant factor quantum advantage is achieved for all values of $\eta$, which, however, becomes insignificant for small path transmissivities $\eta \ll 1$.  For transmissitivities  $\eta \geq 50\%$, the quantum advantage factor is $J_q/J_c \geq 2 \approx 3$dB. This makes our protocol promising for short-range applications where the path losses are small, such as Doppler microscopy for biologicals.

\subsection{Optimal measurement} \label{loss}
As we are estimating only the velocity of the object, there always exists at least one optimal measurement saturating the QFI, but it is not necessarily unique. Quantum estimation theory provides techniques to construct some of these observables, in particular the one related to the symmetric logarithmic derivative (SLD)  $\hat{O}_{\mu} = \mathds{1} \mu + \hat{L}_{\mu}/ J(\mu)$. However, its implementation in a realistic experimental setup is a highly non-trivial task. In the case of a pure-state manifold, the SLD $\hat{L}_\mu$ can be written as $\hat{L}_{\mu} = |\partial_{\mu} \psi_{\mu}\rangle \langle \psi_\mu| + |\psi_\mu \rangle\langle \partial_{\mu} \psi_{\mu} |$ ~\cite{paris2009}. Thus, only $|\partial_{\mu} \psi_{\mu}\rangle$ needs to be calculated, which has been done for the calculation of the QFI and it can be found in Supplementary Material~\ref{App:QFIquantum}. However, a construction of this observable in a lab in an optical setup is far from trivial. Furthermore, it depends on the parameter $\mu$ itself, and we would like to have a measurement working on the whole range of velocities if possible. Otherwise, an adaptive measurement strategy could be followed~\cite{liu2019}.
In the Gaussian formalism, the SLD can be written as a sum of terms that are at most quadratic in the modes~\cite{safranek2018}. In  Supplementary Material~\ref{supp:loss} we have given the explicit expression for the SLD derived in the limit of $N_S \gg 1$ in the high-squeezing regime under photon loss.

Let us analyze a measurement based on frequency-resolved photon-counting of signal and idler photons for the lossless scenario, which is discussed in detail in Supplementary Material~\ref{app:measurement}. This measurement corresponds  to a projection onto the frequency eigenstates $|\bm{\omega} , \tilde{\bm{\omega}} \rangle \equiv \vert \omega_1,\ldots , \omega_n ,\tilde{\omega}_1 ,\ldots \tilde{\omega}_m \rangle \equiv \frac{1}{\sqrt{n! m!}} \otimes_{i=1}^n \otimes_{j=1}^{m} a^\dag (\omega_i) b^\dag (\tilde{\omega}_j) |0\rangle$, where $n,m \in \mathds{N}$ are the signal and idler photon numbers and $\omega_i,\tilde{\omega}_j \in \mathds{R}_{>0}$ are the respective frequencies of each photon.  The corresponding set of POVM operators are $\{| \bm{\omega} , \tilde{\bm{\omega}} \rangle \langle\bm{\omega} ,\tilde{\bm{\omega}}  |  \big| n,m \in \mathds{N}$, \, $\omega_i,\tilde{\omega}_j \in \mathds{R}_{>0} \}$.  We calculate the Fisher information (FI), $\tilde{F}_q$, corresponding to this measurement for a generalization $|\tilde{\psi}_\mu\rangle$ of the probe state $|\psi_\mu \rangle$ given in Eq.~\eqref{refelectedQstate}. This generalized probe state contains phase factors depending on the kinetic properties of the target and a complex squeezing parameter, which were previously omitted in our analysis. We can show that the measurement outcomes do not depend on these phases. Indeed, both states $|\tilde{\psi}_\mu\rangle$ and $|\psi_\mu \rangle$ give rise to the same probability distribution of measurement outcomes. Thus, the POVM $\{|\bm{\omega} , \tilde{\bm{\omega}} \rangle \langle \bm{\omega} ,\tilde{\bm{\omega}}  |\}$ is actually phase insensitive. Finally, we prove that $\tilde{F}_q = J_q$, so this measurement also saturates the QFI in Section~\ref{sec:quantum}. Let us remark that this measurement does not depend on the parameter $\mu$, so it can be used for saturating the QFI for any velocity.  Also, it could in principle be experimentally feasible by using diffraction gratings that map frequency components to distinct locations where photon counters are placed~\cite{gianani2020,davis2020}.  We note, that the POVM is a separate measurement of the signal and idler beam, which further eases the experimental implementation. This also indicates that the idler, and thus the entanglement, solely serves as a state preparation tool. For example, one can check that the idlerless Fock state $\sim (\hat{a}_0^\dag)^{N_{S0}} (\hat{a}_1^\dag)^{N_{S1}} \vert 0\rangle$, which has no frequency entanglement and could be approximately heralded with our probe state, achieves Heisenberg scaling and shows the same behaviour under loss as in Eq.~\eqref{eq:QFIloss1} for the limit $N_S \gg 1$.
\subsection{Further perspectives}
Lastly, we want to emphasize that the protocol can be easily adapted to different frequency and/or bandwidth estimating protocols. Also, the target's trajectory can be generalized to an accelerating one via a Bogoliubov transformation~\cite{gianfelici2017,good2013}, but with an additional complication due to the presence of Casimir radiation. For stationary targets, the protocol can be adapted to estimate the location, which boils down to the estimation of arrival times of the signal beam. The probe state written in the time domain has exactly the same structure as in the frequency domain, where the variances of the double Gaussian change as $\sigma^2/2 \rightarrow 2 \sigma^2$ and $\epsilon^2/2 \rightarrow 2 \epsilon^2$. The Schmidt number remains unaltered under this transformation. Thus, the estimation of time arrival of signal photons is analogous to the estimation of mean frequency of the signal photons. Analogously, a measurement that attains the optimal performance is the measurement of photon arrival times. 

\section{Discussion}

We have proposed a protocol for a quantum Doppler lidar that estimates the radial velocity of a reflecting moving target using a twin beam with frequency entanglement and squeezing as quantum resources. This quantum protocol was benchmarked against a classical one by calculating the QFIs for both strategies. We have identified three different parameter regimes, achieving quantum advantage in two of them. In the high-squeezing regime, where the frequency entanglement becomes less relevant compared to squeezing, the quantum protocol exceeds the standard quantum limit. In the mixed regime, where both quantum resources are comparable, the quantum protocol follows the Heisenberg limit. We have found that frequency-resolved photon counting of signal and idler beam is an optimal measurement in the lossless case. The effect of losses on the performance of the protocol was studied in the high-squeezing regime by modeling the loss channel as  a frequency-independent beam splitter. A constant factor quantum advantage $\geq 3$ dB in the variance of the estimator is achieved given a path transmissivity $\geq 50\%$.

\section{Methods}

\subsection{Quantum estimation theory}\label{sec:estimation}
The objective of quantum estimation theory is to find the ultimate precision limit for the estimation of a parameter $\mu$ that is encoded in a quantum system. In our scenario, the probe state $\rho $ that is emitted by the lidar acquires information about $\mu$ during the reflection off the moving target, which transforms the state as $\rho \rightarrow \rho_\mu$. The classical Fisher information (FI) $F(\mu)$ is a measure of the information about the parameter $\mu$ that can be extracted by a given measurement corresponding to the positive operator-valued measure (POVM) $\lbrace \Pi_z \rbrace$ with $ \int \diff z \, \Pi_z = \mathds{1}$. The FI is given by
\begin{equation}
F(\mu) = \int \diff z \, \frac{1}{p_\mu (z)} \left( \partial_\mu p_\mu (z) \right)^2 ,
\end{equation}
where $ p_\mu (z)=\text{Tr} (\Pi_z \rho_\mu) $ is the probability of having the measurement outcome $z$ given the parameter $\mu$.
The  Cram\'er-Rao bound is given by~\cite{paris2009}
\begin{equation}
\text{Var}(\hat{\mu}) \geqslant \frac{1}{M F(\mu)} \label{CR} ,
\end{equation}
where $\hat{\mu}$ is an unbiased estimator that maps the measurement data of the $M$ experiment repetitions to an estimate of the parameter $\mu$. The bound can be saturated using the maximum likelihood estimator in the limit of large $M$~\cite{fisher1925}. Maximizing the FI over all POVMs $\lbrace \Pi_z \rbrace$ yields the quantum Fisher information $J(\mu) \geqslant F(\mu)$. Eq.~\eqref{CR} for the QFI is called the quantum Cram\'er-Rao bound which sets the absolute precision limit for the estimation of $\mu$. 
In the case of a pure-state manifold, i.e. when $\hat{\rho}_\mu= | \psi_\mu \rangle \langle \psi_\mu|$ for any $\mu$, the QFI is given by~\cite{paris2009}
\begin{equation}
J(\mu) = 4 \left( \langle \partial_\mu \psi_\mu | \partial_\mu \psi_\mu\rangle - | \langle \psi_\mu | \partial_\mu \psi_\mu \rangle|^2 \right) .
\end{equation}
To prove a quantum advantage, we calculate the QFIs $J_q$ and $J_c$ of both the quantum  and classical strategy.   A quantum advantage is achieved, if the ratio is $J_q/J_c >1$ assuming both strategies illuminate the object with the same energy and an optimal measurement is performed. An observable corresponding to the optimal measurement is given by $\hat{O}_\mu = \mathds{1} \mu + \hat{L}_\mu/ J(\mu)$, where $\hat{L}_\mu$ is the symmetric logarithmic derivative (SLD), which satisfies $\hat{L}_\mu \hat{\rho}_\mu + \hat{\rho}_\mu \hat{L}_\mu = 2 \partial_{\mu}\hat{\rho}_\mu$. As the optimal observable generally depends on the parameter itself, a prior guess about the parameter is required to construct the measurement. The measurement can then be adaptively optimized~\cite{liu2019}.

\subsection{Gaussian states}\label{sec:gaussianstates}
A Gaussian state is fully defined by its first two moments $\textbf{d}$ and $\Upsigma$. Their components are defined as
\begin{equation}
    d_m = \text{tr} \left[ \hat{\rho} \hat{\textbf{R}}_m \right] ,
\end{equation}
and
\begin{equation}
    \Sigma_{nm} = \text{tr} \left[ \hat{\rho} \{ \Delta \hat{\textbf{R}}_m , \Delta \hat{\textbf{R}}_n^{\dag} \} \right] ,
\end{equation}
where $\hat{\textbf{R}} = \left( \hat{a}_0, \hat{a}_0^\dag, \hat{a}_1, \hat{a}_1^\dag , \ldots  \right)^T$ and $\Delta \hat{\textbf{R}} = \hat{\textbf{R}} -\hat{\textbf{d}}$.
Gaussian unitaries that transform the state as $\hat{\rho}' =\hat{U}\hat{\rho} \hat{U}^\dag$ transform the first moments as
\begin{equation}
    \textbf{d}' = \textbf{G} \textbf{d} + \textbf{b} ,
\end{equation}
and
\begin{equation}
    \Upsigma' = \textbf{G} \Upsigma \textbf{G}^\dag ,
\end{equation}
where $\textbf{G}$ is the corresponding symplectic matrix, see Ref.~\cite{safranek2018} for more information on how $\textbf{G}$ and $\textbf{b}$ relate to the Gaussian unitary $\hat{U}$. A formula for the QFI of a Gaussian state is given by
\begin{equation*} \label{eq:QFISafranek}
    J(\mu) = \displaystyle{\lim_{\kappa \to 1}}\frac{1}{2} \text{vec} [\partial_\mu \Upsigma]^\dag \mathcal{M}_{\kappa}^{-1} \text{vec} [\partial_\mu \Upsigma] + 2 \partial_\mu \textbf{d}^\dag \Upsigma^{-1} \partial_\mu \textbf{d} ,
\end{equation*} 
where  $\mathcal{M}_{\kappa} = \kappa \Upsigma^\dag \otimes \Upsigma - \textbf{K}\otimes \textbf{K}$ with the symplectic form $\textbf{K}= \text{diag}(1,-1,1,-1,\ldots)$. The operation $\text{vec} [\cdot]$ turns a matrix into a vector as
\begin{equation}
\text{vec} \left[\begin{pmatrix}
a & b \\ c & d
\end{pmatrix} \right] = \begin{pmatrix} a\\b\\c\\d\end{pmatrix} .
\end{equation}
We can also calculate the SLD in this formalism. It is given by
\begin{equation}
\hat{L}_\mu = \Delta \hat{\textbf{R}}^\dag \mathcal{A}_\mu \Delta \hat{\textbf{R}} - \frac{1}{2} \text{tr} [\Upsigma \mathcal{A}_\mu] + 2\Delta \hat{\textbf{R}}^\dag \Upsigma^{-1} \partial_\mu \textbf{d} ,
\end{equation}
where $\text{vec}[\mathcal{A}_\mu] = \displaystyle{\lim_{\kappa \to 1}} \mathcal{M}_{\kappa}^{-1} \text{vec}[\partial_\mu \Upsigma]  $ .

\section{Data availability}
The authors declare that all data supporting the findings of this
study are available within the article and its Supplementary
Material.

\section{Acknowledgements}
We thank Robert Jonsson and G\"oran Johansson for insightful discussions. The authors acknowledge financial support from QMiCS (820505) and OpenSuperQ (820363) of the EU Flagship on Quantum Technologies, the EU FET-Open projects Quromorphic (828826) and EPIQUS (899368), as well as from the QUANTEK project from ELKARTEK program (KK-2021/00070), and the Spanish Ram\'on y Cajal Grant RYC-2020-030503-I. MR acknowledges support from UPV/EHU PhD Grant PIF21/289.
M.W. acknowledges support from the National Science Foundation under Grant CCF-1956211. R.D.C. acknowledges support from the Marie Sk{\l}odowska Curie fellowship number 891517 (MSC-IF Green- MIQUEC), the Alexander von Humboldt Foundation, the Knut and Alice Wallenberg Foundation through the Wallenberg Centre for Quantum Technology (WACQT), and the Academy of Finland, grants no. 353832, 349199.

\section{Author's contribution}
M.R. developed the theoretical formalism and performed the analytic calculations. M.S. suggested the initial idea and supervised the project throughout all stages.  M.R., M.S., R.D.C. and M.W. contributed to the interpretation of the results. M.R. took the lead in writing the manuscript and all authors provided critical feedback.
\subsection{Competing interests}
The authors declare no competing interests.

%\widetext

\begin{widetext}
\pagebreak
\section*{Supplementary Material for Quantum-Enhanced Doppler Lidar}

\section{Classical probe state}
The coherent state $\hat{D}|0\rangle = \exp (\alpha \int \diff \omega \, f(\omega) (\hat{a}(\omega)-\hat{a}^\dag (\omega)))|0\rangle$ is an eigenstate of the continuous annihilation operator $\hat{a}(\omega) \hat{D}|0\rangle = \alpha f(\omega) \hat{D}|0\rangle$. With this, and well-known properties of Gaussian integrals, the photon number and the mean energy can be easily calculated
\begin{equation}
\langle 0 | \hat{D}^\dag  \hat{N}_a \hat{D}|0\rangle = \alpha^2 \int \diff \omega \, f^2 (\omega) = \alpha^2
\end{equation}
and 
\begin{equation}
\langle 0 | \hat{D}^\dag  \hat{H}_a \hat{D}|0\rangle = \hbar \alpha^2 \int \diff \omega \, \omega f^2 (\omega) = \hbar \alpha^2 \omega_c .
\end{equation}
For the reflected state, we find for the photon number $\alpha^2$ and the mean energy $\hbar \alpha^2 \omega_c/\mu$, as expected.

\section{QFI of classical protocol} \label{QFIclassical}
We note that the derivative with respect  $\mu$ of a state of the form $\text{e}^{\hat{A}_\mu} |0\rangle$ is given by $\partial_\mu \text{e}^{\hat{A}_\mu} |0\rangle = (\partial_\mu \hat{A}_\mu) \text{e}^{\hat{A}_\mu} |0\rangle = \text{e}^{\hat{A}_\mu} \partial_\mu \hat{A}_\mu |0\rangle$, if $[\hat{A}_\mu, \partial_\mu \hat{A}_\mu]=0$.

In the case of the classical probe state the exponent is given by $\hat{A}_\mu = \alpha \mu^{1/2} \int \diff \omega \, f (\omega \mu) (\hat{a}(\omega)-\hat{a}^\dag (\omega))$. The commutator is
\begin{align}
[\hat{A}_\mu, \partial_\mu \hat{A}_\mu] &= \int \diff \omega \int \diff \tilde{\omega}\, f (\omega\mu) \partial_\mu (\mu^{1/2} f (\omega\mu)) \left[ \hat{a}(\omega)-\hat{a}^\dag (\omega), \hat{a}(\tilde{\omega})-\hat{a}^\dag (\tilde{\omega}) \right] \\
&= \int \diff \omega \int \diff \tilde{\omega}\, f (\omega\mu) \partial_\mu (\mu^{1/2} f (\omega\mu)) \Big(- \underbrace{\left[ \hat{a}(\omega),\hat{a}^\dag (\tilde{\omega})\right]}_{=\delta(\omega-\tilde{\omega})} + \underbrace{\left[ \hat{a}(\tilde{\omega}),\hat{a}^\dag (\omega) \right]}_{=\delta(\omega-\tilde{\omega})} \Big) \\
&=0 .
\end{align}
Using $\hat{D}^\dag \hat{D} = \mathds{1}$, we find for the inner product
\begin{align}
\langle \partial_\mu \psi_\mu | \partial_\mu \psi_\mu \rangle &= \langle 0 | \partial_\mu \hat{A}^\dag_\mu \hat{D}^\dag \hat{D} \partial_\mu \hat{A}_\mu|0\rangle = \langle 0 | \partial_\mu \hat{A}^\dag_\mu  \partial_\mu \hat{A}_\mu|0\rangle = - \langle 0 | \partial_\mu \hat{A}_\mu  \partial_\mu \hat{A}_\mu|0\rangle \\
&= \alpha^2\int \diff \omega \, \omega \left( \partial_\mu (\mu^{1/2} f (\omega\mu)) \right)^2 =\alpha^2 \int \diff \omega \, \left( \frac{1}{2} \mu^{-1/2} f (\omega\mu) + \mu^{1/2} \omega f' (\omega\mu) \right)^2 \\
&=\frac{\alpha^2}{\mu^2} \int \diff y \left( \frac{1}{2} f (y) +y \partial_{y} f (y) \right)^2  .
\end{align}
This is the general expression for the QFI. Let us now make some approximations to better understand the meaning of this expression. First we use the time domain version $\hat{E}(t)$ and $\hat{E}^\dag(t)$ of the frequency annihilation and creation operators, whose relation is given by $\hat{a}^\dag (\omega) = (2\pi)^{-1/2} \int \diff t \,  \text{e}^{- \text{i} t \omega} \hat{E}^\dag (t)$. The exponent of the coherent probe state in the time-domain basis is then given by
\begin{equation}
\hat{A}_\mu = \alpha\int \diff t \int  \frac{\diff \omega}{\sqrt{2 \pi}} \mu^{1/2}f(\mu \omega) \text{e}^{\text{i}\omega t} \hat{E}(t) - h.c. =  \alpha\mu^{-1/2} \int \diff t \int \frac{\diff \omega}{\sqrt{2\pi}} f(\omega) \text{e}^{\text{i}\frac{\omega}{\mu}t} \hat{E}(t) -h.c..
\end{equation}
Now, we assume $v/c \ll 1$ from which $\mu^{-1} \approx 1 +2 v/c$ follows. Using this approximation leads to
\begin{equation}
   \text{e}^{\text{i}\frac{\omega}{\lambda}t} \approx  \text{e}^{\text{i} \omega t}  \text{e}^{ \text{i} \omega_c t 2v/c } \text{e}^{\text{i} (\omega-\omega_c)) t 2v/c} .
\end{equation}
With the additional approximation $\Delta \omega \Delta T 2 v/c \ll 1$, we can neglect the last term $\text{e}^{\text{i} (\omega-\omega_c) 2  t v/c} \approx 1$. We then find 
\begin{equation}
   \hat{A}_\mu \approx  \alpha\int \diff t \int \frac{\diff \omega}{\sqrt{2\pi}} f(\omega) \text{e}^{\text{i}\omega t} \text{e}^{\text{i} \omega_c 2 t v/c} \hat{E}(t) -h.c. =\alpha \int \diff t \, g(t) \text{e}^{\text{i} \omega_c 2t v/c} \hat{E}(t) -h.c. .
\end{equation}
The derivative of the reflected state with respect to the velocity $v$ is then given by
\begin{align}
   \vert \partial_v \psi \rangle &= \hat{D} \int \diff t \,  \partial_v \left[ \alpha  g(t) \text{e}^{-\text{i} \omega_c 2 t v/c} \right] \hat{E}^\dag (t) \vert 0\rangle  \\
   &= - \hat{D} \int \diff t \,  i 2 \alpha\omega_c  t g(t) \text{e}^{-\text{i}\omega_c 2 t v/c}   \hat{E}^\dag (t) \vert 0\rangle .
\end{align}
The inner product is given by
\begin{align}
    \langle \partial_v \psi \vert \partial_v \psi \rangle =  4 \alpha^2\omega_c^2  \int \diff t \, t^2 \vert g(t)\vert^2 .
\end{align}
Furthermore, we find
\begin{align}
    \langle \psi \vert \partial_v \psi\rangle =  - i 2 \alpha \omega_c  \int \diff t \, t \vert g(t)\vert^2 .
\end{align}
Combining these two terms, we find for the QFI
\begin{align}
   J_c (v) &= 4  \left(  \langle \partial_v \psi \vert \partial_v \psi \rangle - \vert \langle \psi \vert \partial_v \psi\rangle\vert^2 \right) \\
   &= 16\alpha^2 \omega_c^2  \underbrace{\left( \int \diff t \, t^2 g^2(t) - \left(\int \diff t \, t g^2(t) \right)^2 \right)}_{= \Delta T^2} \\
   &= 16 \alpha^2\omega_c^2  \Delta T^2 .
\end{align}
The QFI of the Doppler parameter can then be calculated
\begin{equation}
J_c (\mu) = \frac{4\omega_c^2 \Delta T^2 N_c}{\mu^2} = \frac{\omega_0^2 \Delta T^2 N_S}{\mu^2} ,
\end{equation}
with $\omega_c = \omega_0/2$.
Thus, within the approximation made, the QFI solely depends on the time duration of the pulse and the transmitted energy, not the pulse shape or other parameters of the probe state. Chirping for example, which increases the bandwidth but not the time duration, would have no effect on the QFI.

\section{Double-Gaussian distribution} \label{App:doubleGaussian}
A double Gaussian function
\begin{equation}
f(\omega,\tilde{\omega}) = \sqrt{\frac{2}{\uppi \sigma\epsilon}} \exp \left(- \frac{(\omega+\tilde{\omega} - \omega_0)^2}{2\sigma^2} \right) \exp \left( - \frac{(\omega-\tilde{\omega})^2}{2\epsilon^2} \right) 
\end{equation}
can be decomposed into its Schmidt modes $f(\omega, \tilde{\omega}) = \sum_{n=0}^\infty r_n \psi_n (\omega-\omega_0/2) \psi_n (\tilde{\omega}-\omega_0/2)$. The relative weight $r_n^2$ of each mode is given by $r_n = \frac{2 \sqrt{\sigma\epsilon}}{\sigma+\epsilon}\left( \frac{\sigma-\epsilon}{\sigma+\epsilon} \right)^n$. The number of effective modes is given by the Schmidt number $K= \frac{\sigma^2+\epsilon^2}{2\sigma\epsilon}$. We can write the coefficients completely in terms of the Schmidt number $r_n = (\mp 1)^n \sqrt{\frac{2}{K+1}} \sqrt{\frac{K-1}{K+1}}^n$, where the $(-1)^n$ is for the case $\epsilon > \sigma$. However, for the quantities we calculate like the photon number and QFI, the cases $\sigma > \epsilon$ and $\sigma < \epsilon$ are equivalent as they only depend on $r_n^2$.
The Schmidt modes are given by $\psi_n( \omega-\omega_0/2)= \sqrt{s} \varphi_n (s (\omega-\omega_0/2))$ with $s=\sqrt{ \frac{2}{\sigma\epsilon}}$ and the Hermite functions (harmonic oscillator wavefunction) $\varphi_n (\omega) = (2^n n! \sqrt{\pi}!)^{-1/2} H_{n}(\omega)e^{-\omega^2/2}$, where $H_n(\omega)$ are the Hermite polynomials. The Harmonic oscillator wave functions have the following properties \cite{celeghini2021hermite}
\begin{align}
\varphi_n' (y)&= - \sqrt{\frac{n+1}{2}} \varphi_{n+1} (y) + \sqrt{\frac{n}{2}} \varphi_{n-1} (y)  \label{Hermite1}\\
y \varphi_n(y) &= \sqrt{\frac{n+1}{2}} \varphi_{n+1}(y) + \sqrt{\frac{n}{2}} \varphi_{n-1} (y) , \label{Hermite2}
\end{align}
with $\varphi_n' (y) =\frac{\diff}{\diff y}\varphi_n (y) $.
The center frequency is independent of $n$ and given by 
\begin{align}
\int_{-\infty}^{\infty} \diff \omega \, \omega \psi^2_n(\omega-\omega_0/2) &= \int^\infty_{-\infty} \diff \omega \, \omega s \varphi^2_n (s (\omega-\omega_0/2)) = \int^\infty_{-\infty} \frac{\diff y}{s} \left( \frac{y}{s}+ \frac{\omega_0}{2} \right) s \varphi^2_n(y) \\
&=  \int^\infty_{-\infty} \diff y  \left( \frac{1}{s} \sqrt{\frac{n+1}{2}} \varphi_{n+1}(y) \varphi_n(y) + \frac{1}{s} \sqrt{\frac{n}{2}} \varphi_{n-1} (y) \varphi_n(y) + \frac{\omega_0}{2} \varphi_n^2(y) \right) \\ &= \frac{\omega_0}{2} ,
\end{align} 
using the properties in Eq.~\eqref{Hermite2}. 
For the Doppler shifted mode $\mu^{1/2} \psi_n (\omega\mu-\omega_0/2)$ we obtain a mean frequency of $\omega_0/(2\mu)$.
The second moment is given by
\begin{align}
\int \diff \omega \, \omega^2  \psi_n^2( \omega - \omega_0/2) &= \int \diff \omega \, \omega^2 s  \varphi_n^2 ( \omega - \omega_0/2)  \\
&= \int \frac{\diff y}{s} \left( \frac{y}{s} + \frac{\omega_0}{2} \right)^2 s \varphi^2_n (y) = \int \diff y\left(  \frac{y^2}{s^2}+\frac{y\omega_0}{s} + \frac{\omega_0^2}{4}  \right)^2 \varphi^2_n (y) \\
&= \int \diff y \frac{1}{s^2} \left( \sqrt{\frac{n+1}{2}} \varphi_{n+1}(y) + \sqrt{\frac{n}{2}} \varphi_{n-1} (y) \right)^2 + \frac{\omega_0^2}{4} = \frac{1}{s^2} \left( \frac{n+1}{2} + \frac{n}{2} \right) + \frac{\omega^2_0}{4} \\
&= \frac{\sigma \epsilon}{2} \left( n + \frac{1}{2} \right) + \frac{\omega_0^2}{4} . \label{eq:Schmidtmodevariance}
\end{align}
Thus, the frequency's variance is $\frac{\sigma\epsilon}{2} (n+1/2)$ and for the Doppler shifted mode $\frac{\sigma\epsilon}{2 \mu^2} (n+1/2)$.

\section{Quantum probe state} \label{App:twinbeam}
The twin-beam multimode squeezed vacuum state can be written as a tensor product of individual squeezing operators
$
\hat{S} = \bigotimes_{n=0}^\infty \hat{S}_n =\bigotimes_{n=0}^\infty \exp ( r_n \xi (\hat{a}_n \hat{b}_n-\hat{a}^\dag_n \hat{b}^\dag_n))
$. Also the vacuum state, that is independent of $\mu$, can be written as a product of the individual vacua of each mode $|0\rangle = \bigotimes_n |0\rangle_n$. The squeezing operator acting on the vacuum thus yields
\begin{align}
\hat{S}|0\rangle = \bigotimes_n \hat{S}_n \bigotimes_m |0\rangle_m = \bigotimes_n \hat{S}_n |0\rangle_n &= \bigotimes_n \frac{1}{\cosh(\xi r_n)} \exp \left( - \tanh (\xi r_n) \hat{a}_{n\mu}^\dag \hat{b}_n^\dag \right) |0\rangle_n \\
&= \mathcal{N} \exp \left(- \sum_n \tanh (\xi r_n) \hat{a}_{n\mu}^\dag \hat{b}_n^\dag \right) |0\rangle , \label{reflectedQstate}
\end{align}
where $\mathcal{N}= \prod_n \frac{1}{\cosh (\xi r_n)}$.
 \subsubsection{Transformation rules}\label{App:transformationrules}
The transformation of the annihilation and creation operators due to squeezing  reduces to the well-known transformation of the two-mode squeezed vacuum state due to the tensor product structure of the squeezing operator
\begin{equation}
\hat{S}^\dag \hat{a}_{n\mu} \hat{S} = \bigotimes_r \hat{S}^\dag_r \hat{a}_n \bigotimes_q \hat{S}_q = \hat{S}_n^\dag \hat{a}_{n\mu} \hat{S}_n  .
\end{equation}
With this, we find the relations
 \begin{align}
 \hat{S}^\dagger \hat{a}_{n\mu} \hat{S} &= \hat{a}_{n\mu} \cosh (\xi r_n) - \hat{b}_n^\dagger \sinh (\xi r_n) \\
  \hat{S}^\dagger \hat{b}_{n} \hat{S} &= \hat{b}_{n} \cosh (\xi r_n) - \hat{a}_{n\mu}^\dagger \sinh (\xi r_n)\\
   \hat{S}^\dagger \hat{a}_{n\mu}^\dagger \hat{S} &= \hat{a}_{n\mu}^\dagger \cosh (\xi r_n) - \hat{b}_n \sinh (\xi r_n) \\
  \hat{S}^\dagger \hat{b}_{n}^\dagger \hat{S} &= \hat{b}_{n}^\dagger \cosh (\xi r_n) - \hat{a}_{n\mu} \sinh (\xi r_n) .
 \end{align}
\subsubsection{Photon number}\label{App:photonnumber}
Next, let us calculate the photon number  for both beams. The photon-number operator is given as
\begin{equation}
\hat{N} = \hat{N}_a + \hat{N}_b = \int \diff \omega \, \hat{a}^\dag(\omega) \hat{a}(\omega) + \int \diff \omega \, \hat{b}^\dag (\omega) \hat{b}(\omega)
\end{equation}
in the continuous formalism, where $\hat{N}_a$ ($\hat{N}_b$) is the photon-number operator of the signal (idler). Using $\hat{a}(\omega) = \sum_n \psi_n (\omega-\omega_0/2) \hat{a}_n$ and $\hat{b}(\omega) = \sum_n \psi_n (\omega-\omega_0/2) \hat{b}_n$, we can rewrite $\hat{N}$ in terms of discrete operators
\begin{align}
\hat{N} &= \int \diff \omega \, \sum_{n=0}^\infty \psi_n(\omega-\omega_0/2) \hat{a}^\dag_n  \sum_{m=0}^\infty \psi_m(\omega-\omega_0/2) \hat{a}_m + \int \diff \omega \, \sum_{n=0}^\infty \psi_n(\omega-\omega_0/2) \hat{b}^\dag_n  \sum_{m=0}^\infty \psi_m(\omega-\omega_0/2) \hat{b}_m \\
&= \sum_n \hat{a}^\dag_n \hat{a}_n + \sum_n \hat{b}^\dag_n \hat{b}_n,
\end{align} 
where $\hat{a}^\dag_n \hat{a}_n $ ($\hat{b}^\dag_n \hat{b}_n$) is the photon-number operator of the signal (idler) beam of the mode $n$. 
With the transformation rules, the photon number of the twin-beam multimode squeezed vacuum can be calculated
\begin{multline}
\langle 0 | \hat{S}^\dag \hat{U}^\dag_\mu \hat{N} \hat{U}_\mu \hat{S} |0\rangle = \langle 0 | \sum_n \left( \hat{a}_{k\mu}^\dagger \cosh (\xi r_k) - \hat{b}_k \sinh (\xi r_k) \right)\left( \hat{a}_{k\mu} \cosh (\xi r_k) - \hat{b}^\dag_k \sinh (\xi r_k)  \right) |0\rangle \\
+\langle 0 | \sum_n \left( \hat{b}_{k}^\dagger \cosh (\xi r_k) - \hat{a}_{k\mu} \sinh (\xi r_k )\right)\left( \hat{b}_{k} \cosh (\xi r_k) - \hat{a}^\dag_{k\mu} \sinh (\xi r_k)  \right) |0\rangle \\= \sum_n \sinh^2(\xi r_n)+\sum_n \sinh^2(\xi r_n) 
= 2 \sum_n \sinh^2(\xi r_n).
\end{multline}
In both signal and idler, the photon number is equal $N_S=N_I$. Furthermore, the number of photons is independent of $\mu$, thus the photon number is conserved in the case of Doppler reflection for a target of constant velocity. By expanding the hyperbolic sine and using the geometric series, we find 
\begin{equation}
N_S=\sum_{n=0}^\infty \sinh^2 (\xi r_n) = \xi^2 \left( 1 + \frac{\xi^2}{3 K} + \frac{2 }{45} \frac{4\xi^4}{3K^2+1}+ \ldots \right) .
\end{equation}
The average photon number of mode $n$ is $N_{Sn} = \sinh^2(\xi r_n)$.
\subsubsection{Energy}
Now, let us write the Hamiltonian in terms of discrete modes and calculate the energy of the twin-beam multimode squeezed vacuum. The Hamiltonian in the continuous formalism is 
\begin{equation}
\hat{H} = \hat{H}_a+\hat{H}_b = \hbar  \int \diff \omega \, \omega \hat{a}^\dag (\omega) \hat{a}(\omega) +  \hbar \int \diff \omega \, \omega \hat{b}^\dag (\omega) \hat{b}(\omega) .
\end{equation}
For the Hamiltonian operator of the signal beam we find
\begin{align}
\hat{H}_a &=\hbar  \sum_{n,m} \int \diff \omega \, \omega \mu^{1/2} \psi_n (\omega\mu-\omega_0/2)  \mu^{1/2} \psi_m (\omega\mu-\omega_0/2)  \hat{a}^\dag_n \hat{a}_m \\ 
&= \hbar \mu s \sum_{n,m} \int \diff \omega \, \omega \varphi_n (s\mu (\omega-\omega/2\mu) \varphi_m (s\mu (\omega-\omega_0/2\mu)) \hat{a}_n^\dag \hat{a}_m \\
&= \hbar \mu s \sum_{n,m} \int \frac{\diff y}{\mu s } \left( \frac{y}{s\mu}+ \frac{\omega_0}{2\mu} \right) \varphi_n (y) \varphi_m (y) \hat{a}_n^\dag \hat{a}_m \\
&= \hbar \sum_{n,m} \int \diff y \left( \sqrt{\frac{n}{2}} \varphi_{n-1} (y)+\sqrt{\frac{n+1}{2}} \varphi_{n+1} (y) \right) \frac{ \varphi_m (y)}{s\mu} \hat{a}_n^\dag \hat{a}_m +\hbar  \frac{\omega_0}{2\mu} \sum_n \hat{a}_n^\dag \hat{a}_n \\
&= \hbar \sum_n \frac{\omega_0}{2\mu} \hat{a}_n^\dag \hat{a}_n + \hbar \frac{1}{s\mu} \sqrt{\frac{n}{2}} \hat{a}^\dag_n \hat{a}_{n-1} + \hbar \frac{1}{s\mu} \sqrt{\frac{n+1}{2}} \hat{a}_n^\dag \hat{a}_{n+1} .
\end{align}
An analogous calculation for the idler mode yields
\begin{equation}
\hat{H}_b =  \hbar \sum_n \frac{\omega_0}{2} \hat{b}_n^\dag \hat{b}_n + \hbar \frac{1}{s} \sqrt{\frac{n}{2}} \hat{b}^\dag_n \hat{b}_{n-1} + \hbar \frac{1}{s} \sqrt{\frac{n+1}{2}} \hat{b}_n^\dag \hat{b}_{n+1} .
\end{equation}
Thus, in this basis the Hamiltonian operator is not diagonal. However, the expectation value of the non-diagonal terms is zero and we find for the energy
\begin{equation}
\langle 0 | \hat{S}^\dag \hat{U}_\mu^\dag \hat{H} \hat{U}_\mu \hat{S} |0\rangle = \hbar  \frac{\omega_0}{2\mu} N_S +\hbar   \frac{\omega_0}{2} N_I = \hbar  \frac{\omega_0}{2} \left( \frac{1}{\mu} + 1 \right) \sum_n \sinh^2( \xi r_n) .
\end{equation}

\section{QFI of the quantum protocol} \label{App:QFIquantum}
The only component of the probe state that depends on $\mu$ is the operator $a_{n\mu}^\dag$. So let us calculate its derivative using the properties of the Hermite functions
\begin{align}
\partial_\mu \left( \mu^{1/2} \psi_{n} (\mu \omega-\omega_0/2) \right)& = \partial_\mu \left( \mu^{1/2} s^{1/2}\varphi_{n} (\mu s \omega- s\omega_0/2) \right) \\
&= \frac{1}{2\mu} \mu^{1/2}s^{1/2} \varphi_n(\mu s\omega-s\omega_0/2) + \mu^{1/2} s^{1/2}  \varphi_n' (\mu s \omega-s\omega_0/2) s \omega .
\end{align}
Let us set $y = \mu s \omega- s \omega_0/2$, then $s\omega =\frac{y}{\mu}+\frac{s\omega_0}{2\mu}$. We find for the second term
\begin{align}
\varphi_n' (y)\cdot (y+ s \omega_0/2) &= \left(- \sqrt{\frac{n+1}{2}} \varphi_{n+1} (y) + \sqrt{\frac{n}{2}} \varphi_{n-1} (y) \right) (y+ s \omega_0/2) \\
&= - \sqrt{\frac{n+1}{2}} \left( \sqrt{\frac{n+2}{2}} \varphi_{n+2}(y) + \sqrt{\frac{n+1}{2}} \varphi_{n} (y)  \right) + \sqrt{\frac{n}{2}} \left( \sqrt{\frac{n}{2}} \varphi_{n}(x) + \sqrt{\frac{n-1}{2}} \varphi_{n-2} (y) \right) \\ &+\left(- \sqrt{\frac{n+1}{2}} \varphi_{n+1} (y)  + \sqrt{\frac{n}{2}} \varphi_{n-1} (y) \right)  s \omega_0/2 \\
&= \left( - \frac{n+1}{2} + \frac{n}{2} \right)\varphi_n(y) + \frac{\omega_0 s}{2} \sqrt{\frac{n}{2}} \varphi_{n-1} (y) - \frac{\omega_0 s}{2} \sqrt{\frac{n+1}{2}}  \varphi_{n+1}(y) \\ &+ \sqrt{\frac{n(n-1)}{4}} \varphi_{n-2}(y) - \sqrt{\frac{(n+1)(n+2))}{4}} \varphi_{n+2}(y) .
\end{align}
With this
\begin{align*}
&\partial_\mu \left( \mu^{1/2} \psi_n (\mu\omega-\omega_0/2) \right) = \frac{1}{2\mu} \mu^{1/2}s^{1/2} \varphi_n(y) + \mu^{1/2} s^{1/2}  \varphi_n' (y) \frac{y+s\omega_0/2}{\mu} \\
&= \frac{\mu^{1/2} s^{1/2}}{\mu} \left( \frac{\omega_0 s}{2} \sqrt{\frac{n}{2}} \varphi_{n-1} (y) - \frac{\omega_0 s}{2} \sqrt{\frac{n+1}{2}}  \varphi_{n+1}(y) + \sqrt{\frac{n(n-1)}{4}} \varphi_{n-2}(y) - \sqrt{\frac{(n+1)(n+2))}{4}} \varphi_{n+2}(y) \right) .
\end{align*}
We thus find for the derivative of the creation operator
 \begin{align}
 \partial_\mu \hat{a}_{n\mu}^\dagger & = \frac{1}{\mu} \left( \frac{\omega_0 s}{2} \sqrt{\frac{n}{2}} \hat{a}_{n-1\mu}^\dagger - \frac{\omega_0 s}{2} \sqrt{\frac{n+1}{2}} \hat{a}_{n+1\mu}^\dagger + \sqrt{\frac{n(n-1)}{4}} \hat{a}_{n-2\mu}^\dagger - \sqrt{\frac{(n+1)(n+2)}{4}} \hat{a}_{n+2\mu}^\dagger \right) \\
 &= \frac{1}{\mu} \left( \alpha_{n} \hat{a}_{n-1\mu}^\dagger + \beta_n \hat{a}_{n+1\mu}^\dagger +\gamma_n \hat{a}_{n-2\mu}^\dagger +\delta_{n} \hat{a}_{n+2\mu}^\dagger \right),
 \end{align}
 with $\alpha_n = \frac{\omega_0 s}{2} \sqrt{\frac{n}{2}}$, $\beta_n = - \frac{\omega_0 s }{2} \sqrt{\frac{n+1}{2}}$, $\gamma_n = \sqrt{\frac{n(n-1)}{4}}$ and $\delta_n = - \sqrt{\frac{(n+1)(n+2)}{4}}$.
 
Because the reflected state in Eq.~\eqref{reflectedQstate} can be written as $|\psi_\mu \rangle =  \mathcal{N} e^{\hat{B}_\mu}|0\rangle$ and the derivative of the creation operator $\partial_\mu \hat{a}_{n\mu}^\dag$ is a linear combination of creation operators and thus $[\partial_\mu \hat{B}_\mu, \hat{B}_\mu] = 0$, the derivative is  given as $|\partial_\mu \psi_\mu\rangle=  \mathcal{N} (\partial_\mu \hat{B}_\mu)  e^{\hat{B}_\mu}|0\rangle$. The transformation rules imply $\langle \psi_\mu | \partial_\mu \psi_\mu \rangle =0$. Then, the QFI is given by $4\langle \partial_\mu \psi |\partial_\mu \psi\rangle$. Therefore, we only need to calculate the scalar product of the derivative of the state. We start by examining
\begin{align}
|\partial_\mu \psi \rangle &= \partial_\mu \hat{S} |0\rangle = \mathcal{N} \partial_\mu \exp \left( -\sum_{n} \tanh (\xi r_n) \hat{a}_{n\mu}^\dagger \hat{b}_n^\dagger \right) |0\rangle \\
&= \left(- \sum_n \tanh (\xi r_n) \partial_{\mu} \hat{a}_{n\mu}^\dagger \hat{b}_n^\dagger\right) \mathcal{N} \exp \left( -\sum_{m} \tanh (\xi r_m) \hat{a}_{m\mu}^\dagger \hat{b}_m^\dagger \right) |0\rangle \\
&= -  \sum_n \tanh (\xi r_n) \partial_{\mu} \hat{a}_{n\mu}^\dagger \hat{b}_n^\dagger \hat{S} |0\rangle =  - \hat{S} \sum_n \tanh (\xi r_n)\hat{S}^\dagger  \partial_{\mu}  \hat{a}_{n\mu}^\dagger \hat{S} \hat{S}^\dagger \hat{b}_n^\dagger \hat{S} |0\rangle \\
&= - \hat{S} \sum_n \tanh(\xi r_n) \hat{S}^\dagger \partial_\mu \hat{a}_{n\mu}^\dagger \hat{S} \left( \hat{b}^\dagger_n \cosh (\xi r_n) -\sinh (\xi r_n) \hat{a}_{n\mu}  \right) |0\rangle \\
&= - \hat{S} \sum_n \sinh(\xi r_n) \hat{S}^\dagger \partial_\mu \hat{a}_{n\mu}^\dagger \hat{S}  \hat{b}^\dagger_n |0\rangle \\
&= -\frac{1}{\mu} \hat{S}\sum_{n} \sinh (\xi r_n) \hat{S}^\dagger \left(  \alpha_{n} \hat{a}_{n-1 \mu}^\dagger + \beta_n \hat{a}_{n+1 \mu}^\dagger +\gamma_n \hat{a}_{n-2\mu}^\dagger +\delta_{n} \hat{a}_{n+2\mu}^\dagger  \right) \hat{S} \hat{b}_n^\dagger |0\rangle \\
&= -\frac{1}{\mu} \hat{S} \sum_{n} \sinh (\xi r_n) \Big( \alpha_n \cosh (\xi r_{n-1}) \hat{a}_{n-1\mu}^\dagger + \beta_n \cosh (\xi r_{n+1}) \hat{a}_{n+1\mu}^\dagger \\
&\hphantom{= -\frac{1}{\mu} \hat{S} \sum_{n} \sinh (\xi r_n) \Big( \alpha_n \cosh }+ \gamma_{n} \cosh (\xi r_{n-2}) \hat{a}_{n-2 \mu}^\dagger + \delta_{n} \cosh (\xi r_{n+2}) \hat{a}_{n+2\mu}^\dagger \Big) \hat{b}_n^\dagger |0\rangle .
\end{align}
We thus find for the QFI
\begin{align}
J_q(\mu) = 4 \langle \partial_\mu \psi_\mu |\partial_\mu \psi_\mu\rangle 
&= \frac{4}{\mu^2}\sum_{n} \sinh^2 (\xi r_n) \left(  \alpha_n^2 \cosh^2 (\xi r_{n-1})  + \beta_n^2 \cosh^2 (\xi r_{n+1})  + \gamma_{n}^2 \cosh^2 (\xi r_{n-2})  + \delta_{n}^2 \cosh^2 (\xi r_{n+2}) \right) \\
&=  \frac{4}{\mu^2}\sum_{n} \sinh^2 (\xi r_n) \big(  \frac{\omega_0^2 s^2}{4} \frac{n}{2} \cosh^2 (\xi r_{n-1})  + \frac{\omega_0^2s^2}{4} \frac{n+1}{2} \cosh^2 (\xi r_{n+1}) \\&  + \frac{n(n-1)}{4} \cosh^2 (\xi r_{n-2})  + \frac{(n+1)(n+2)}{4} \cosh^2 (\xi r_{n+2}) \big) .
\end{align}
The QFI splits into two parts, the part proportional to $\omega_0^2 s^2$ due to the frequency shift, and the remaining terms that are due to the bandwidth shift.
This can be seen by writing the modes  $\psi_n$ as  functions of the mean frequency $\overline{\omega} = \omega_0/2\mu$ and  the variable related to the bandwidth $\overline{\sigma} = (s \mu)^{-1} = \sqrt{\sigma\epsilon / 2 \mu^2}$. With this
\begin{equation}
\partial_\mu \hat{a}_n^\dag = \frac{\partial\overline{\omega}}{\partial \mu}  \partial_{\overline{\omega}} \hat{a}_{n\mu}^\dag + \frac{\partial \overline{\sigma}}{\partial \mu} \partial_{\overline{\sigma}} \hat{a}_{n\mu}^\dag ,
\end{equation}
where
\begin{equation}
\partial_{\overline{\omega}} \hat{a}_{n\mu}^\dag = - \mu \sqrt{\frac{2}{\sigma\epsilon}} \left( \sqrt{\frac{n}{2}} \hat{a}_{n-1\mu}^\dag - \sqrt{\frac{n+1}{2}} \hat{a}_{n+1 \mu}^\dag\right) ,
\end{equation}
and 
\begin{equation}
\partial_{\overline{\sigma}} \hat{a}_{n\mu}^\dag = -\mu \sqrt{\frac{2}{\sigma\epsilon}} \left( \sqrt{\frac{n(n-1)}{4}} \hat{a}_{n-2\mu}^\dag - \sqrt{\frac{(n+1)(n+2)}{4}} \hat{a}_{n+2\mu}^\dag \right) .
\end{equation}
With this, the QFI splits up into a frequency and a bandwidth part
\begin{equation}
J_q(\mu) = \left( \frac{\partial \overline{\omega}}{\partial \mu} \right)^2 J_q (\overline{\omega}) +  \left( \frac{\partial \overline{\sigma}}{\partial \mu} \right)^2 J_q (\overline{\sigma}) .
\end{equation}

\subsection{Calculating the time duration of the quantum signal beam} \label{supp:time}
We define the normalized power of the signal beam as
\begin{equation}
    \vert s(t) \vert^2 = \frac{\langle \psi \vert \hat{E}^\dag(t) \hat{E}(t) \vert \psi \rangle}{ N_S} \label{eq:power} .
\end{equation}
The time duration  $\Delta T$ of the signal beam is defined via 
\begin{equation}
\Delta T^2 = \int \diff t \, t^2 \vert s(t)\vert^2 - \left( \int \diff t \, t \vert s(t)\vert^2 \right)^2.
\end{equation}
So far we have mainly worked in the frequency domain, but for calculating the time duration of the signal pulse, it is more convenient to switch to the time-domain basis. The state $E^\dag (t) \vert 0\rangle$ can be interpreted as a photon at time $t$ at the detector.  The two bases are related via
\begin{align}
\hat{E}^\dag (t) &= \frac{1}{\sqrt{2\pi}} \int \diff \omega \, \text{e}^{\text{i}\omega t} \hat{a}^\dag (\omega)\\
&= \frac{1}{\sqrt{2\pi}} \int \diff \omega\,  \text{e}^{\text{i}\omega t} \sum_n \psi_n (\omega-\omega_0/2) \hat{a}_n^\dag \\
&   = \sum_n \underbrace{ \frac{1}{\sqrt{2\pi}} \int \diff \omega \, \text{e}^{\text{i}\omega t} \sum_n \psi_n (\omega-\omega_0/2)}_{=\tilde{\psi}_n (t)} \hat{a}_n^\dag  ,
\end{align}
where $\tilde{\psi}_n (t)= (\sigma \epsilon/2)^{1/4} \varphi_n (t\sqrt{\sigma \epsilon/2}) \text{e}^{\text{i}t\omega_0/2 }$ is the Fourier transform of the Schmidt-modes $\psi_n (\omega-\omega_0/2)$ and $\varphi_n$ are the Harmonic oscillator wave function that were introduced earlier. The Fourier transform causes the change $\sigma\rightarrow 2/\sigma$ and $\epsilon\rightarrow 2/\epsilon$. Now, let us expand the operator in Eq.~\eqref{eq:power}. We find for the enumerator
\begin{align}
    \langle 0\vert \hat{S}^\dag \hat{E}^\dag (t) \hat{E}(t) \hat{S}\vert0 \rangle &= \sum_{n,m} \tilde{\psi}_n (t) \tilde{\psi}_m^* (t) \langle 0 \vert \hat{S}^\dag \hat{a}_n^\dag \hat{a}_m  \hat{S}\vert 0 \rangle  \\
    &= \sum_{n,m} \tilde{\psi}_n (t) \tilde{\psi}_m^* (t) \langle 0 \vert (\hat{a}_n^\dag \cosh(\xi r_n) - \hat{b}_n \sinh(\xi r_n) (\hat{a}_m \cosh(\xi r_m) - \hat{b}_m \sinh(\xi r_m))\vert 0 \rangle \\
    &=\sum_n \sinh^2(\xi r_n) \vert \tilde{\psi}_n (t)\vert^2  .
    \end{align}
For the time duration, we have to calculate the term
    \begin{align}
    \int d t \, t^2 \vert s(t)\vert^2   = \frac{\sum_n \sinh^2(\xi r_n) \int d t \, t^2 \vert \tilde{\psi}_n(t)\vert^2 }{\sum_m \sinh^2 (\xi r_m)} ,
\end{align}
where
\begin{equation}
    \int dt \, t^2 \vert \tilde{\psi}_n(t) \vert^2 = \frac{2}{\sigma\epsilon} \left( n + \frac{1}{2} \right) ,
\end{equation}
which directly follows from the previous result in Eq.~\eqref{eq:Schmidtmodevariance}. For the other term we have
\begin{equation}
    \int dt \, t \vert \tilde{\psi}_n(t) \vert^2 = 0 .
\end{equation}
 We finally arrive at
 \begin{equation}
     \Delta T^2 = \frac{2}{\sigma\epsilon} \left( \frac{\sum_n \sinh^2 (\xi r_n) n}{\sum_m \sinh^2 (\xi r_m)} + \frac{1}{2} \right) .
 \end{equation}
In the following, we will derive analytic approximations for the time duration in three different regimes.

\subsection{QFI for no frequency entanglement} \label{App:QFInoEntanglement}
In the case of no frequency entanglement $\sigma=\epsilon$, we have $K =1$ and $r_n = 0$ for $n \neq 0$ and $r_0 = 1$. We thus find
\begin{align}
J_q = 4 \langle \partial_\mu \psi |\partial_\mu \psi\rangle  = \frac{4}{\mu^2} \sinh^2 (\xi) \left( \frac{\omega_0^2 s^2}{4} \frac{1}{2}+ \frac{1}{2} \right) &= \frac{4}{\mu^2} \sinh^2 (\xi) \left( \frac{\omega_0^2}{4 \sigma^2} + \frac{1}{2}\right) \\
&= \frac{\sinh^2(\xi)}{\mu^2}  \left( \frac{\omega_0^2}{\sigma^2} +2 \right) .
\end{align}
As $\omega_0 \gg \sigma$, the bandwidth contribution can be neglected.  For the time duration we straightforwardly find $\Delta T ^2 = 1/\sigma \epsilon$.  

\subsection{The high frequency-entanglement regime}
In the case of $\xi \sqrt{\frac{2}{K+1}} \ll 1 \Leftrightarrow \xi \ll K^{1/2}$, we have $\sinh^2(\xi r_n) \approx \xi^2 \frac{2}{K+1} (\frac{\sigma-\epsilon}{\sigma+\epsilon})^n$ and $\cosh^2 (\xi r_n) \approx 1$. With this, we find
\begin{align}
J_q(\mu) = 4 \langle \partial_\mu \psi |\partial_\mu \psi\rangle  &= 4\frac{\xi^2}{\mu^2} \frac{2}{K+1} \sum_{n} \left( \frac{\sigma-\epsilon}{\sigma+\epsilon} \right)^{2n} \left( \frac{\omega_0^2 s^2}{4} \left( \frac{n}{2}+\frac{n+1}{2} \right) + \frac{n(n-1)}{4} +\frac{(n+1)(n+2)}{4} \right)\\
 &= 4 \frac{\xi^2	}{\mu^2} \frac{2}{K+1} \sum_{n} \left( \frac{\sigma-\epsilon}{\sigma+\epsilon} \right)^{2n} \left( \frac{\omega_0^2 s^2}{4} \left( n+\frac{1}{2} \right) +\frac{n^2+n+1}{2} \right) .
\end{align}
With $\sum_{n=0}^\infty (\frac{\sigma-\epsilon}{\sigma+\epsilon})^{2n}=\frac{K+1}{2}$,  $\sum_{n=0}^\infty n(\frac{\sigma-\epsilon}{\sigma+\epsilon})^{2n}=\frac{K^2-1}{4}$,  $\sum_{n=0}^\infty n^2(\frac{\sigma-\epsilon}{\sigma+\epsilon})^{2n}=\frac{K^3-K}{4}$, we find
\begin{align}
J_q(\mu)=  \frac{\xi^2}{\mu^2} \frac{K}{K+1} \left( \frac{\omega^2_0}{\sigma\epsilon} (K+1) + K^2+K+1+\frac{2}{K}\right) .
\end{align}
In the case of $K \gg 1$, we obtain
\begin{equation}
J_q(\mu) \approx  \frac{\xi^2}{\mu^2} K\left( \frac{\omega^2_0}{\sigma\epsilon}  + K\right) .
\end{equation}
For $\frac{\omega_0^2}{\sigma \epsilon} > K$, the frequency contribution is dominant. Conversely, for $\frac{\omega_0^2}{\sigma \epsilon}< K$ the bandwidth contribution is dominant. 
  Let us come to the time duration of the signal quantum pulse. In the high-entanglement regime, we have $\sinh^2(\xi r_n) \approx \xi^2 r_n^2$. So the sum can be evaluated straightforwardly
 \begin{align}
    \sum_{n=0}^{\infty} \sinh^2 (\xi r_n) n &\approx \xi^2 \frac{2}{K+1} \sum_{n=0}^{\infty} n (q^2)^n = \xi^2 \frac{2}{K+1} \frac{q^2}{(1-q^2)^2} = \xi^2 \frac{2}{K+1} \frac{\frac{K-1}{K+1}}{(1-\frac{K-1}{K+1})^2} \\
    &= \xi^2\frac{K-1}{2} \approx \frac{\xi^2 K}{2} .
 \end{align}
 With this we obtain
 \begin{equation}
 \Delta T^2 \approx \frac{K}{\sigma \epsilon} .
 \end{equation}

\subsection{High-squeezing regime}\label{App:highsqueezing}
We define the high-squeezing regime in which the squeezing parameter is much larger than the Schmidt number, more accurately $\xi \gg K^{3/2}$.  Furthermore, we require $K>1.5$, which will help us to put the QFI in an analytic form as we will show in the following. This requirement also helps distinguishing the high-squeezing regime from the no-entanglement case with $K=1$. Let us look at the relative photon number $N_{Sn}/N_{S}$ of the first two modes, where $N_{Sn}= \sinh^2 (\xi r_n)$ and $N_S = \sum_n \sinh^2 (\xi r_n)$. We find
\begin{equation}
\frac{N_{S1}/N_S}{N_{S0}/N_S} =\frac{N_{S1}}{N_{S0}} = \frac{\sinh^2 \left(\xi \sqrt{\frac{2}{K+1}} \sqrt{\frac{K-1}{K+1}}\right)}{\sinh^2 \left(\xi \sqrt{\frac{2}{K+1}} \right)} . 
\end{equation}
Both hyperbolic sines can be approximated as exponential functions $\sinh(x) \approx \exp(x)/2$, if $\xi \sqrt{\frac{2}{K+1}} \sqrt{\frac{K-1}{K+1}} \gg 1$. This condition is satisfied in the case of $\xi \gg K^{3/2}$ and $K>1.5$, which are the conditions for the high-squeezing regime. We thus have $N_{Sn} = \sinh^2 (\xi r_n) \approx \cosh^2 (\xi r_n) \approx \exp (2\xi r_n) /4$. The ratio of relative photon numbers per mode is then approximately given by
\begin{align}
\frac{N_{S1}}{N_{S0}} \approx \frac{\exp \left(2\xi \sqrt{\frac{2}{K+1}} \sqrt{\frac{K-1}{K+1}}\right)}{\exp \left(2\xi \sqrt{\frac{2}{K+1}} \right)} = \exp \left( 2\xi \sqrt{\frac{2}{K+1}} \left(\sqrt{\frac{K-1}{K+1}} - 1 \right)\right) \ll 1,
\end{align}
which is small because the absolute value of the argument is small, that is $2\xi \sqrt{\frac{2}{K+1}} \left(1-\sqrt{\frac{K-1}{K+1}}  \right) \gg 1$, which is  equivalent to the condition $\xi \gg K^{3/2}$, the condition for the high-squeezing regime. The  mode $n=0$ in the high-squeezing regime contains the majority of photons and the signal photon number is thus approximately given by 
\begin{equation}
N_S = \sum_n \sinh^2 \left( \xi r_n \right) \approx N_{S0} = \sinh^2(\xi r_0) \approx \frac{\exp \left( 2\xi \sqrt{\frac{K-1}{K+1}} \right)}{4} .
\end{equation}
As the mode $n=0$ yields the biggest contribution in terms of photon number, and the higher modes  contain relatively few photons, we only consider the terms of the QFI due to the first few modes
\begin{multline}
J_q (\mu) = \frac{1}{\mu^2} \Bigg( \frac{\omega_0^2 s^2}{2} \left[ \sinh^2(\xi r_0)\cosh^2(\xi r_1) + \sinh^2 (\xi r_1) \cosh^2 (\xi r_0) \right] \\+ 2 \sinh^2 (\xi r_0) \cosh^2 (\xi r_2) + 6 \sinh^2 (\xi r_1) \cosh^2 (\xi r_3) + \ldots \Bigg) .
\end{multline}
The first two terms stem from the frequency contribution, the last two terms stem from the bandwidth contribution which is strongly suppressed because of $\omega_0^2 s^2 \gg 1$. Additionally, $N_{S2}< N_{S1} \ll N_{S0} $, which  justifies neglecting the bandwidth terms,  as they do not contain terms of order $\sinh^2 (\xi r_0) \cosh^2(\xi r_1)\approx \sinh^2 (\xi r_1) \cosh^2(\xi r_0) \approx N_{S0}N_{S1}$. Only the frequency contribution contains two terms containing $N_{S0} N_{S1}$, thus, these terms are the main ones contributing and we approximate
\begin{equation} \label{eq:QFIhighsqueezmode}
J_q \approx \frac{\omega_0^2}{\sigma\epsilon \mu^2} 2 N_{S0} N_{S1}, 
\end{equation}
which clearly indicates a scaling better than the SQL because $N_{S1}>1$.
 We can rewrite the QFI in terms of the total signal photon number using the fact that $N_{S1}\sim N_S^{\sqrt{(K-1)/(K+1)}}$. We find 
\begin{equation}
J_q(\mu) \approx \frac{\omega_0^2}{8 \mu^2 \sigma \epsilon} \left( 4 N_S\right)^{1+ \sqrt{\frac{K-1}{K+1}}} .
\end{equation}
 For $K$ big enough, one nearly attains  the Heisenberg limit.

Let us now come to the time duration of the signal pulse. We have $\sinh^2 (\xi r_1) /N_S \ll 1$ in the high-squeezing regime. With this we find
 \begin{equation}
     (\Delta T)^2 \approx \frac{N_{S0} \left(  \frac{2}{\sigma\epsilon} \left( 0+\frac{1}{2}\right)\right)+ N_{S1} \left(  \frac{2}{\sigma\epsilon} \left( 1+\frac{1}{2}\right)\right)}{N_{S0}+N_{S1}}  = \frac{1}{\sigma\epsilon}\frac{N_{S0}+3N_{S1}}{N_{S0}+N_{S1}} \approx \frac{1}{\sigma\epsilon} ,
 \end{equation} 
 where we have neglected all modes with $n>1$, because their contribution is relatively small.

\subsection{The mixed regime}\label{App:mixed}
In the mixed regime, we require $\xi \gg K^{1/2}$ and $\xi \ll K^{3/2}$, from which follows $\xi \gg 1$ and $K\gg 1$. We want to relate the photon number to the QFI and to do so we derive asymptotic analytic expressions of sums containing hyperbolic trigonometric functions.
We introduce the parameters $x = \xi \sqrt{2/(K+1)} \gg 1$ and $q = \sqrt{(K-1)/(K+1)}$ satisfying $0< 1-q \ll 1$ as convenient parameters for the derivations that follow. Let us start with
\begin{align}
   \sum_{n=0}^{\infty} \sinh^2 \left( x q^n \right)  &=    \sum_{n=0}^{\infty}   \sum_{m=1}^{\infty}  \frac{2^{-1+2m} x^{2m}}{(2m)!} (q^n)^{2m} =    \sum_{m=1}^{\infty}  \frac{2^{-1+2m} x^{2m}}{(2m)!} \sum_{n=0}^{\infty}  (q^{2m})^{n}  \label{ir1}  \\
   & =  \sum_{m=1}^{\infty}  \frac{2^{-1+2m} x^{2m}}{(2m)!} \frac{1}{1-q^{2m}} \approx \sum_{m=1}^{\infty}  \frac{2^{-1+2m} x^{2m}}{(2m)!} \frac{1}{2m(1-q)} \label{ir2} \\ %https://www.wolframalpha.com/input?i=expansion+1%2F%281-x%5Em%29+at+x%3D1
   &=\sum_{m=1}^{\infty}  \frac{2^{-1+2m+1} x^{2m+1}}{2m(2m)!} \frac{1}{2x} \frac{1}{1-q}  \approx \sinh^2 (x) \frac{1}{2x} \frac{1}{1-q}  ,
\end{align}
where we changed the summation order in Eq.~\eqref{ir1},  allowed by Fubini's theorem because the series is absolutely convergent. The approximation in Eq.~\eqref{ir2} is asymptotically valid if $q \approx 1$.
We remind that the above sum in Eq.~\eqref{ir1} is the equal to the expected number of signal photons $ N_S = \sum_n \sinh^2 (\xi \sqrt{\frac{2}{K+1}} \sqrt{\frac{K-1}{K+1}}^n)$. In this regime, we can approximate $\sinh^2(x q) \approx \exp(2xq)/4$ because of $\xi \gg K^{1/2}$. We derive at the asymptotic limit for the photon number 
\begin{align}
   N_S \approx \frac{\exp \left( 2 \xi \sqrt{\frac{2}{K+1}} \right)}{4 \cdot 2 \xi \sqrt{\frac{2}{K+1}} \left(1- \sqrt{\frac{K-1}{K+1}}\right)} .
\end{align}  
Now, let us examine the frequency contribution of the QFI $J_q$. To do so, we examine the expression
\begin{align}
 \sum_{n=0}^{\infty} n^k \sinh^2 (x q^n) \cosh^2 (x q^n) &= \sum_{n=0}^{\infty} n^k \frac{1}{8}\left( \cosh (4x q^n) -1\right) = \frac{1}{8} \sum_{n=0}^{\infty} n^k \sum_{m=1}^{\infty} \frac{(4xq^n)^{2m}}{(2m)!} \\
 &= \frac{1}{8} \sum_{m=1}^{\infty} \frac{(4x)^{2m}}{(2m)!} \sum_{n=0}^{\infty} n^k (q^{2m})^n ,
 \end{align}
where $k=0,1$. Let us start with $k=0$
\begin{align}
\frac{1}{8} \sum_{m=1}^{\infty} \frac{(4x)^{2m}}{(2m)!} \sum_{n=0}^{\infty}  (q^{2m})^n &= \frac{1}{8} \sum_{m=1}^{\infty} \frac{(4x)^{2m}}{(2m)!} \frac{1}{1- q^{2m}} \approx \frac{1}{8} \sum_{m=1}^{\infty} \frac{(4x)^{2m}}{(2m)!} \frac{1}{2m(1- q)} \\
&= \frac{1}{8\cdot 4 x (1-q)} \sum_{m=1}^{\infty} \frac{(4 x)^{2m+1}}{2m (2m)!} \approx \frac{1}{8\cdot 4 x (1-q)} \cosh (4x) .
\end{align}
For $k=1$ we have 
\begin{align}
\frac{1}{8} \sum_{m=1}^{\infty} \frac{(4x)^{2m}}{(2m)!} \sum_{n=0}^{\infty} n (q^{2m})^n &= \frac{1}{8} \sum_{m=1}^{\infty} \frac{(4x)^{2m}}{(2m)!} \frac{q^{2m}}{(1- q^{2m})^2} \approx \frac{1}{8} \sum_{m=1}^{\infty} \frac{(4x)^{2m}}{(2m)!} \frac{1}{(2m)^2(1- q)^2} \\
&= \frac{1}{8\cdot (4 x)^2 (1-q)^2} \sum_{m=1}^{\infty} \frac{(4 x)^{2m+2}}{2m \cdot 2m \cdot (2m)!} \approx \frac{1}{8\cdot (4 x)^2 (1-q)^2} \cosh (4x) .
\end{align}
In the mixed regime we have the condition $\xi \ll K^{3/2}$, which implies $\cosh^2(\xi  r_{n \pm 1} ) \approx \sinh^2 (\xi r_n)$ asymptotically. With this, we find that for the frequency contribution with $x =  \xi \sqrt{\frac{2}{K+1}}$ and $q=\sqrt{\frac{K-1}{K+1}}$
\begin{align}
 Z_{\overline{\omega}} &= \sum_n \sinh^2 \left(\xi r_n\right) \left[ n  \cosh^2 \left(\xi r_{n-1}\right) + (n+1) \cosh^2 \left(\xi r_{n+1} \right) \right] \\
 &\approx 2 \sum_{n=0}^{\infty} n \sinh^2 (\xi r_n)  \cosh^2 (\xi r_{n})   +  \sum_{n=0}^{\infty}  \sinh^2 (\xi r_n)  \cosh^2 (\xi r_{n})  \\
 &\approx 2  \frac{1}{8\cdot \left(4  \xi \sqrt{\frac{2}{K+1}} \right)^2 \left(1-\sqrt{\frac{K-1}{K+1}}\right)^2} \cosh \left(4  \xi \sqrt{\frac{2}{K+1}} \right) + \frac{1}{8\cdot 4  \xi \sqrt{\frac{2}{K+1}} \left(1-\sqrt{\frac{K-1}{K+1}}\right)} \cosh \left(4  \xi \sqrt{\frac{2}{K+1}} \right)  \\
 &\approx 2 \frac{1}{8\cdot \left(4  \xi \sqrt{\frac{2}{K+1}} \right)^2 \left(1-\sqrt{\frac{K-1}{K+1}}\right)^2} \frac{\exp \left(4  \xi \sqrt{\frac{2}{K+1}} \right)}{2}\\
 &= \frac{\exp \left( 4 \xi \sqrt{\frac{2}{K+1}} \right)}{2 \cdot 4^2 \cdot (2 \xi \sqrt{\frac{2}{K+1}})^2 \left(1- \sqrt{\frac{K-1}{K+1}}\right)^2}  =  \frac{N_S^2}{2} .
\end{align}
Finally, let us examine the bandwidth contribution to the QFI for which we study the series
\begin{align}
 \sum_{n=0}^{\infty} n^2 \sinh^2 (x q^n) \cosh^2 (x q^n) &= \frac{1}{8} \sum_{m=1}^{\infty} \frac{(4x)^{2m}}{(2m)!} \sum_{n=0}^{\infty} n^2 (q^{2m})^n = \frac{1}{8} \sum_{m=1}^{\infty} \frac{(4x)^{2m}}{(2m)!} \frac{q^{2m}(q^{2m} +1)}{(1-q^{2m})^3} \\
 &\approx \frac{1}{8} \sum_{m=1}^{\infty} \frac{(4x)^{2m}}{(2m)!} \frac{2}{(2m)^3 (1-q)^3} = \frac{2}{8 (4x)^3 (1-q)^3} \sum_{m=1}^{\infty} \frac{(4x)^{2m+3}}{2m\cdot 2m\cdot 2m \cdot (2m)!} \\
 &\approx  \frac{2}{8 (4x)^3 (1-q)^3} \cosh (4x) .
 \end{align}
The bandwidth contribution is given by
\begin{align}
 Z_{\overline{\sigma}}=&\sum_n \sinh^2 (\xi r_n) \left[ n (n-1) \cosh^2 (\xi r_{n-2}) + (n+1)(n+2) \cosh^2 (\xi r_{n+2}) \right] \\
& \approx \sum_n \sinh^2 (\xi r_n)  \left[ (n^2-n) \cosh^2 (\xi r_{n-2})+ (n^2+3n+2) \cosh^2 (\xi r_{n+2}) \right] \\ 
 & \approx 2 \sum_n n^2 \sinh^2 (\xi r_n) \cosh^2 (\xi r_n) = \frac{2\cdot 2}{8} \frac{1}{\xi \sqrt{ \frac{2}{K+1}} \left( 1- \sqrt{\frac{K-1}{K+1}}\right)} \frac{\cosh \left( 4 \xi \sqrt{\frac{2}{K+1}} \right)}{4^3 \left( \xi \sqrt{ \frac{2}{K+1}} \right)^2 \left( 1- \sqrt{\frac{K-1}{K+1}} \right)^2}\\
&\approx \frac{1}{4 \xi \sqrt{\frac{2}{K+1}} \left(1- \sqrt{\frac{K-1}{K+1}}\right)} N_S^2 \approx \frac{K^{3/2}}{2^{5/2} \xi}N_S^2 .  \label{eq:bandwidthcontribution}
\end{align} 
Here we only kept the term of leading order in $1/(1-q)$ because we only consider the asymptotic limit. We also used $\cosh^2 (\xi r_{n\pm 2}) \approx \sinh^2 (\xi r_n)$, which is valid for $\xi \ll K^{3/2}$.

\begin{figure}[h!]
\includegraphics[width=0.4 \linewidth]{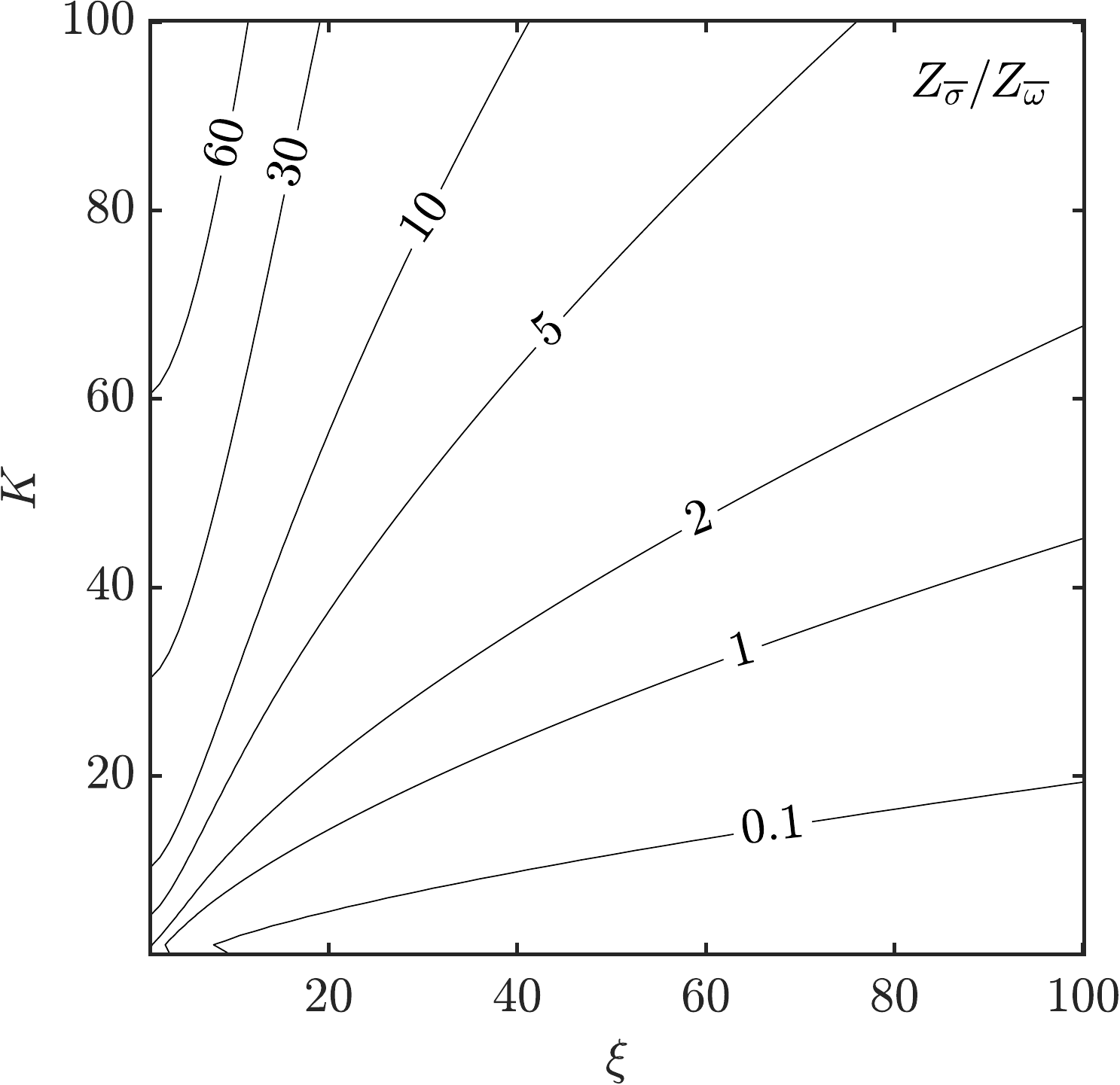}% Here is how to import EPS art
\caption{\label{fig:comparison} The two terms $Z_{\overline{\sigma}}$ and $Z_{\overline{\omega}}$  that make up the QFI are compared by considering the contours of their ratio. The bandwidth term $Z_{\overline{\sigma}}$ is up to $100$ times bigger than its counterpart $Z_{\overline{\omega}}$ in the mixed regime in the considered parameter range. Because the frequency contribution is further suppressed by the factor $\sigma\epsilon/\omega_0^2 (\sim 10^{-4}$ typically), the bandwidth contribution can be neglected in most experimental scenarios. }
\end{figure} In the mixed regime, the prefactor in Eq.~\eqref{eq:bandwidthcontribution} is big, that is  $K^{3/2}/2^{5/2 }\xi \gg 1$. However, the bandwidth contribution is suppressed by the factor $\sigma \epsilon / \omega_0^2$, which is in most experimental setups small, of the order of $10^{-4}$.  A direct comparison between frequency and bandwidth contribution is made in Fig.~\ref{fig:comparison}. The bandwidth contribution can be safely neglected in the considered parameter range.

Now, let us calculate the asymptotic limit of the time duration for the signal beam in the mixed regime. We start by considering the sum
 \begin{align}
     \sum_{n=0}^{\infty} \sinh^2 (x q^n) n &= \sum_{m=1}^\infty \frac{2^{-1+2m}x^{2m}}{(2m)!} \sum_{n=0}^\infty n (q^{2m})^{n} = \sum_{m=1}^\infty \frac{2^{-1+2m}x^{2m}}{(2m)!} \frac{q^{2m}}{(1-q^{2m})^2} \\
     &\approx  \sum_{m=1}^\infty \frac{2^{-1+2m}x^{2m}}{(2m)!} \frac{1}{(2m)^2 (1-q)^2} =  \sum_{m=1}^\infty \frac{2^{-1+2m+2}x^{2m+2}}{2m\cdot 2m\cdot(2m)!} \frac{1}{2^2 x^2 (1-q)^2} \\
     &\approx \frac{\sinh^2 (x)}{2^2 x^2 (1-q)^2} = \frac{\sinh^2 (x)}{2 x (1-q)} \frac{1}{2 x (1-q)} ,
 \end{align}
 with $x = \xi \sqrt{\frac{2}{K+1}}$ and $q= \sqrt{\frac{K-1}{K+1}}$, we find 
 \begin{align}
     \sum_{n=0}^{\infty} \sinh^2 (\xi r_n) n \approx N_S \frac{1}{2\xi \sqrt{\frac{2}{K+1}} \left(1- \sqrt{\frac{K-1}{K+1}} \right)}\approx N_S \frac{K^{3/2}}{2^{3/2} \xi} ,
 \end{align}
 where the last approximation is valid in the asymptotic limit $K \gg 1$.
 With this we obtain
 \begin{equation}
     \Delta T^2 \approx \frac{1}{\sigma\epsilon} \frac{K^{3/2}}{2^{1/2}\xi } .
 \end{equation}
 We note that all of these expressions are the asymptotic limits. For values $\xi\approx K\approx 100$, these expressions already give reasonable approximations, with relative errors around $10-20\%$.

\section{Photon loss analysis}\label{supp:loss}
Let us now discuss photon loss for the  high-squeezing regime. We assume that the idler beam suffers no losses. The signal beam, however, can suffer losses on the way from the emitter to the target, at the target itself due to imperfect reflection, and on the way from the target to the receiver.  The loss is modelled as a frequency independent beam splitter 
\begin{equation}
\hat{U}_B \hat{a}(\omega) \hat{U}_B^\dag =  \sqrt{\eta} \hat{a}(\omega) + \sqrt{1-\eta} \hat{c}(\omega) ,
\end{equation}
which couples signal mode $\hat{a}(\omega)$ to an auxiliary mode $\hat{c}(\omega)$ which the experimentator has no access to and that will be traced out at the end.
Now, let us examine how this transformation acts on the Schmidt modes
\begin{align}
\hat{U}_B \hat{a}_n \hat{U}_B^\dag &=   \int \diff \omega \psi_n (\omega-\omega_0/2) \hat{U}_B \hat{a}(\omega) \hat{U}_B^\dag =    \int \diff \omega \psi_n (\omega-\omega_0/2) \left(\sqrt{\eta} \hat{a}(\omega) + \sqrt{1-\eta} \hat{c}(\omega) \right)  \\
&=\sqrt{\eta} \hat{a}_n+ \sqrt{1-\eta} \hat{c}_n ,
\end{align}
where $\hat{c}_n = \int \diff \omega \psi_n (\omega-\omega_0/2)  \hat{c}(\omega) $ is the auxiliary mode in the Schmidt mode basis. It is apparent that the beam splitter operation commutes with the Doppler reflection.

Because the beam splitter operator commutes with the reflection operator $\hat{U}_\mu$, we can introduce an effective round-trip transmissivity parameter $\eta$. In the upcoming calculation, we will introduce the loss after the reflection.
As there are only a couple of relevant modes in the high-squeezing regime, we switch to the Gaussian formalism, which allows us to fully describe the probe state by its covariance matrix. We introduce the basis
\begin{equation}
\hat{\textbf{R}} = \left( \hat{a}_{0}, \hat{a}_{0}^\dag,\hat{a}_{1}, \hat{a}_{1}^\dag,\hat{a}_{2}, \hat{a}_{2}^\dag,\hat{b}_{0}, \hat{b}_{0}^\dag,\hat{b}_{1}, \hat{b}_{1}^\dag,\hat{c}_{0}, \hat{c}_{0}^\dag,\hat{c}_{1}, \hat{c}_{1}^\dag,\hat{c}_{2}, \hat{c}_{2}^\dag \right)^{T} .
\end{equation}
All higher modes are neglected, which corresponds to tracing them out. By neglecting these higher modes, the end result of the QFI will be a lower bound. The initial probe state before the Doppler reflection and photon loss is in the chosen basis a product state of squeezed vacua given by
\begin{equation}
\Upsigma = \textbf{S}_{\hat{a}_1 \hat{b}_1}   \textbf{S}_{\hat{a}_0 \hat{b}_0}  \textbf{I}  \textbf{S}_{\hat{a}_0 \hat{b}_0}^\dag  \textbf{S}_{\hat{a}_1 \hat{b}_1}^\dag ,
 \end{equation}
 where $ \textbf{I}$ is the identity and
 \begin{equation}
  \textbf{S}_{\hat{a}_n \hat{b}_n} =  \begin{pmatrix}
\cosh (r_n) & 0 & 0 &  -\sinh (r_n) \\
0 & \cosh (r_n) & - \sinh (r_n) & 0 \\
0 & -\sinh (r_n) & \cosh (r_n) & 0 \\
-\sinh (r_n)& 0 & 0  & \cosh (r_n)  
\end{pmatrix} .
 \end{equation}
 Note that the matrix is a $16\times 16$ matrix, but we we have only given the relevant elements of the matrix.
 
Now, let us derive the corresponding matrix that describes the Doppler reflection. The Doppler reflection transforms the modes as $\hat{U}_\mu \hat{a}_n \hat{U}_\mu^\dag = \hat{a}_{n\mu}$. Therefore, the parameter dependence is fully contained within the basis, i.e. the modes $\hat{a}_{n\mu}$, and there is no parameter dependence in the elements of the covariance matrix. To use the formula for the QFI in Eq.\eqref{eq:QFISafranek}, all of the parameter dependence has to be in the elements of the covariance matrix. The basis has to be parameter independent. To achieve parameter independence of the basis, i.e. the modes, we have to introduce a prior known estimate of the parameter $\mu_0$, and expand the modes around this prior. We assume that the true value of the parameter is given by $\mu = \mu_0 + \delta$, where $\delta$ is a small deviation. We will only consider the expansion up to first order in $\delta$, to keep the amount of modes that have to be considered small, and we obtain using properties of the Hermite functions
\begin{align}
\hat{a}_{0\mu} & \approx \hat{a}_{0\mu_0} - \delta \frac{\omega_0}{2\mu \sqrt{\sigma\epsilon}} \hat{a}_{1\mu_0} \\
\hat{a}_{1\mu} & \approx \hat{a}_{1\mu_0} + \delta \frac{\omega_0}{2\mu \sqrt{\sigma\epsilon}}  \left( \hat{a}_{0\mu_0} - \sqrt{2} \hat{a}_{2\mu_0} \right) .
\end{align} The symplectic matrix that corresponds to this infinitesimal change of basis is
 \begin{equation}
\Uplambda =  \begin{pmatrix}
1 & 0 & \delta \frac{\omega_0}{2\mu_0 \sqrt{\sigma\epsilon}}& 0  & 0 & 0\\
0 & 1 & 0 &\delta \frac{\omega_0}{2\mu_0 \sqrt{\sigma\epsilon}} & 0 & 0\\
-\delta \frac{\omega_0}{2\mu_0 \sqrt{\sigma\epsilon}}& 0 & 1 & 0  & \sqrt{2} \delta \frac{\omega_0}{2\mu_0 \sqrt{\sigma\epsilon}} & 0\\
0 & -\delta \frac{\omega_0}{2\mu_0 \sqrt{\sigma\epsilon}}& 0 & 1  & 0 & \sqrt{2}\delta \frac{\omega_0}{2\mu_0 \sqrt{\sigma\epsilon}} \\
0 & 0 & -\sqrt{2}\delta \frac{\omega_0}{2\mu_0 \sqrt{\sigma\epsilon}} & 0  & 1 & 0\\
0 & 0 & 0 & -\sqrt{2}\delta \frac{\omega_0}{2\mu_0 \sqrt{\sigma\epsilon}}& 0 & 1\\
\end{pmatrix}  ,
 \end{equation}
 where again we only show the matrix' relevant entries. The covariance matrix after the Doppler reflection is given by
 \begin{equation}
\Upsigma_\delta = \Uplambda \Upsigma \Uplambda^\dag .
 \end{equation}
 At the last step we introduce the photon loss operation, which can be modelled as the beam splitter transformation $\hat{U}_B \hat{a}_n \hat{U}_B^\dag =\sqrt{\eta} \hat{a}_n+ \sqrt{1-\eta} \hat{c}_n$. This transformation in the Gaussian formalism corresponds to the symplectic matrix given by
 \begin{equation}
\textbf{B}_\eta =  \begin{pmatrix}
\sqrt{\eta} & 0 & \sqrt{1-\eta} & 0 \\
0 & \sqrt{\eta} & 0 & \sqrt{1-\eta} \\
-\sqrt{1-\eta} & 0 & \sqrt{\eta} & 0 \\
0 & - \sqrt{1-\eta} & 0  & \sqrt{\eta} 
\end{pmatrix} ,
 \end{equation}
again we only show the relevant entries of the matrix. The state after photon loss and after tracing out the auxiliary modes is then given by
 \begin{equation}
\Upsigma_{\delta,\eta} = \text{tr}_c  \left[ \textbf{B}_\eta \Upsigma_\delta  \textbf{B}_\eta^\dag \right] .
 \end{equation}
The tracing operation is simply done by removing the rows and columns corresponding to $\hat{c}_n$ and $\hat{c}_n^\dag$. We find
\begin{equation*}
\resizebox{\hsize}{!}{ $
\begin{aligned}
\Upsigma_{\mu,\eta} &=
\left(\begin{matrix}
 2 \eta  N_{S0}+1 & 0 & -2 \delta  \frac{\omega_0}{2\mu_0 \sqrt{\sigma\epsilon}}  \eta ( N_{S0}-  N_{S1}) & 0 & 0 \\
 0 & 2 \eta  N_{S0}+1 & 0 & -2 \delta \frac{\omega_0}{2\mu_0 \sqrt{\sigma\epsilon}}  \eta  (  N_{S0}-  N_{S1}) & 0 \\
 -2 \delta \frac{\omega_0}{2\mu_0 \sqrt{\sigma\epsilon}}  \eta ( N_{S0}- N_{S1}) & 0 & 2 \eta  N_{S1}+1 & 0 & -2 \sqrt{2} \frac{\omega_0}{2\mu_0 \sqrt{\sigma\epsilon}}  \eta  \delta N_{S1}    \\
  0 & -2 \delta\frac{\omega_0}{2\mu_0 \sqrt{\sigma\epsilon}}  \eta  (  N_{S0}-  N_{S1}) & 0 & 2 \eta  N_{S1}+1 & 0 \\
   0 & 0 & -2 \sqrt{2} \frac{\omega_0}{2\mu_0 \sqrt{\sigma\epsilon}}  \eta  \delta N_{S1} & 0 & 1\\
    0 & 0 & 0 & -2 \sqrt{2} \frac{\omega_0}{2\mu_0 \sqrt{\sigma\epsilon}}  \eta  \delta N_{S1} & 0 \\     
     0 & -2 \sqrt{\eta } \sqrt{N_{S0}} \sqrt{N_{S0}+1} & 0 & 2 \frac{\omega_0}{2\mu_0 \sqrt{\sigma\epsilon}}  \sqrt{\eta } \delta \sqrt{N_{S0}} \sqrt{N_{S0}+1} & 0 \\
 -2 \sqrt{\eta } \sqrt{N_{S0}} \sqrt{N_{S0}+1} & 0 & 2 \frac{\omega_0}{2\mu_0 \sqrt{\sigma\epsilon}}  \sqrt{\eta } \delta \sqrt{N_{S0}} \sqrt{N_{S0}+1} & 0 & 0 \\
 0 & -2 \frac{\omega_0}{2\mu_0 \sqrt{\sigma\epsilon}}  \sqrt{\eta } \delta \sqrt{N_{S1}} \sqrt{N_{S1}+1} & 0 & -2 \sqrt{\eta } \sqrt{N_{S1}} \sqrt{N_{S1}+1} & 0\\
 -2 \frac{\omega_0}{2\mu_0 \sqrt{\sigma\epsilon}}  \sqrt{\eta } \delta \sqrt{N_{S1}} \sqrt{N_{S1}+1} & 0 & -2 \sqrt{\eta } \sqrt{N_{S1}} \sqrt{N_{S1}+1} & 0 & 2 \sqrt{2} \frac{\omega_0}{2\mu_0 \sqrt{\sigma\epsilon}}  \sqrt{\eta } \delta
   \sqrt{N_{S1}} \sqrt{N_{S1}+1} 
     \end{matrix}\right.\\
&\qquad\qquad
\left.\begin{matrix}
 0 & 0 & -2 \sqrt{\eta } \sqrt{N_{S0}} \sqrt{N_{S0}+1} & 0 & -2 \frac{\omega_0}{2\mu_0 \sqrt{\sigma\epsilon}} 
   \sqrt{\eta } \delta \sqrt{N_{S1}} \sqrt{N_{S1}+1} \\
 0 & -2 \sqrt{\eta } \sqrt{N_{S0}} \sqrt{N_{S0}+1} & 0 & -2 \frac{\omega_0}{2\mu_0 \sqrt{\sigma\epsilon}} 
   \sqrt{\eta } \delta \sqrt{N_{S1}} \sqrt{N_{S1}+1} & 0 \\
     0 & 0 & 2 \frac{\omega_0}{2\mu_0 \sqrt{\sigma\epsilon}}  \sqrt{\eta } \delta
   \sqrt{N_{S0}} \sqrt{N_{S0}+1} & 0 & -2 \sqrt{\eta } \sqrt{N_{S1}} \sqrt{N_{S1}+1} \\
    -2 \sqrt{2} \frac{\omega_0}{2\mu_0 \sqrt{\sigma\epsilon}}  \eta  \delta N_{S1} & 2 \frac{\omega_0}{2\mu_0 \sqrt{\sigma\epsilon}}  \sqrt{\eta } \delta
   \sqrt{N_{S0}} \sqrt{N_{S0}+1} & 0 & -2 \sqrt{\eta } \sqrt{N_{S1}} \sqrt{N_{S1}+1} & 0 \\
     0 & 0 & 0 & 0 & 2 \sqrt{2} \frac{\omega_0}{2\mu_0 \sqrt{\sigma\epsilon}}  \sqrt{\eta } \delta \sqrt{N_{S1}} \sqrt{N_{S1}+1} \\
      1 & 0 & 0 & 2 \sqrt{2} \frac{\omega_0}{2\mu_0 \sqrt{\sigma\epsilon}}  \sqrt{\eta } \delta \sqrt{N_{S1}} \sqrt{N_{S1}+1} & 0 \\
      0 & 2 N_{S0}+1 & 0 & 0 & 0 \\
      0 & 0 & 2 N_{S0}+1 & 0 & 0 \\
       2 \sqrt{2} \frac{\omega_0}{2\mu_0 \sqrt{\sigma\epsilon}}  \sqrt{\eta } \delta  \sqrt{N_{S1}} \sqrt{N_{S1}+1} & 0 & 0 & 2 N_{S1}+1 & 0 \\
        0 & 0 & 0 & 0 & 2 N_{S1}+1 \\
\end{matrix}\right) .
\end{aligned}
$ } 
\end{equation*}
With the formula for the QFI given in Section~\ref{sec:gaussianstates} and Ref.~\cite{safranek2018}, we obtain
\begin{align}
J_q (\delta) \Big| _{\delta =0} \equiv J_q (\mu_0) &= \frac{\omega_0^2}{\mu_0^2 \sigma\epsilon} \eta \frac{  \left(N_{S0}^2 (2 N_{S1}+1)+2 N_{S0} N_{S1} ((3-2 \eta ) N_{S1}+2)+3 N_{S1}^2\right)}{2 (1-\eta) N_{S0}
   N_{S1}+N_{S0}+N_{S1}} \\
   &=  \frac{\omega_0^2  \Delta T^2\eta }{\mu_0^2} \frac{N_{S0}+N_{S1}}{N_{S0}+3N_{S1}}  \frac{  \left(N_{S0}^2 (2 N_{S1}+1)+2 N_{S0} N_{S1} ((3-2 \eta ) N_{S1}+2)+3 N_{S1}^2\right)}{2 (1-\eta) N_{S0}
   N_{S1}+N_{S0}+N_{S1}}  . \label{eq:QFIloss}
\end{align}
For the lossless case, that is $\eta = 1$, we find using $N_{S0} \gg N_{S1}$
\begin{equation}
J_{q}(\mu_0) \approx \frac{\omega_0^2}{\mu_0^2 \sigma \epsilon} 2 N_{S0} N_{S1},
\end{equation}
which is the same result we had previously obtained in Eq.~\eqref{eq:QFIhighsqueezmode} with QFI formula for pure states, thus confirming the validity of our approach.

For the case with losses, that is $\eta \neq 1$, and the conditions $N_{S0} \gg N_{S1}$ and $(1-\eta) N_{S1} \gg 1$, we find
\begin{equation}
J_q (\mu_0) \approx  \frac{\omega_0^2  \Delta T^2}{\mu_0^2} N_{S0} \frac{\eta}{1-\eta} .
\end{equation}
The QFI of the classical strategy is given by $J_c \approx \eta \omega_0^2 N_S \Delta T^2$, and with this we find for the ratio
\begin{equation}
\frac{J_q}{J_c} \approx \frac{1}{1-\eta} .
\end{equation}
For the SLD we find with the formula given in Section~\ref{sec:gaussianstates} and Ref.~\cite{safranek2018}, and in the limit $N_{S1}(1-\eta) \gg 1$, the expression
\begin{equation}
\hat{L}_{\mu_0} =\frac{\omega_0}{\mu_0 \sqrt{\sigma\epsilon}}  \frac{ \sqrt{\eta }}{1-\eta}\left(-\hat{a}_{0\mu_0} \hat{b}_{1}+\hat{a}_{1\mu_0} \left(\sqrt{2} \hat{a}_{2\mu_0}^\dag \sqrt{\eta }+\hat{b}_{0}\right)+\sqrt{2} \hat{a}_{2\mu_0} \hat{b}_{1} +h.c. \right) .
\end{equation}
The eigenbasis of this operator constitutes an optimal measurement for the lossy scenario.

\section{Frequency-resolved photon counting }\label{app:measurement}
We want to prove that  frequency-resolved  photon counting attains the accuracy given by the QFI $J_q$ that was derived in the main text for the lossless case. Our proof starts as follows: We first show that $F(\mu) = 4 \langle \partial_\mu \psi_\mu | \partial_\mu \psi_\mu \rangle = J_q (\mu)$, if the POVM and the state satisfy the conditions
\begin{enumerate}
\item[1)] The POVM elements are  one dimensional projectors, $\Pi_z = |z\rangle \langle z|$.
\item[2)] The measurement does not depend on the parameter $\mu$, $\partial_\mu |z\rangle = 0$. 
\item[3)] \label{item3} The amplitudes are real, $\langle z | \psi_\mu \rangle = \langle \psi_\mu | z \rangle$.
\end{enumerate}
Using these conditions, we obtain for the classical Fisher information
\begin{align}
F(\mu) &= \int \diff z \frac{1}{\langle z|\psi_\mu \rangle^2} \left( \partial_\mu \langle z|\psi_\mu \rangle^2  \right)^2 =  \int \diff z \frac{1}{\langle z|\psi_\mu \rangle^2} \left( 2 \langle z|\psi_\mu \rangle \langle z| \partial_\mu \psi_\mu \rangle \right)^2 = 4 \int \diff z \langle z |\partial_\mu \psi_\mu \rangle^2 \\
&=  4 \int \diff z \langle \partial_\mu \psi_\mu | z \rangle \langle z | \partial_\mu \psi_\mu \rangle =4 \langle \partial_\mu \psi_\mu |\partial_\mu \psi_\mu \rangle ,
\end{align}
which is indeed equal to the QFI.

Now, let us look at the frequency-resolved photon counting corresponding to projection onto the state
\begin{equation}
|\bm{\nu} ,  \tilde{\bm{\nu}} \rangle= \frac{1}{\sqrt{n!}} \frac{1}{\sqrt{m!}} \otimes_{i=1}^n \otimes_{j=1}^{m} a^\dag(\nu_i) b^\dag (\tilde{\nu}_j) \vert 0 \rangle,
\end{equation}
where $n,m\in \mathds{R}$ gives the number of detected signal and idler photons respectively and $ \nu_i, \tilde{\nu}_j \in \mathds{R}_{>0}$ are the detected frequencies of each photon. The corresponding set of POVM operators are $\{| \bm{\nu} , \tilde{\bm{\nu}} \rangle \langle\bm{\nu} ,\tilde{\bm{\nu}}  |  \big| n,m \in \mathds{N}$, \, $\nu_i,\tilde{\nu}_j \in \mathds{R}_{>0} \}$Due to the assumed absence of photon loss and thermal noise, the number of signal and idler photons is the same $n=m$ for each measurement outcome.  Conditions 1) and 2) are clearly satisfied. Condition 3) is also satisfied if we take the state $|\psi_\mu \rangle$ from the main text and thus, for this particular state the measurement is optimal. In the main text, we neglected the phase factors the state acquires due to its propagation. However, when a specific measurement is considered such as $\{|\bm{\nu },  \tilde{\bm{\nu}} \rangle \langle \bm{\nu} ,  \tilde{\bm{\nu}}| \}$, the phase factors could play a role. This is why we introduce them here, to keep the analysis as general as possible
\begin{equation}
|\tilde{\psi}_\mu \rangle = \exp \left(- |\xi| \text{e}^{ \text{i} \zeta} \int \diff \omega \int \diff \tilde{\omega} \mu^{1/2} f(\mu \omega, \tilde{\omega}) \hat{a}(\omega) \text{e}^{\text{i} \omega \varphi} \hat{b}(\tilde{\omega}) \text{e}^{\text{i} \tilde{\omega} \vartheta}  + h.c. \right) |0\rangle , 
\end{equation}
where we now also allow for a complex squeezing parameter $\xi = |\xi| \text{e}^{ \text{i} \zeta}$.
The phase $\varphi = \varphi (\mu, x_m)$ of the signal mode generally depends on the parameter $\mu$ and the distance $x_m$ between the moving object and the emitter.  We also note that if multiple runs of the experiment are performed, the phase generally changes due to the movement of the object.
With this more general state, it is not immediately clear that  3) is satisfied, most certainly it is not. So let us now study the performance of this state for the given measurement. First we introduce the discrete operators

\begin{align}
\hat{\tilde{a}}_{n\mu} &= -\int \diff \omega \, \mu^{1/2} \psi_n (\mu\omega-\omega_0/2) \text{e}^{\text{i} \omega \varphi}\hat{a} (\omega)   \\
\hat{\tilde{b}}_n &= e^{i\zeta} \int \diff \tilde{\omega} \, \psi_n (\tilde{\omega}-\omega_0/2)\text{e}^{\text{i} \tilde{\omega} \vartheta}  \hat{b} (\tilde{\omega}) ,
\end{align}
where we have absorbed the complex phase of the squeezing parameter into $\tilde{\hat{b}}_n$.
With this, the state can be written as
\begin{equation}
|\tilde{\psi}_\mu \rangle =  \mathcal{N} \exp \left( - \sum_n \tanh (|\xi| r_n) \hat{\tilde{a}}_n^\dag \hat{\tilde{b}}_n^\dag \right) |0\rangle ,
\end{equation}
where $ \mathcal{N}= \prod_n 1/\cosh(|\xi | r_n)$.
Let us now  write this state  in terms of the continuous frequencies
\begin{align}
|\tilde{\psi}_\mu \rangle &= \mathcal{N} \exp \Bigg( -\int \diff \omega \int \diff \tilde{\omega}  \underbrace{\sum_{n} \tanh (|\xi| r_{n}) \mu^{1/2} \psi_{n} ( \mu \omega-\omega_0/2) \psi_{n} (\tilde{\omega}-\omega_0/2)}_{\equiv t_\mu (\omega, \tilde{\omega})} \text{e}^{-\text{i} \omega \varphi} \text{e}^{-\text{i} \tilde{\omega} \vartheta} \text{e}^{-\text{i} \zeta} \hat{a}^\dag(\omega) \hat{b}^\dag (\tilde{\omega} )\Bigg) |0\rangle \\
&= \mathcal{N}\exp \Bigg( -\int \diff \omega \int \diff \tilde{\omega} \,  t_\mu (\omega, \tilde{\omega}) \text{e}^{-\text{i} \omega \varphi} \text{e}^{-\text{i} \tilde{\omega} \vartheta} \text{e}^{-\text{i} \zeta} \hat{a}^\dag(\omega) \hat{b}^\dag (\tilde{\omega} )\Bigg) |0\rangle .
\end{align}
We have written the state in this form to easily find the $2m$-photon contributions.
So let us project onto the $2m$-photon state $\langle \bm{\nu},\tilde{\bm{\nu}} | = \frac{1}{m!}\langle 0| \hat{a}(\nu_1) \hat{b}(\tilde{\nu}_1) \cdot \ldots \cdot \hat{a}(\nu_m) \hat{b}(\tilde{\nu}_m) $
\begin{linenomath}
\begin{multline}
 \langle \bm{\nu},\tilde{\bm{\nu}} | \tilde{\psi}_\mu \rangle =  \frac{ \mathcal{N}(-1)^m}{(m!)^2} \int \diff \omega_1 \int \diff \tilde{\omega}_1 \cdot \ldots \cdot  \int \diff \omega_m \int \diff \tilde{\omega}_m t_\mu (\omega_1,\tilde{\omega}_1) \cdot \ldots \cdot t_\mu (\omega_m,\tilde{\omega}_m) \text{e}^{-\text{i} \varphi (\omega_1+\ldots + \omega_m)} \text{e}^{-\text{i} \vartheta (\tilde{\omega}_1+\ldots + \tilde{\omega}_m)} \text{e}^{-\text{i} m \zeta}\\
 \langle 0| \hat{a}(\nu_1) \hat{b}(\tilde{\nu}_1) \cdot \ldots \cdot \hat{a}(\nu_m) \hat{b}(\tilde{\nu}_m) \hat{a}^\dag(\omega_1) \hat{b}^\dag(\tilde{\omega}_1) \cdot \ldots \cdot \hat{a}^\dag(\omega_m) \hat{b}^\dag(\tilde{\omega}_m)|0\rangle .
\end{multline}
\end{linenomath}
For the expectation value of the ladder operators we find
\begin{linenomath}
\begin{multline}
 \langle 0| \hat{a}(\nu_1) \hat{b}(\tilde{\nu}_1) \cdot \ldots \cdot \hat{a}(\nu_m) \hat{b}(\tilde{\nu}_m) \hat{a}^\dag(\omega_1) \hat{b}^\dag(\tilde{\omega}_1) \cdot \ldots \cdot \hat{a}^\dag(\omega_m) \hat{b}^\dag(\tilde{\omega}_m)|0\rangle = \sum_{\gamma} \delta (\nu_1 - \omega_{\gamma_{1}}) \cdot \ldots \cdot \delta (\nu_m - \omega_{\gamma_{m}}) \\ \times \sum_{\gamma} \delta (\tilde{\nu}_1 - \tilde{\omega}_{\gamma_{1}}) \cdot \ldots \cdot \delta (\tilde{\nu}_m - \tilde{\omega}_{\gamma_{m}}) ,
\end{multline}
\end{linenomath}
where $\sum_\gamma$ means that we sum over all permutations. There are no permutations between signal and idler frequencies $\nu_n \nleftrightarrow \tilde{\nu}_m$, the permutations are only of the form $\nu_n \leftrightarrow \nu_m$ and $\tilde{\nu}_n \leftrightarrow \tilde{\nu}_m$. For these kind of permutations the phases $\text{e}^{-\text{i} \varphi (\omega_1+\ldots + \omega_m)} \text{e}^{-\text{i} \vartheta (\tilde{\omega}_1+\ldots + \tilde{\omega}_m)}$ are invariant and thus for every summand the same. Therefore, they can be factored out and we find
\begin{align}
\langle \bm{\nu},\tilde{\bm{\nu}} | \tilde{\psi}_\mu \rangle \sim \text{e}^{-\text{i} \varphi (\omega_1+\ldots + \omega_m)} \text{e}^{-\text{i} \vartheta (\tilde{\omega}_1+\ldots + \tilde{\omega}_m)} \text{e}^{-\text{i} m \zeta}\sum_{\gamma} t_\mu(\omega_1, \tilde{\omega}_{\gamma_1})\cdot \ldots \cdot  t_\mu(\omega_m, \tilde{\omega}_{\gamma_m}) .
\end{align}
We thus find $|\langle \bm{\nu},\tilde{\bm{\nu}} | \tilde{\psi}_\mu \rangle|^2 = |\langle \bm{\nu},\tilde{\bm{\nu}} | \psi_\mu \rangle|^2$. The proposed measurement is phase insensitive. Thus, both states $|\psi_\mu\rangle$ and $|\tilde{\psi}_\mu\rangle$ give rise to the same probability distribution of measurement outcomes. With this, it immediately follows that $\tilde{F}_q (\mu) = 4 \langle \partial_\mu \psi_\mu | \partial_\mu \psi_\mu \rangle = J_q (\mu)$, where $\tilde{F}_{\mu}(\mu)$ is the FI of the state $|\tilde{\psi}_\mu\rangle$ and the measurement $ \{ |\bm{\nu},\tilde{\bm{\nu}}\rangle\langle \bm{\nu},\tilde{\bm{\nu}} | \}$. Thus, the proposed measurement attains the accuracy given by the QFI $J_q$, however, it does not attain the accuracy given by the QFI $\tilde{J}_q = 4 ( \langle \partial_\mu \tilde{\psi}_\mu | \partial_\mu \tilde{\psi}_\mu \rangle- |\langle \tilde{\psi}_\mu | \partial_\mu \tilde{\psi}_\mu\rangle |^2 )$ due to its phase insensitivity. \\

To demonstrate the validity of the above result, we look at the special case of high frequency entanglement $K\gg 1$ and  low squeezing $|\xi| \ll 1$, where we can explicitly calculate the FI. The state can be approximated as the superposition of the vacuum and a two-photon state 
\begin{equation}
|\psi_\mu \rangle \approx |0\rangle + | \xi | \text{e}^{-\text{i}\zeta} \int \diff \omega \int \diff \tilde{\omega} \mu^{1/2} f(\mu \omega , \tilde{\omega}) \hat{a}^\dag (\omega) \text{e}^{-\text{i}\omega \varphi} \hat{b}^\dag (\tilde{\omega}) \text{e}^{-\text{i}\tilde{\omega} \vartheta} |0\rangle .
\end{equation}
Thus, only zero and two-photon events are relevant, and the corresponding POVM operators are $\lbrace |0\rangle \langle 0|,   \hat{a}^\dag (\omega) \hat{b}^\dag (\tilde{\omega}) |0\rangle \langle 0| \hat{b}(\tilde{\omega}) \hat{a}(\omega )\rbrace $.
For the calculation of the FI, the vacuum term does not contribute, as it does not depend on the parameter. The measurement of the signal and idler frequency corresponds to the probability distribution, the joint spectrum
\begin{equation}
p_\mu (\omega, \tilde{\omega}) = | \langle 0 | \hat{a}(\omega) \hat{b}(\tilde{\omega}) |\psi_\mu \rangle|^2 = \mu \xi^2  f^2 (\mu \omega, \tilde{\omega}) .
\end{equation}
Plugging this into the definition of the FI yields
\begin{align}
F(\mu) &= \int \diff \omega \int \diff \tilde{\omega} \frac{1}{\mu \xi^2 f^2(\mu\omega, \tilde{\omega})} \left( \partial_\mu \left( \mu^{1/2} \xi f(\mu \omega, \tilde{\omega}) \right)^2 \right)^2 \\
&= \xi^2 \int \diff \omega \int \diff \tilde{\omega} \frac{1}{\mu f^2(\mu\omega, \tilde{\omega})} \left( 2 \left( \mu^{1/2} f(\mu \omega,\tilde{\omega}) \right) \partial_\mu \left( \mu^{1/2} f(\mu\omega,\tilde{\omega}) \right) \right)^2 \\
&=  4\xi^2 \int \diff \omega \int \diff \tilde{\omega} \left( \partial_\mu \left( \mu^{1/2} f(\mu\omega, \tilde{\omega}) \right) \right)^2 \\
&= 4\xi^2  \int \diff \omega \int \diff \tilde{\omega}\left( \frac{1}{2\mu^{1/2}}  f (\mu \omega,\tilde{\omega}) + \mu^{1/2} \partial_\mu f (\mu \omega , \tilde{\omega}) \right)^2\\
&=4 \xi^2\int \diff \omega \int \diff \tilde{\omega} f^2(\mu\omega,\tilde{\omega}) \left( \frac{1}{2\mu^{1/2}} - \mu^{1/2} \omega \left( \frac{\omega\mu+\tilde{\omega}- \omega_0}{\sigma^2} + \frac{\omega \mu-\tilde{\omega}}{\epsilon^2}\right) \right)^2 \\
&= \frac{4\xi^2}{\mu^2} \int \diff \omega \int \diff \tilde{\omega} f^2 (\omega,\tilde{\omega}) \left( \frac{1}{4} - \omega \left( \frac{\omega+\tilde{\omega}-\omega_0}{\sigma^2} + \frac{\omega-\tilde{\omega}}{\epsilon^2} \right) + \omega^2 \left( \frac{\omega+\tilde{\omega}-\omega_0}{\sigma^2} + \frac{\omega-\tilde{\omega}}{\epsilon^2} \right)^2 \right) .
\end{align}
Using the substitution $u = \omega +\tilde{\omega}$ , $v= \omega-\tilde{\omega}$, with  $2 \omega = u +v$, the value of the Jacobi determinant $1/2$ and the fact that $f^2(\omega (u,v), \tilde{\omega}(u,v)) = \frac{2}{\uppi \sigma \epsilon} \exp (- (u-\omega_0)^2/\sigma^2) \exp (- v^2/\sigma^2)$  we can compute the integral using well-known identities for Gaussian integrals 
\begin{align*}
F(\mu)&= \frac{4\xi^2}{\mu^2} \int \diff u \int \diff v  \frac{1}{\uppi \sigma\epsilon} \exp \left(- \frac{\left( u - \omega_0\right)^2}{\sigma^2} \right) \exp \left( - \frac{v^2}{\epsilon^2} \right) \left( \frac{1}{4}- \frac{u+v}{2} \left( \frac{u-\omega_0}{\sigma^2} + \frac{v}{\epsilon^2} \right) + \frac{(u+v)^2}{4}\left(\frac{u-\omega_0}{\sigma^2} + \frac{v}{\epsilon^2} \right)^2 \right) \\
&= \frac{4\xi^2}{\mu^2} \left(\left( \frac{1}{4} \right) + \left(- \frac{1}{2} \right) + \left( \frac{5}{2\cdot 4}+ \frac{\omega_0^2 K}{4 \sigma\epsilon} + \frac{K^2}{4} - \frac{1}{2\cdot4} \right) \right) \\
&=\frac{\xi^2}{\mu^2} \left( \frac{\omega_0^2 K}{\sigma\epsilon}+ K^2 +1 \right) \\
&\approx \frac{\xi^2 K}{\mu^2}  \left( \frac{\omega_0^2}{\sigma\epsilon} + K \right) .
\end{align*}
The FI indeed coincides with $J_q (\mu)$, as expected.

\end{widetext} 

\end{document}